\documentclass[twocol]{ametsoc}
\usepackage[mathscr]{eucal}
\usepackage{ulem}

\journal{jpo}

\bibpunct{(}{)}{;}{a}{}{,}

\newcommand{\beq}{\begin{equation}}
\newcommand{\eeq}{\end{equation}}

\newcommand{\defn}{\ensuremath{\stackrel{\mathrm{def}}{=}}}

\newcommand{\com}{\, ,}
\newcommand{\per}{\, .}

\newcommand{\ybjp}{YBJ$^{+}$}

\newcommand{\p}{\partial}

\newcommand{\WE}{\mathrm{WE}}

\newcommand{\laa}{\left \langle}
\newcommand{\raa}{\right \rangle}

\newcommand{\half}{\tfrac{1}{2}}

\newcommand{\bnabla}{\boldsymbol{\nabla}}
\newcommand{\bcdot}{\boldsymbol{\cdot}}

\newcommand{\grad}{\bnabla \hspace{-0.1em}}

\newcommand{\lap}{\triangle}

\newcommand{\bU}{\mathbf{U}}

\newcommand{\bk}{\boldsymbol{k}}

\newcommand{\ii}{\mathrm{i}}
\newcommand{\dd}{\mathrm{d}}
\newcommand{\ddg}{\mathrm{d}_g}
\newcommand{\ee}{\mathrm{e}}

\newcommand{\Um}{U}

\newcommand{\LL}{\mathsf{L}}
\newcommand{\LLp}{\mathsf{L}^{\!\!\scriptscriptstyle+}\!}

\newcommand{\pss}{\text{s$^{-2}$}}

\newcommand{\halfi}{\tfrac{\ii}{2}}

\usepackage{float}
\usepackage{mwe,tikz}\usepackage[percent]{overpic}

\title{Refraction and straining of wind-generated near-inertial waves by barotropic eddies}

\authors{Olivier Asselin\correspondingauthor{Olivier Asselin, Keck 254, Scripps Institution of Oceanography, University of California San Diego, La Jolla, CA 90293-0213, USA}$^1$, Leif N. Thomas$^2$, William R. Young$^1$ and Luc Rainville$^3$}

\affiliation{
$^1$Scripps Institution of Oceanography, University of California San Diego \\
$^2$Department of Earth System Science, Stanford University \\
$^3$Applied Physics Laboratory, University of Washington}
\email{oasselin@ucsd.edu}

\abstract{We analyze the distortion of wind-generated near-inertial waves by steady and unsteady barotropic quasi-geostrophic eddies, with a focus on the evolution of the horizontal wavevector $\bk$ under the effects of mesoscale strain and refraction. The model is initialized with a horizontally-uniform ($\bk=0$) surface-confined near-inertial wave which then evolves according to the phase-averaged model of Young and Ben Jelloul.  A steady barotropic vortex dipole is first considered. Nearly monochromatic shear bands appear in the jet region as wave energy propagate downwards and towards anticyclone. As a result of refraction,  both  horizontal and vertical wavenumbers grow linearly with the time $t$ elapsed since generation such that their ratio, the slope of wave bands, is time-indepedent. Analogy with passive scalar dynamics suggests that strain should result in the exponential growth of $|\bk|$. Here instead, strain is ineffective not only at the jet center, but also at its confluent and diffluent regions. Low modes rapidly escape below the anticyclonic core such that the weakly-dispersive high modes are dominant in the mixed layer. In the weakly-dispersive limit, $\bk=- t \grad \zeta(x,y,t)/2$ provided that ($i$) the eddy vertical vorticity $\zeta$ evolves according to the barotropic quasi-geostrophic equation; and ($ii$) $\bk=0$ initially, as is typically assumed for near-inertial waves generated by large-scale atmospheric storms. In steady flows, strain is ineffective because $\bk$ is always perpendicular to the flow. In unsteady flows, straining modifies the vorticity gradient and hence $\bk$, and may account for significant energy transfers.}

\begin{document}

\maketitle

\section{Introduction}

Atmospheric storms sweeping across the ocean resonantly excite near-inertial waves, or internal waves oscillating at a frequency close to $f$, the Coriolis frequency \citep{alford2016review}. These waves originate within a shallow, $O(100)$-m deep surface mixed-layer, but with the large $O(1000)$-km horizontal scale characteristic of synoptic pressure systems \citep{pollard1980,thomson1981wind}. Given such anisotropic primordial scales, wind-generated near-inertial waves are inefficient at radiating their energy and shear into the ocean interior \citep{gill1984}. A reduction in  the horizontal length  scale of the wave is necessary to increase the vertical group velocity and enable penetration into the ocean interior.

Three contender scale-reduction mechanisms are summarized on the right hand side of the ray-tracing formula for the evolution of the horizontal wavenumber $\bk=(k,l)$:
\begin{align}
\frac{\ddg}{\dd t} \begin{pmatrix} k\\ l \end{pmatrix} & = 
	\!\!-\!\!\! \underbrace{\beta \begin{pmatrix}0 \\ 1
		\end{pmatrix} }_{\text{$\beta$--refraction}}  
	\!\!- \!\! \underbrace{\half \begin{pmatrix} \zeta_x \\ \zeta_y
		\end{pmatrix} }_{\text{$\zeta$--refraction}} 
\!\!-  \underbrace{\begin{pmatrix}  U_x & V_x \\ U_y & V_y\end{pmatrix} \begin{pmatrix} k \\ l \end{pmatrix}}_{\text{strain}}\per 
\label{eq1}  
\end{align}
In \eqref{eq1},  $\ddg/\dd t$ is a derivative following the group velocity. The three processes on the right-hand side  --- $\beta$-refraction, $\zeta$-refraction and strain --- tend to increase $|\bk|$, \textit{i.e.}, decrease the horizontal length scale of a freshly generated near-inertial wave.

The Ocean Storms Experiment \citep{dasaro1989,dasaro1995} provided observational evidence that the latitudinal variation in the Coriolis frequency, $\beta = \dd f/\dd y$, leads to a linear growth of the meridional wavenumber $l$. In \eqref{eq1}, this corresponds to a dominant balance in which $\beta$-refraction is the main term on the right, implying $l=l_0 - \beta t$. In physical terms, the southernmost and northernmost portions of the primordial wave experience slightly  different inertial frequencies. Over time, this frequency shift results in phase decoherence in the meridional direction, which is equivalent to an increase of $l$.

Gradients in the vertical vorticity, $\zeta = V_x-U_y$, with $(U,V)$ the horizontal velocity of  mesoscale eddies,  can also reduce the initial horizontal wave scale \citep{kunze1985,ybj}.   Near-inertial waves embedded in a field of eddies experience different rotation rates in cyclonic and anticyclonic regions, such that the wave phase  acquires the 10-100 km scale of eddies. 
This   process is $\zeta$-refraction in \eqref{eq1}. For spatially-uniform and steady vorticity gradients, $\zeta$-refraction also produces a linear growth of the wavevector, $\bk=\bk_0 - \half t \nabla \zeta$. Because mesoscale eddies typically have  $|\grad \zeta| \gg \beta$, one expects that $\zeta$-refraction is more rapid than $\beta$-refraction \citep{van1998interactions}.  For this reason we shall assume constant planetary vorticity $f$ and neglect $\beta$-refraction throughout this paper. 


Differential advection by mesoscale eddies stretches and rotates the  wavevector $\bk$. This process is captured by  the strain term in \eqref{eq1}. By analogy with passive-scalar advection, one expects strain to cause an exponential growth of $|\bk|$ \citep{jones1969}. In strain-dominated regions, and with sufficient  vertical shear, the horizontal and vertical wavenumbers both grow exponentially and the group velocity goes to zero. Provided uniform velocity gradients along a ray, waves are captured and strained into oblivion \citep{bm2005}. \cite{polzin2008mesoscale} argues that the vertical profiles of horizontal velocity observed during  the Mid-Ocean Dynamics Experiment are consistent with wave capture by mesoscale strain.

What are the respective roles of strain and $\zeta$-refraction in shaping $\bk$  in a generic geostrophic flow in which the two processes are at play? Naively, one would expect strong-enough strain to  dominate $\zeta$-refraction because strain produces  exponential--in--time growth of $|\bk|$, while refraction results only in  linear--in--time growth. We show here, however, that this expectation may be violated in the important case of a wind-generated near-inertial wave, for which the initial horizontal scale is large.

This insight emerged from the Near-Inertial Shear and Kinetic Energy experiment (NIKSINe), a research initiative funded by the US Office of Naval Research with a field work component focused in a region located about 500 km south of Iceland. In a companion paper, \cite{leifobs} provide observational evidence of the $\zeta$-refraction of a wind-generated near-inertial wave in a barotropic vortex dipole. A few inertial periods after the wind event, the phase differences in the inertial velocities are consistent with a linear growth of $\bk$ at a rate comparable to half the local vorticity gradient. This is despite the fact that ballpark estimates indicate that  strain is strong enough to produce an exponential growth of $|\bk|$. Why does strain appear to be ineffective in the NISKINe dipole? Furthermore, mesoscale vorticity in the NISKINe region undergoes significant changes over a few inertial periods. How do these rapid changes in mesoscale vorticity  variations affect $\zeta$-refraction and strain?

\section{Problem formulation} \label{sec:prob}

\begin{figure*}
\centering
\includegraphics[trim = 60 0 50 0, clip, width=.4\textwidth]{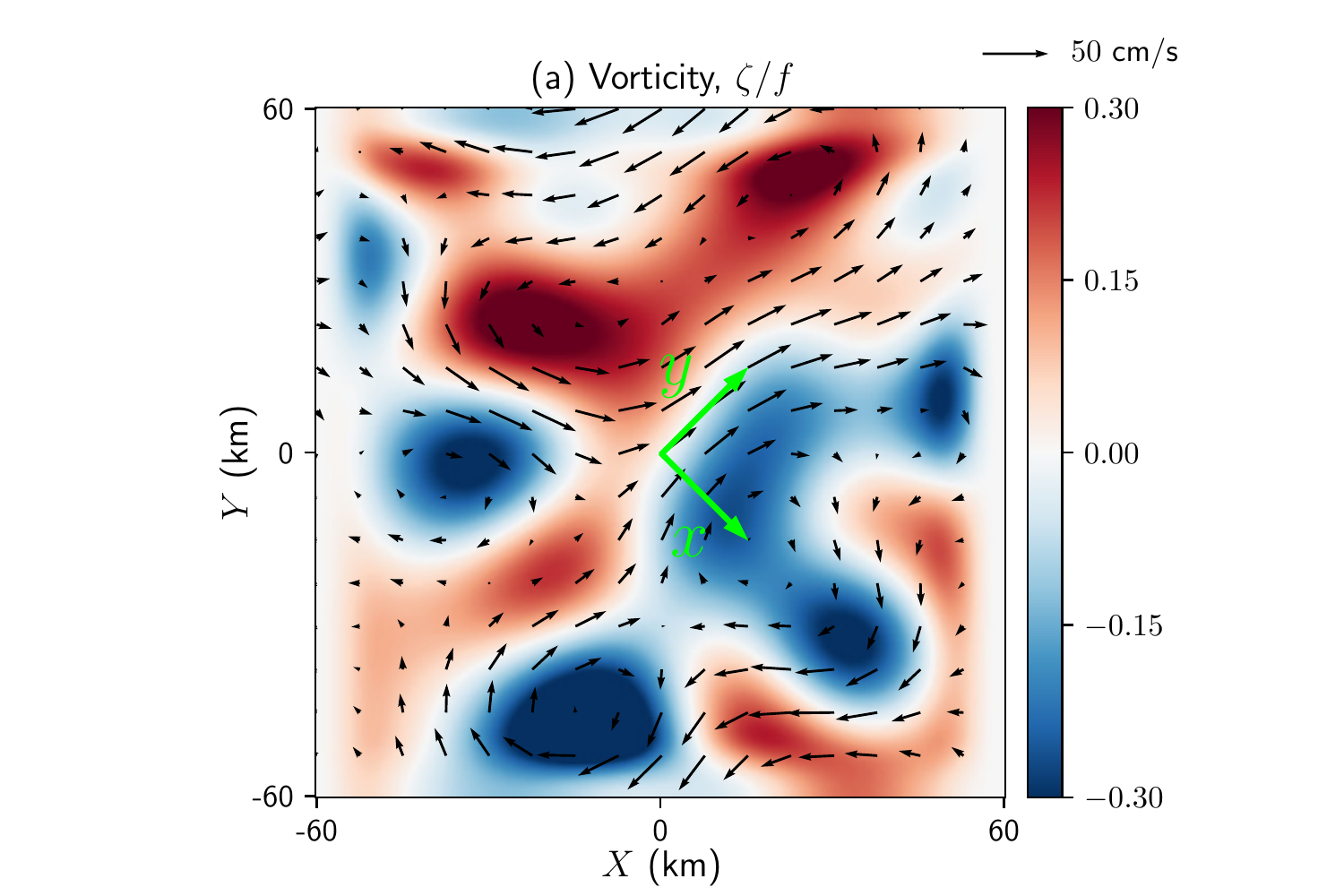} \qquad
\includegraphics[trim = 60 0 50 0, clip, width=.4\textwidth]{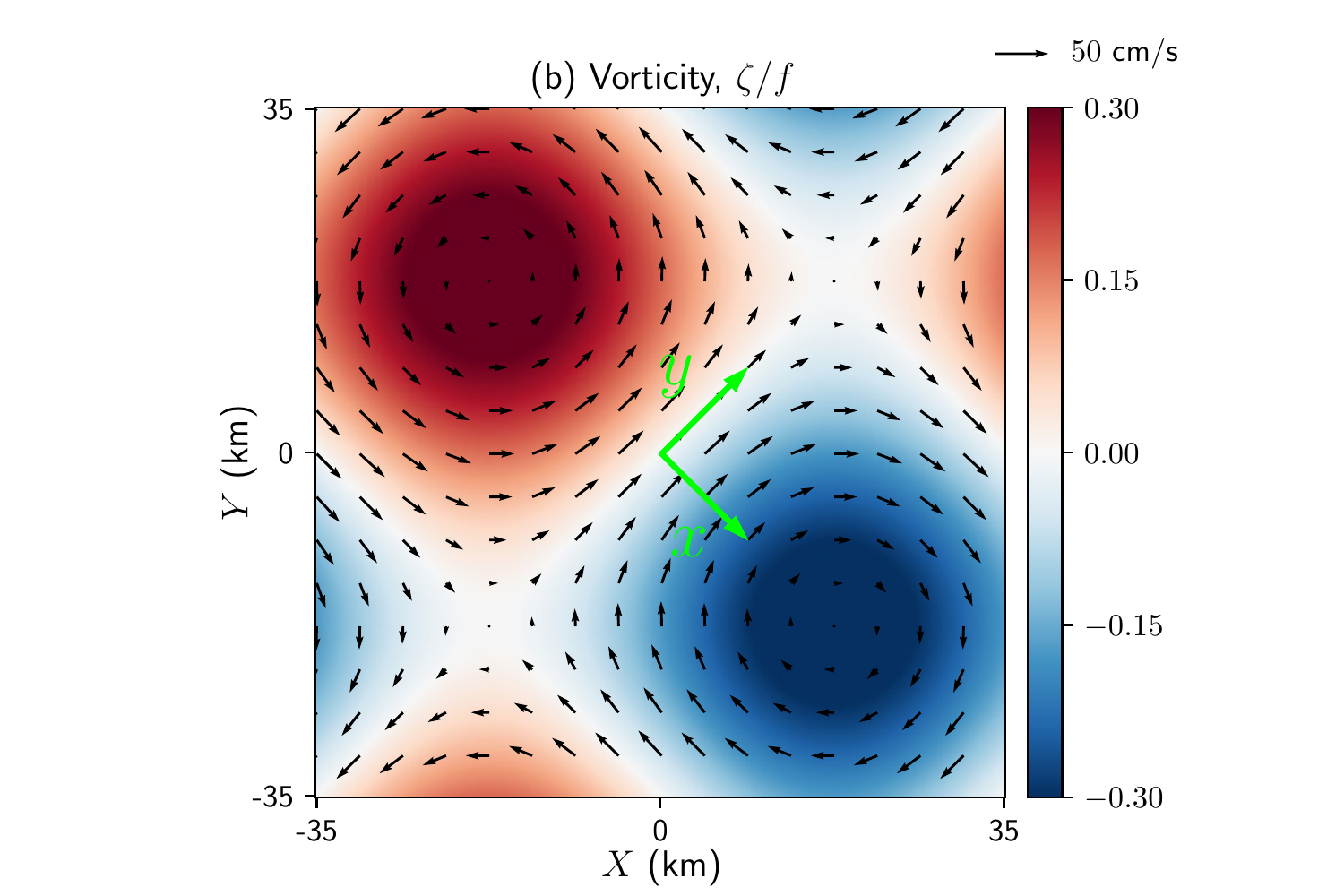}
\caption{Normalized vertical relative vorticity map from NISKINe (left) and its idealized, dipole version (right). The rotated axes, $(x,y)$,  used for analysis are shown in green. The relation between the two coordinate  sets is $(X,Y)= (x+y,y-x)/\sqrt{2}$.}
        \label{fig:vort}
\end{figure*}

\subsection{Flow setup}

Figure \ref{fig:vort}a  shows an estimate of the horizontally non-divergent flow observed during the NISKINe cruise based on satellite altimetry refined by in-situ velocity measurements from both ship-mounted and drifting instruments \citep{leifobs}. Vertical profiles reveal a surprisingly uniform flow vertical structure in the top several hundred meters of the ocean, motivating the assumption of barotropic flow throughout this paper. Analysis of wave evolution in this complex flow is confined to section \ref{sec:all} of this paper and the companion paper \citep{leifobs}. In the next few sections, we limit attention to an idealized, dipole model of the NISKINe flow shown in  figure \ref{fig:vort}b. 

The dipole  is best defined using a coordinate system that is rotated by 45 degrees relative to the cardinal directions, represented by the $(X,Y)$-axes in figure \ref{fig:vort}. This rotated $(x,y)$-coordinate system is shown by the green axes in figure  \ref{fig:vort}. The $x$-axis is anti-parallel to the vorticity gradient at the origin. In terms of the rotated coordinates  the dipole streamfunction is 
\beq
\psi = \Um \kappa^{-1} \sin \kappa x \, \cos \kappa  y\per
\eeq
The vorticity is $\zeta = - 2 \kappa^2 \psi$, or
\beq
\zeta =-  \gamma \kappa^{-1}  \sin \kappa x \, \cos \kappa  y\com \label{vort}
\eeq
where $\gamma \defn 2 \kappa^2 \Um$.
The scales of the dipole flow are set to fit both the observed velocity maximum at the center of the NISKINe jet, $\Um =  33.5$ cm/s, and the observed maximum  vorticity gradient, $\gamma = 2.7 \times 10^{-9}$ (ms)$^{-1}$. This  requires $\kappa = \sqrt{2} \pi/ 70$ km.

NISKINe is  near 58.5$^\circ$N, such that $\beta \sim 10^{-11}$ m$^{-1}$ s$^{-1}$ is two orders of magnitude smaller than the maximum relative vorticity gradient, $\gamma$. We thus neglect the latitudinal dependence of the Coriolis frequency and set $f=1.24 \times 10^{-4}$ s$^{-1}$. In the $(X,Y)$-coordinate system used for the axes in figure \ref{fig:vort}  the   domain is horizontally-periodic with equal east-west ($X$) and north-south ($Y$) dimensions of $70$ km. The flow is  barotropic and extends down to the ocean bottom at $H = 3$ km depth. Stratification is  vertically uniform. Unless otherwise specified, we use $N^2 = 10^{-5}$ s$^{-2}$.


%
%
%
%
%
%

\subsection{Waves}

To setup the wave part of the model, we assume that the horizontal scale of the atmospheric storm is much larger than the domain. Furthermore, the momentum imparted by the storm is assumed to be rapidly distributed over the depth of the mixed layer, $\sigma = 30$ meters. Mathematically this translates into an initial value problem for the waves, with a perfect inertial oscillation initially confined to the mixed layer:
\beq
u(t=0) = u_0 \exp{(-z^2/\sigma^2}), \qquad v(t=0) = 0. \label{wave_IC}
\eeq
Without loss of generality, the inertial motion begins with an eastward velocity $u_0 = 10$ cm s$^{-1}$. As we are dealing with waves near the inertial frequency, it is insightful to express wave variables in terms of the back rotated velocity,
\beq
\LL A = (u+iv) \, \ee^{ift},
\eeq 
where $A(x,y,z,t)$ is a space- and time-dependent complex field and 
\beq
\LL = \frac{\partial}{\partial_z} \left(  \frac{f^2}{N^2} \frac{\partial}{\partial_z}  \right)
\eeq
is a frequently-occurring operator. Thus, $\LL A$ is the slowly-evolving envelope of the near-inertial wave, and its leading-order evolution is dictated by the YBJ equation \citep{ybj},
\beq
\p_t\LL A + J(\psi,\LL A) +  i \left( \beta y+ \frac{\zeta}{2}  \right) \LL A+ \frac{\ii f}{2} \lap A   =0 \com 
\label{ybj}
\eeq
where $J(a,b)\defn a_x b_y - a_y b_x$ is the Jacobian and $\lap \defn \partial_x^2 + \partial_y^2$ is the horizontal Laplacian. From left to right, the YBJ equation expresses the changes to the near-inertial wave envelope due to advection by the mean flow, refraction by planetary and relative vorticity, and dispersion. 

\subsection{Numerics}

Although the analysis is carried in the classic YBJ framework, numerical integrations are performed using the \ybjp{} equation \citep{ybjp}. This equation is identical to \eqref{ybj}, except that the operator $\LL$ is replaced with
\beq
\LLp \defn \LL + \tfrac{1}{4} \lap.
\eeq
This tweak in the definition of the wave envelope brings the twin advantages of higher accuracy and lower computational effort whilst maintaining ease of implementation \citep{ybjp}. 

The \ybjp{} model is pseudo-spectral in the $x$ and $y$ directions, allowing horizontal derivatives to be computed with spectral accuracy. The 2/3 rule is used to remove aliased modes \citep{durran2013numerical}. Vertical derivatives are approximated with second-order centered finite differences. Time integration is accomplished with the leap-frog scheme with weak time diffusion \citep{asselin1972}. Weak horizontal hyperdiffusion is applied to filter out sub-grid horizontal wave scales.

\section{The dipole solution} \label{sec:dipole}

\begin{figure*}[h!]
\centering
\includegraphics[trim = 60 0 60 0, clip, width=.32\textwidth]{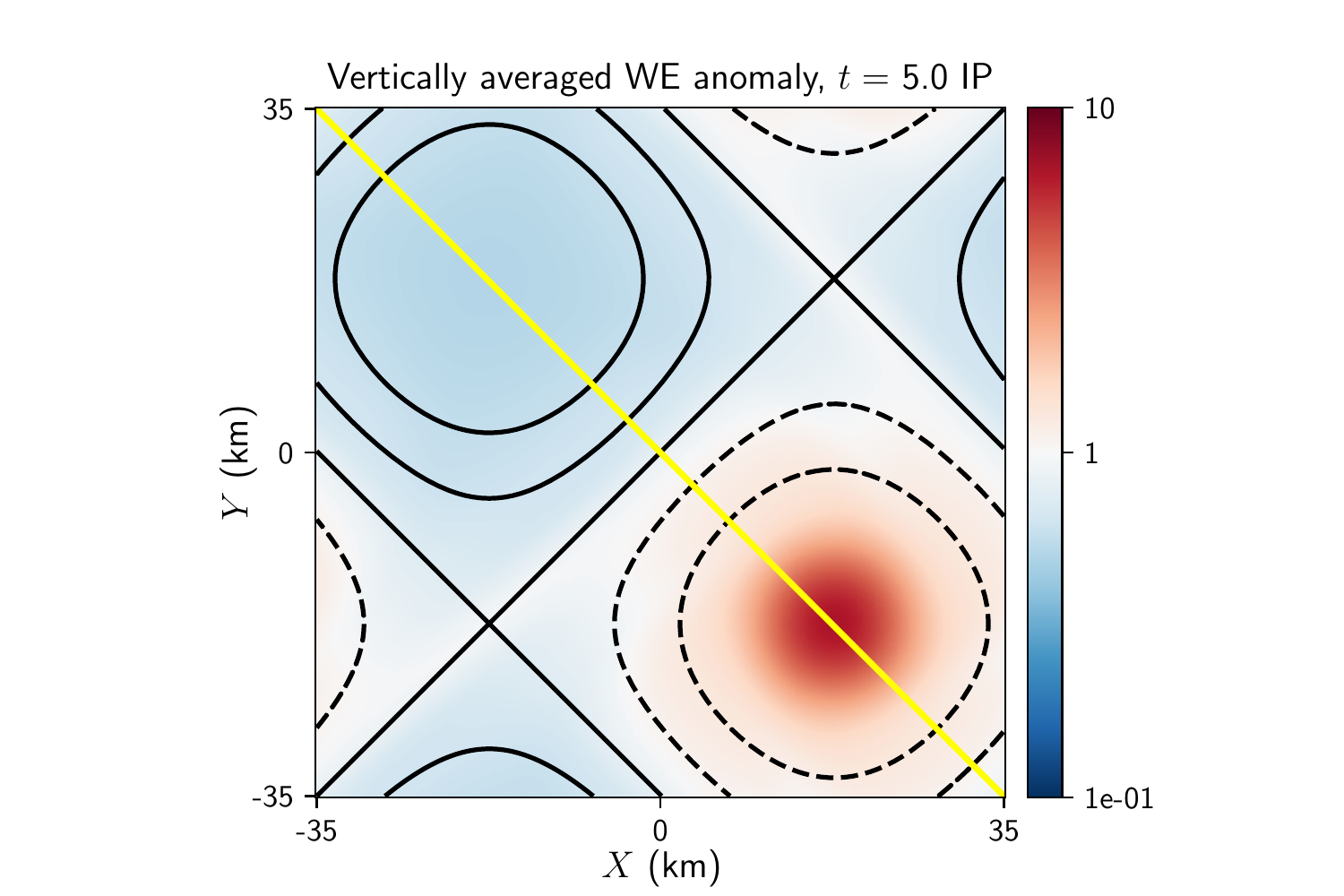}
\includegraphics[trim = 60 0 60 0, clip, width=.32\textwidth]{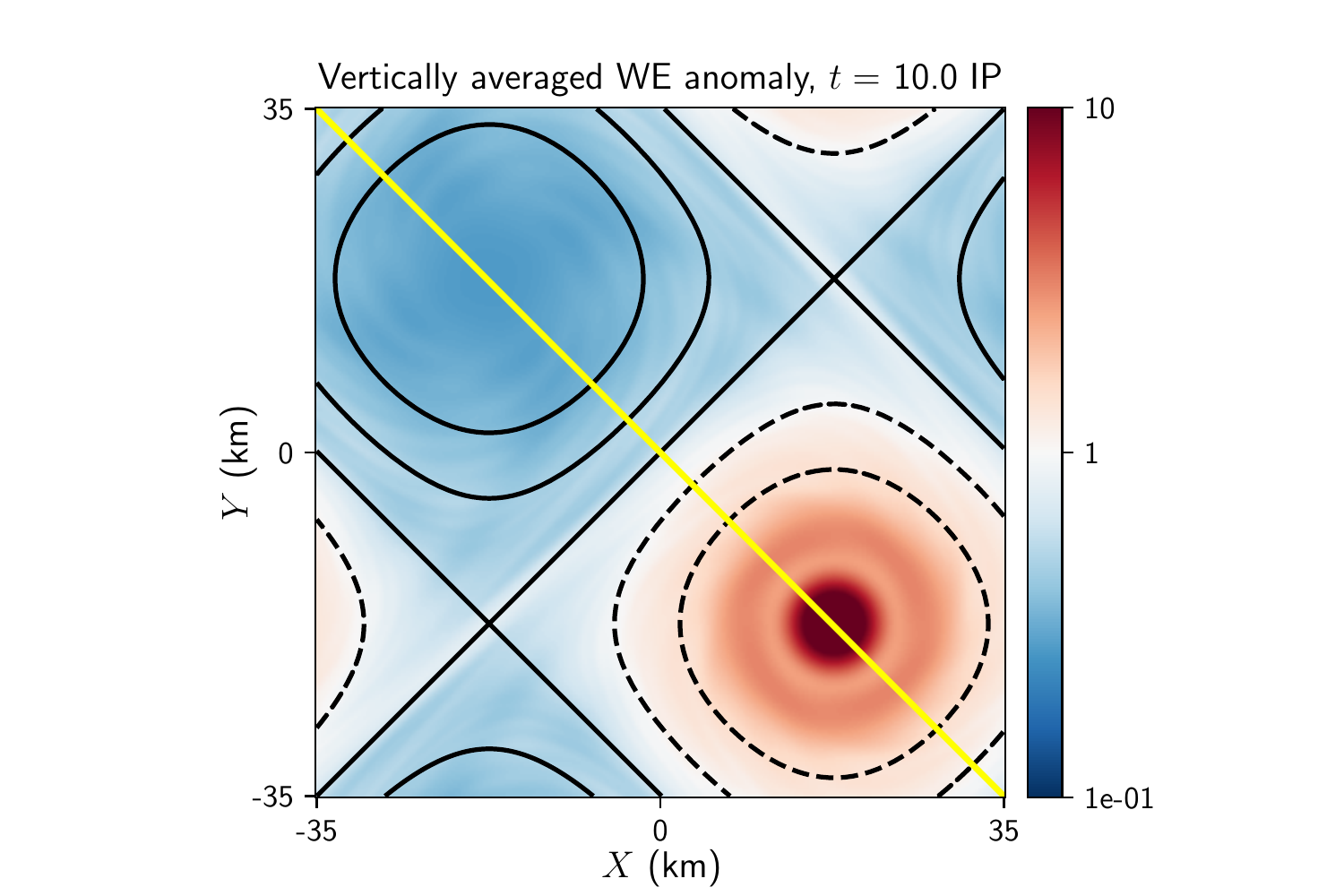}
\includegraphics[trim = 60 0 60 0, clip, width=.32\textwidth]{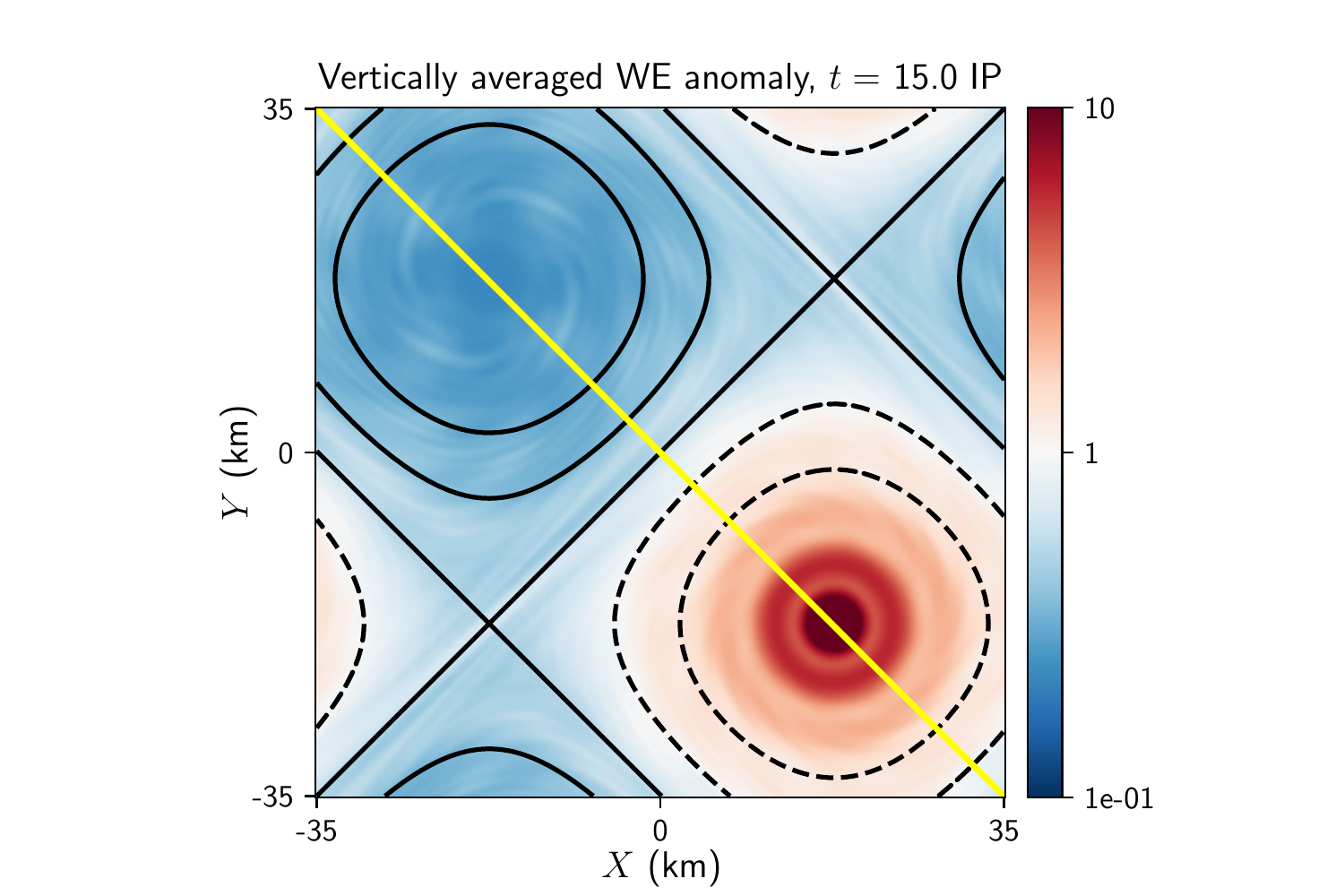}
\includegraphics[width=.32\textwidth]{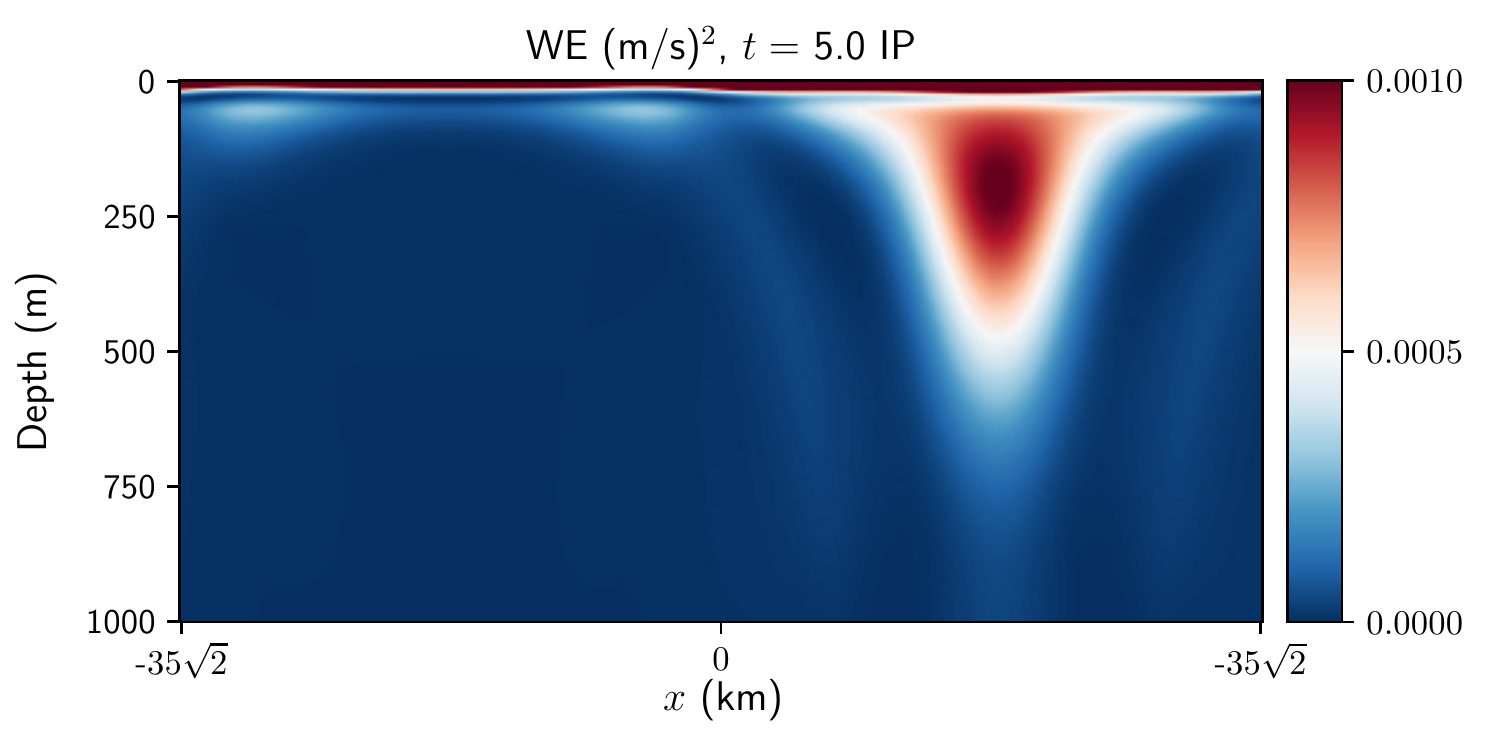}
\includegraphics[width=.32\textwidth]{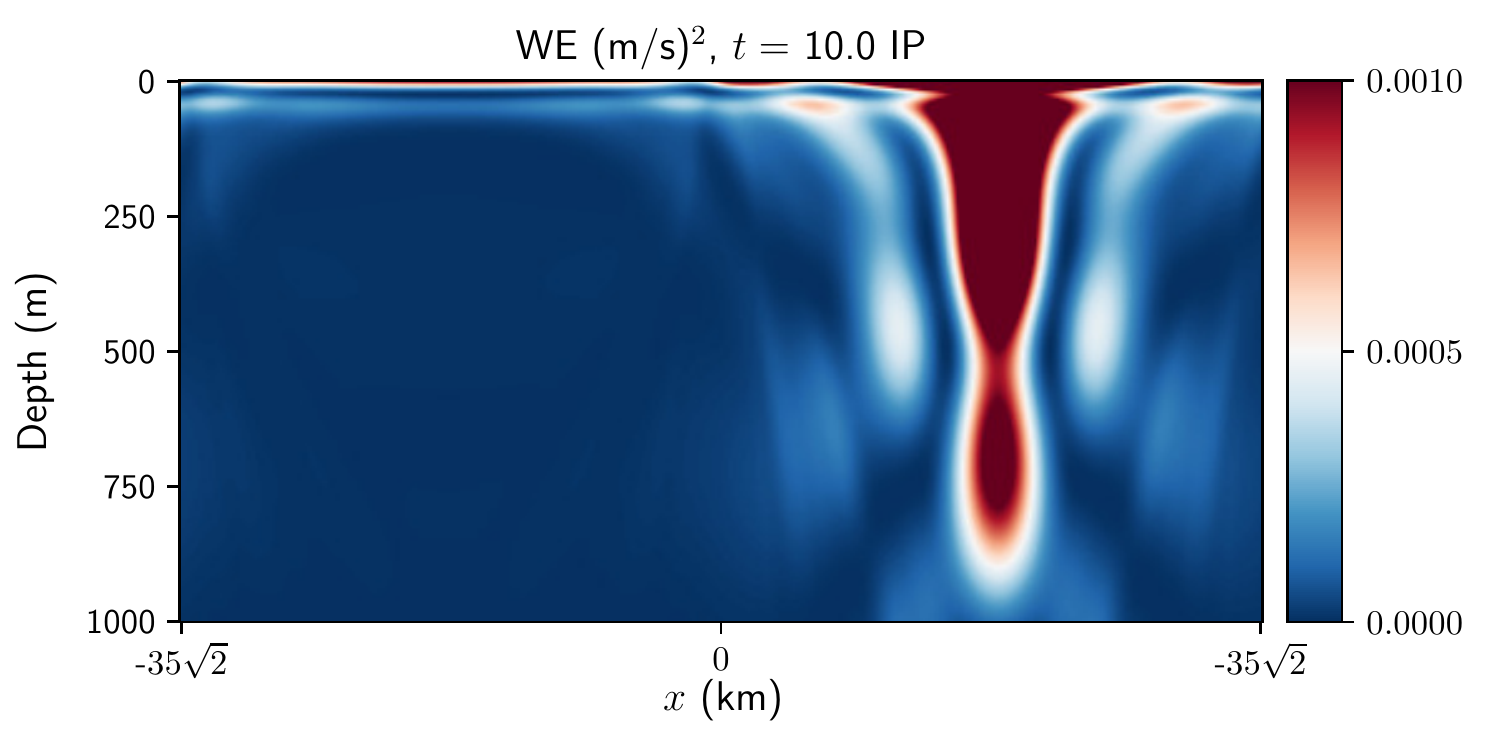}
\includegraphics[width=.32\textwidth]{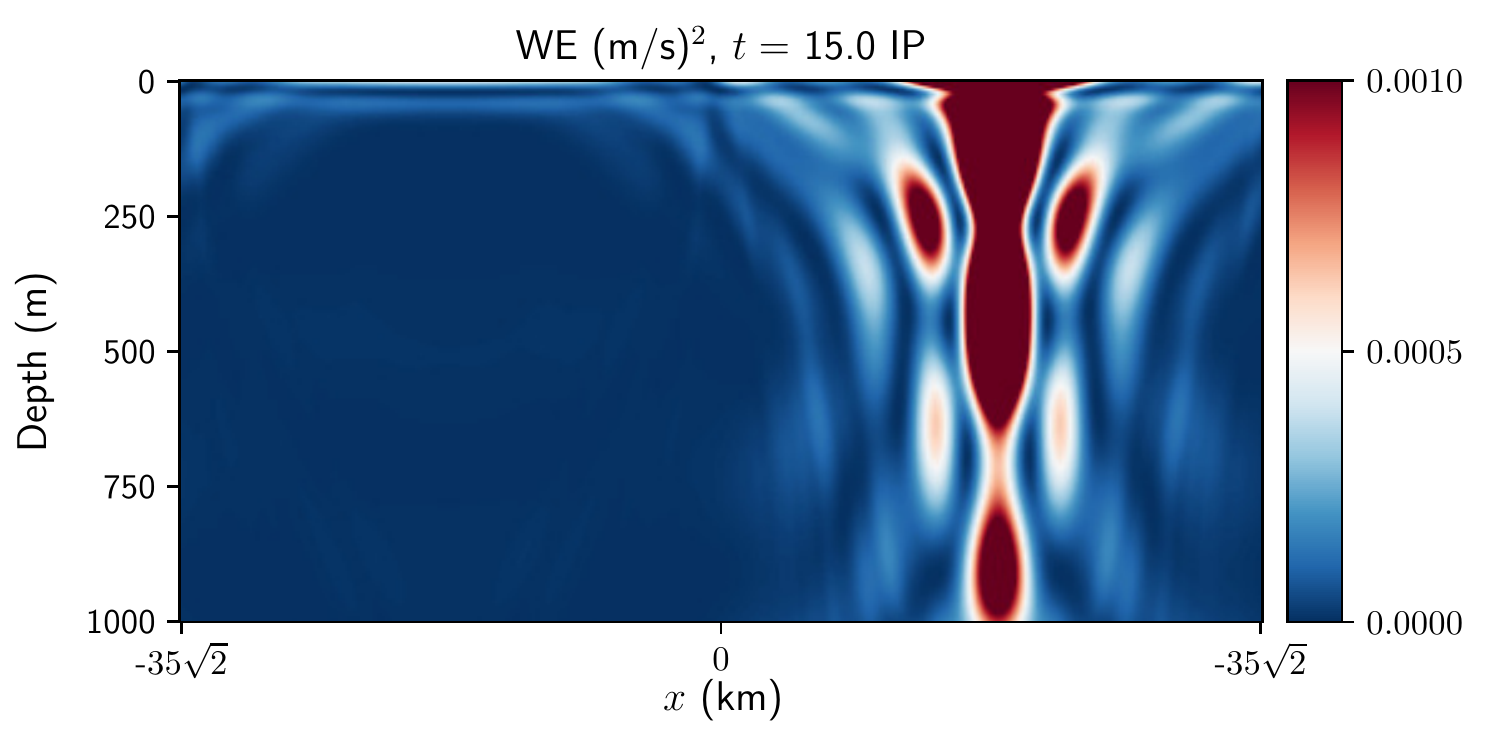}
\includegraphics[trim = 0 0 5 0, clip, width=.32\textwidth]{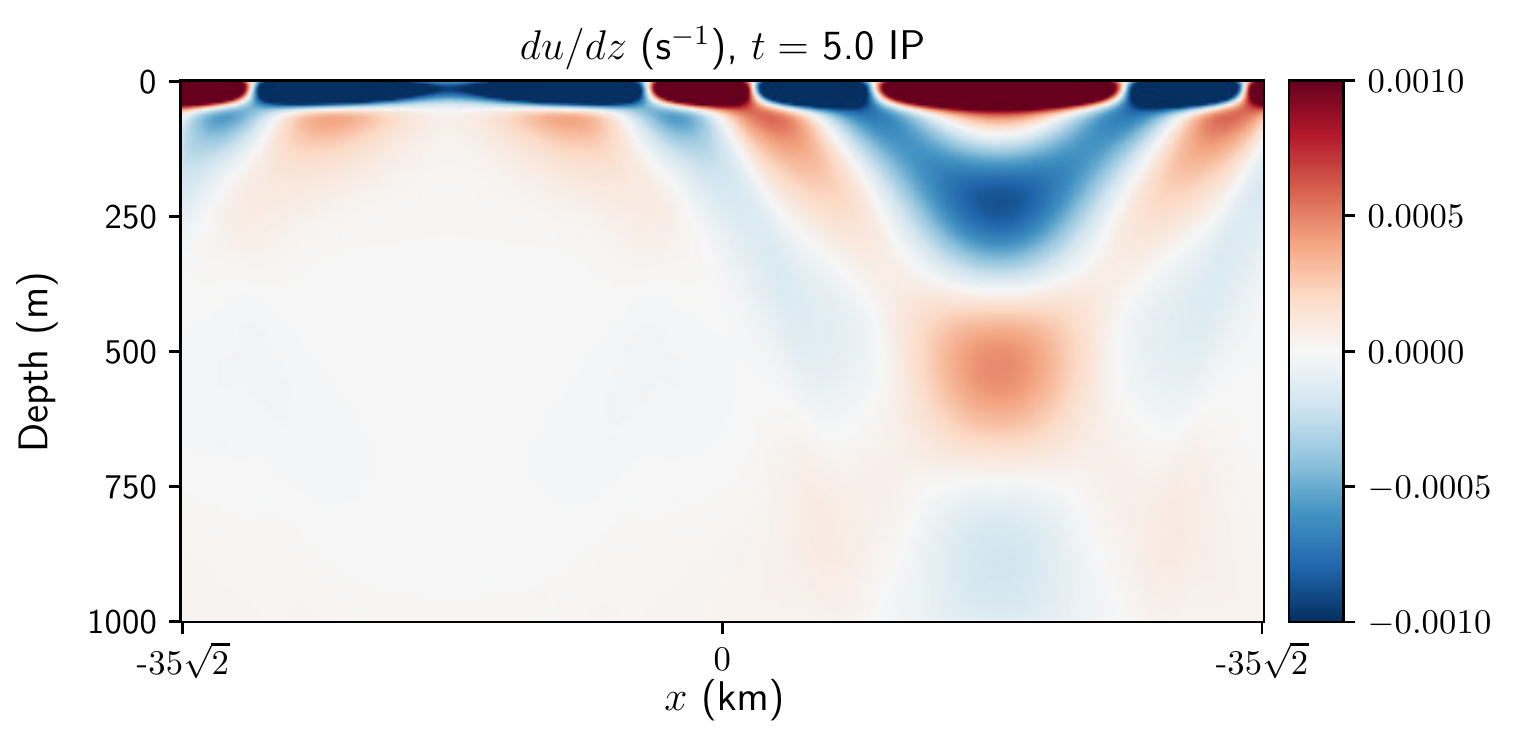}
\includegraphics[trim = 0 0 5 0, clip, width=.32\textwidth]{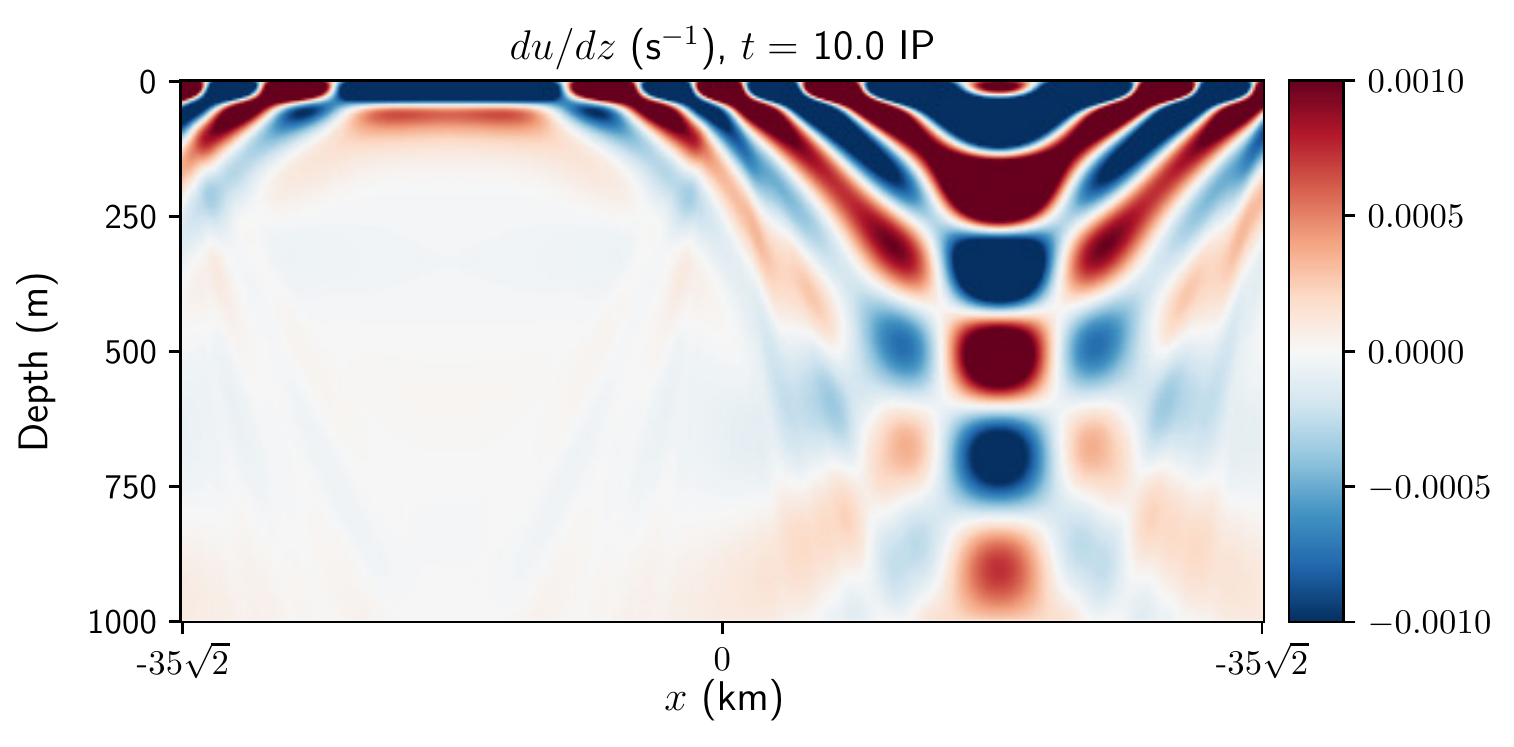}
\includegraphics[trim = 0 0 5 0, clip, width=.32\textwidth]{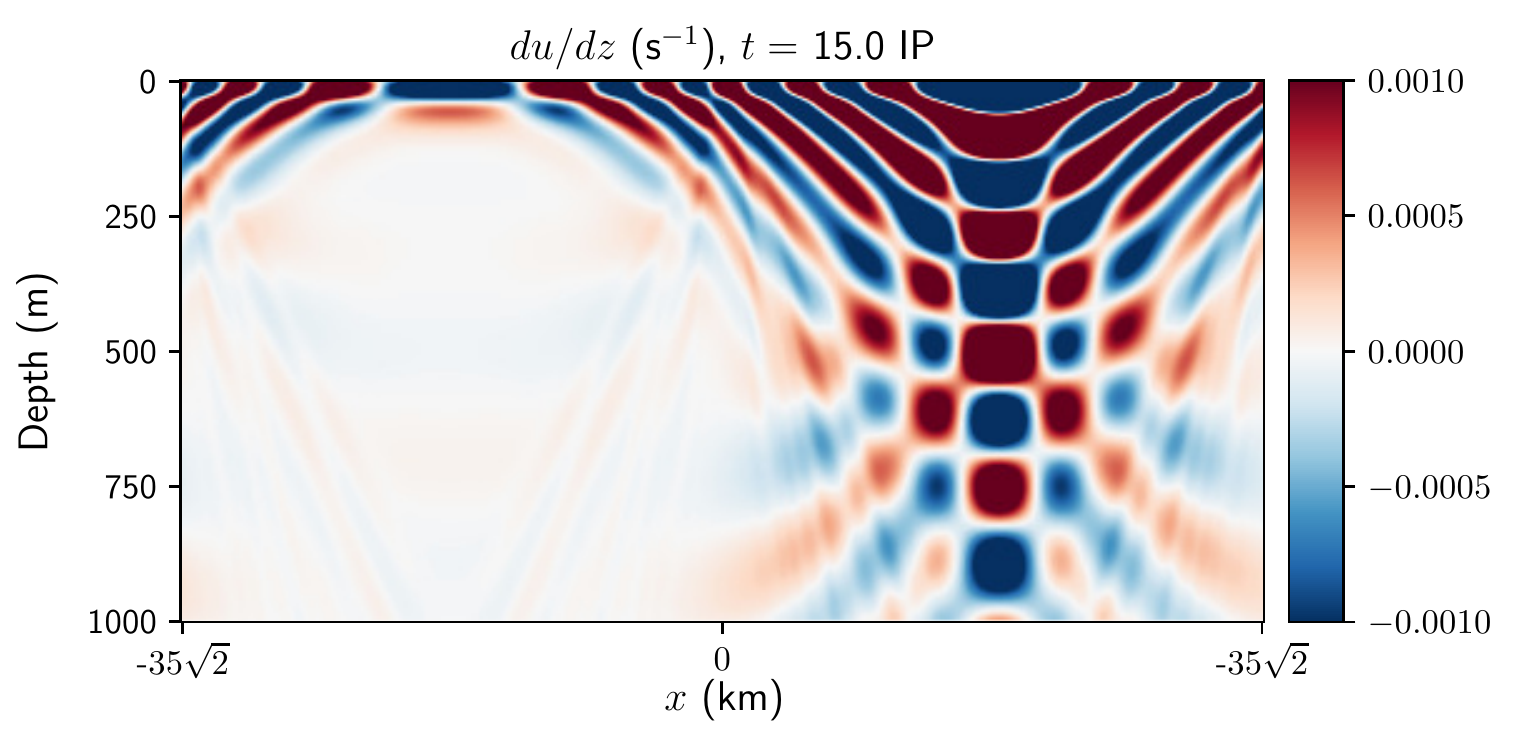}
\caption{Case: dipole with $N^2= 10^{-5}$ s$^{-2}$. Top: vertically-averaged wave energy (WE) anomaly in logarithmic scale with vorticity contours. Middle: vertical cuts of WE along the yellow line drawn in top panels ($y=0$). Bottom: vertical cuts of eastward wave velocity shear, also along the yellow transect. Time increases from left to right: 5, 10 and 15 inertial periods.}
\label{fig:dipole_overview}
\end{figure*}


We begin by presenting the general appearance of the dipole solution. 
The vertically-averaged wave energy anomaly is shown in the top panels of figure \ref{fig:dipole_overview}.
Contours denote vorticity, and columns are associated with wave fields after 5, 10 and 15 inertial periods (IP). The most striking feature is the strong accumulation of wave energy in the negative vortex core (note the logarithmic scale). After a few inertial periods there is little energy left in the cyclone (solid contours). This rapid attraction of wave energy by anticyclones has been observed repeatedly \citep{weller1985,kunze1986,kunze1995,elipot}. Explanations have been proposed relying on the broadening of the allowable frequency band in negative vorticity regions \citep{kunze1985}, appeals to the quantum analogy between energy wells and negative vorticity \citep{balmforth1998,cesar}, and a conservation law for steady barotropic flows \citep{DVB2015}. 

The middle panels of figure \ref{fig:dipole_overview} show vertical slices of wave energy density in the upper kilometer. Transects are along the $y=0$ axis, marked by a yellow line in the upper panels. Again, wave energy gathers along the anticyclone axis, whose core is located at $ x \approx 25$ km. As time passes (left to right panels), wave energy is drained down the anticyclonic pipe \citep{leeniiler,ay2020}.

The lower panels show the vertical shear of horizontal wave velocity along the same vertical slices as above. These shear plots reveal the presence of nearly monochromatic wave bands which are symmetric about the anticyclonic axis. Phase lines propagate upwards as wave energy propagation down and towards the anticyclone. The bands get narrower with time while keeping a time-independent slope. They also appear to steepen with depth. 


\begin{figure*}[h!]
\centering
\includegraphics[trim = 0 0 5 0, clip, width=.32\textwidth]{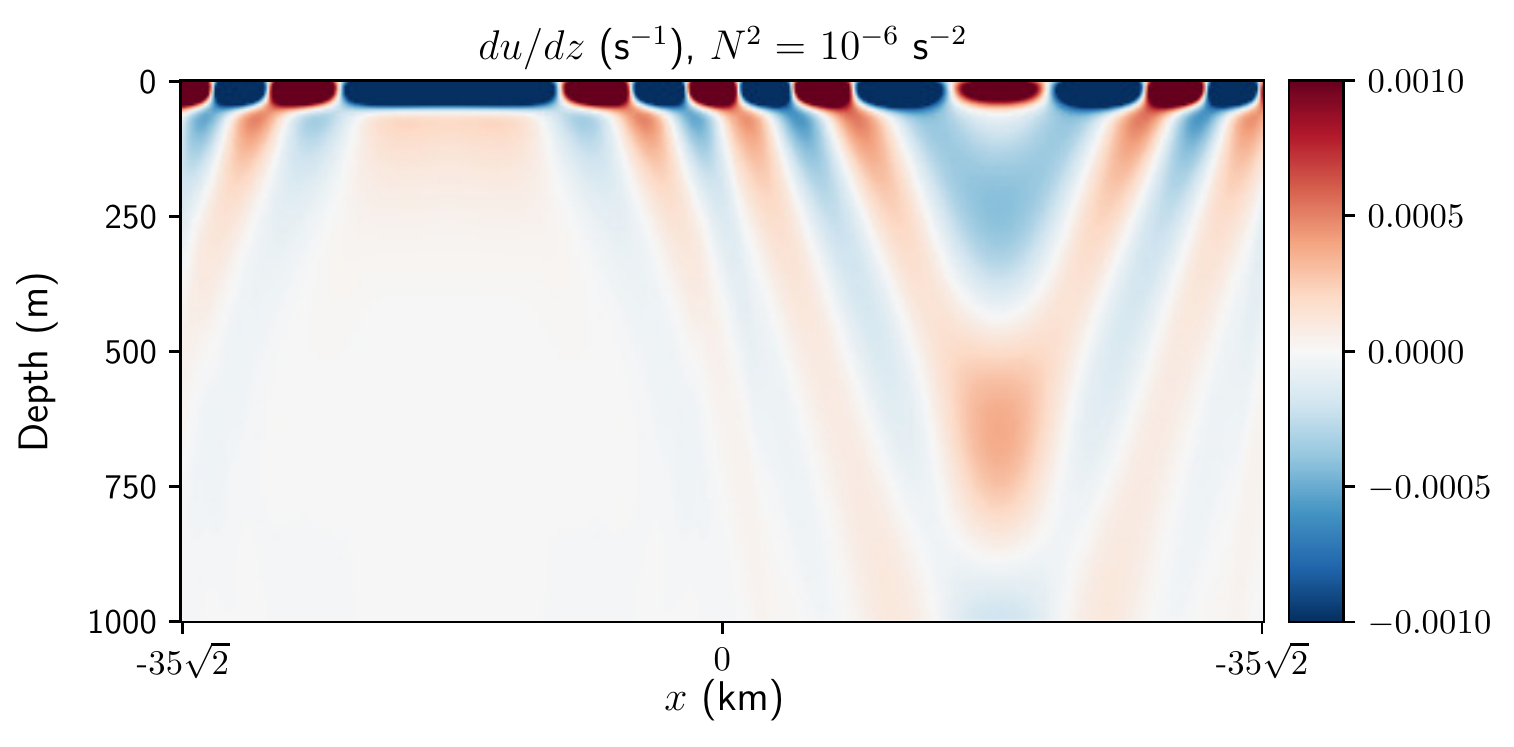}
\includegraphics[trim = 0 0 5 0, clip, width=.32\textwidth]{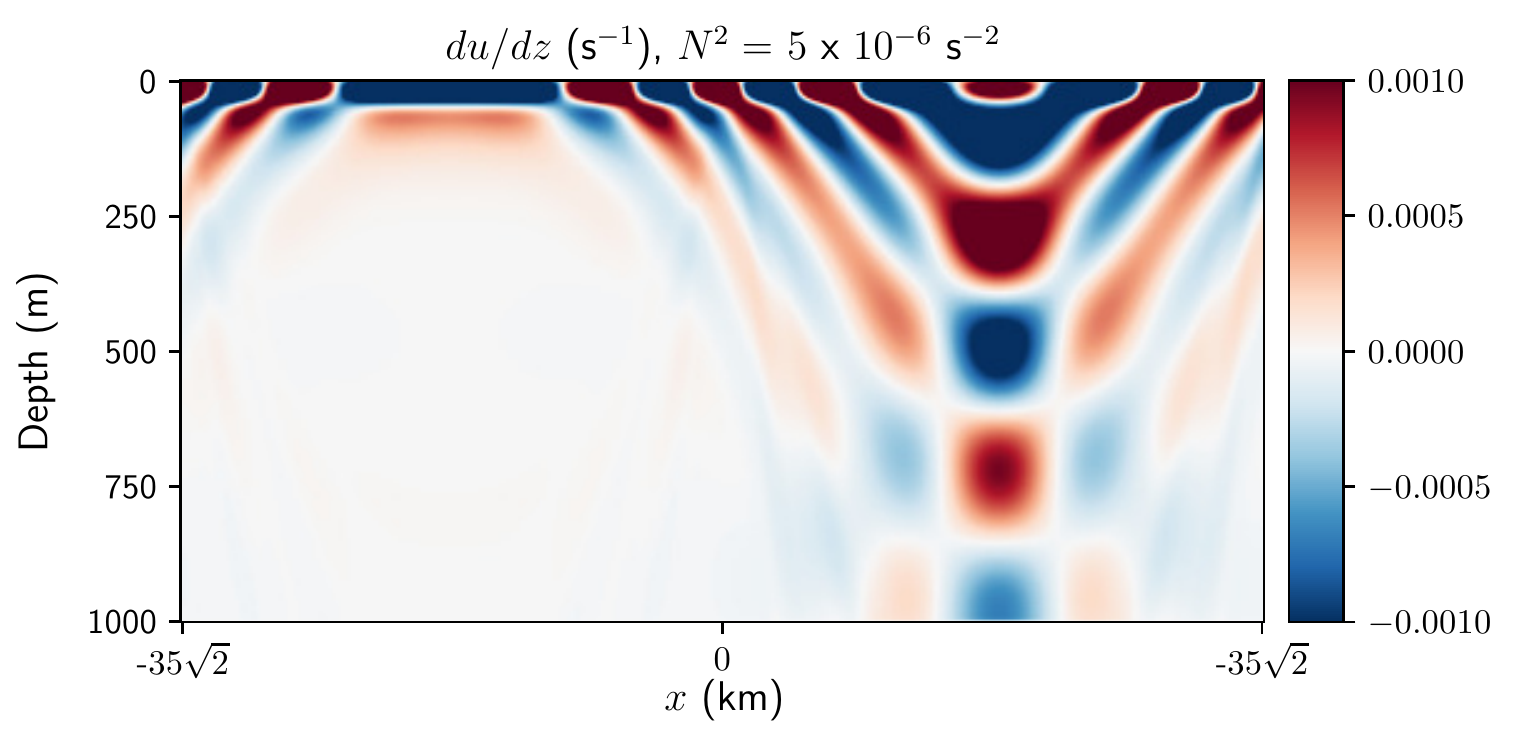}
\includegraphics[trim = 0 0 5 0, clip, width=.32\textwidth]{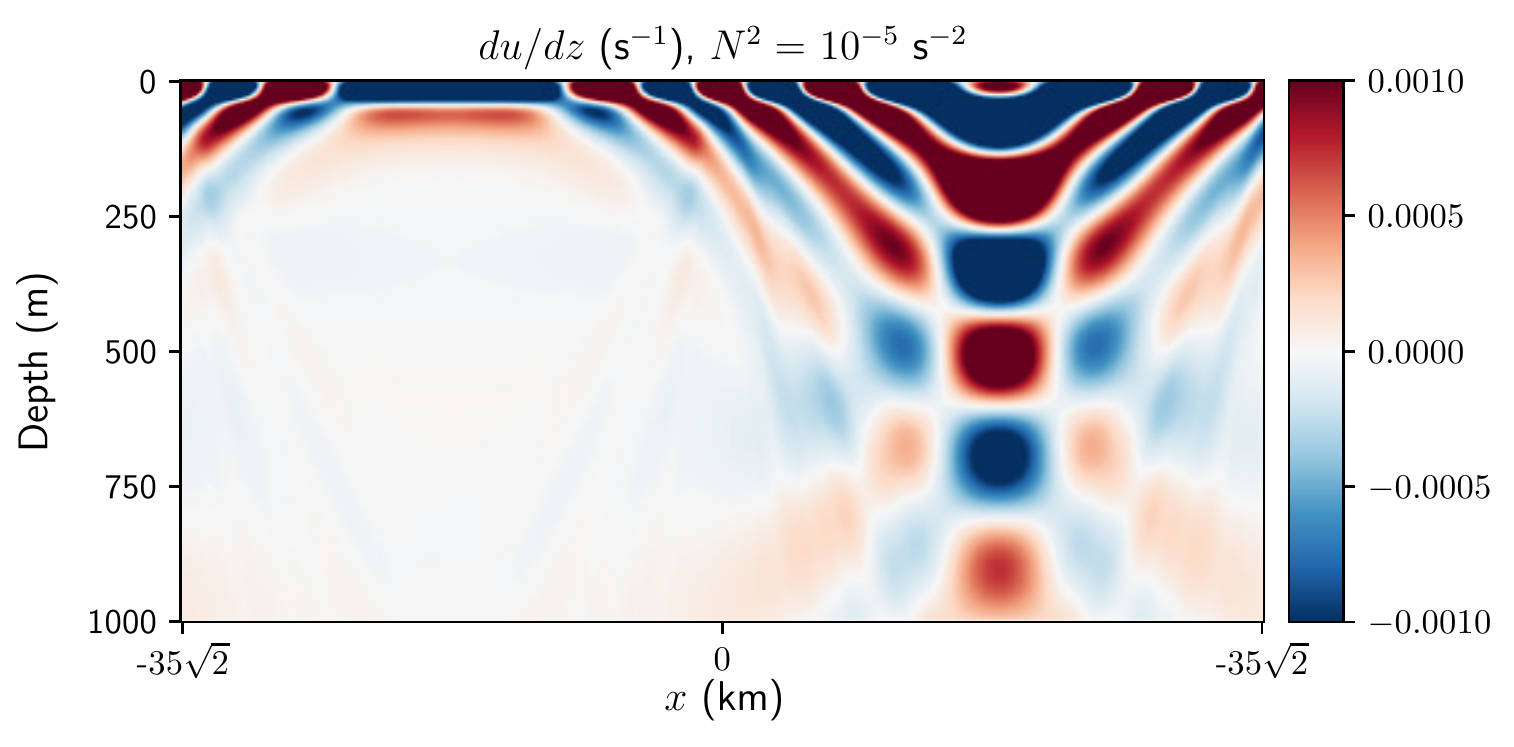}
\caption{Wave shear bands after 10 inertial periods for different values of (constant) stratification. The vertical slices are along the yellow line drawn in top panels of figure \ref{fig:dipole_overview} ($y=0$).}
\label{fig:band_strat}
\end{figure*}

Are these features dependent on stratification? Figure \ref{fig:band_strat} shows vertical slices of wave shear in the same region, but with fixed time (10 inertial periods) and  stratification ranging from $N^2 = 10^{-6}$ s$^{-2}$ (left) to $N^2 = 10^{-5}$ s$^{-2}$ (right). As stratification is increased, shear bands become shallower. The vertical wavenumber increases with $N$ while the horizontal wavenumber seems largely unaffected.



What explains the shape of the bands --- nearly monochromatic, nearly straight but steepening with depth, shallowing with increasing stratification --- and why is their slope constant in time? These questions are the focus of the next section. 

\section{Refraction} \label{sec:refraction}

As a starting point, we consider the wave evolution in the presence of $\zeta$-refraction alone, \textit{i.e.}, neglecting strain in \eqref{eq1}. While this premise is  strictly valid only near the jet center, its implications turn out to be insightful and apply over a broader region than anticipated. This section begins with a heuristic approach to quantifying the shear bands properties reported in section \ref{sec:dipole}. These heuristic results are confirmed in Appendix I, where a 
 derivation of the linear wave solution is presented for an arbitrary, albeit slowly-varying vorticity gradient profile. 


\subsection{Wavevector evolution at the jet center} \label{heuristic}

Let us consider the wavevector evolution near the center of dipole jet. For simplicity, we assume a uniform vorticity gradient, $\grad \zeta = -\gamma \hat{\mathbf{x}}$ with $\gamma>0$.  At $t=0$, a storm generates a horizontally-uniform inertial wave in the mixed-layer \eqref{wave_IC}. Then,  $\zeta$-refraction causes an initial  linear growth of the wavevector along the vorticity gradient:
\beq
k = \tfrac{1}{2} \gamma t,  \qquad l =0\per\label{kt} 
\eeq
Although we are considering  the center of the jet, where velocity is maximum, $\bU=(0,U)$, the wavevector \eqref{kt} is perpendicular to the flow and the Doppler shift, $\bU\cdot \bk$, is  zero. This  justifies the neglect of advection throughout this section. We show in section \ref{sec:all} that the neglect of advection is much less restrictive than one might at first suppose.

 The inertial wave is initially confined to a shallow surface mixed layer, and thus projects on a broad spectrum of vertical wavenumbers, $m$. This compact initial disturbance then disperses as a wavetrain. Because the flow is barotropic, $m$ is constant along a ray. And because there is no Doppler shifting, wave energy  propagates at the intrinsic group velocity,
\beq
c^{x}_g =\frac{N^2 }{f m^2} \frac{\gamma t}{2}\com \quad \,  c^y_g = 0, \quad \, c^z_g = -\frac{N^2}{fm^3} \left( \frac{\gamma t}{2}\right)^2\com  \label{cg}
\eeq
where $\gamma t/2$ is  the horizontal wavenumber in \eqref{kt}.
Wave energy propagates downwards  at a vertical group velocity  $c^z_g\sim m^{-3}$ and thus there is  a dominant vertical wavelength at a given depth and time. To quantify this, we integrate $c^z_g$ with respect to time:
\beq
z(t) = \int_0^t c_g^z(t')dt' = - \frac{N^2 \gamma^2}{12 f m^3} t^3,
\eeq
where we have applied the initial condition $z(0)=0$. Inverting the above expression for $m$, one finds
\beq
m = \left( \frac{N^2 \gamma^2}{12 f|z|}\right)^{1/3} t, \label{mt}
\eeq
where $|z|$ is the (positive) depth. Expression \eqref{mt} predicts  the dominant vertical wavenumber $m$ found at depth $z$ after a time $t$.

From \eqref{kt} and \eqref{mt} we conclude that at a fixed depth $|z|$, both $k$ and $m$ grow linearly with $t$; this is consistent with figure \ref{fig:dipole_overview}. Furthermore, as highlighted in section \ref{sec:dipole},  the wave band slope, defined as
\beq
\frac{dz}{dx} =-\frac{k}{m}= - \left( \frac{3f\gamma |z| }{2N^2} \right) ^{1/3} \label{dzdx_pred},
\eeq
is time-independent. From \eqref{dzdx_pred}, the band slope increases proportionally to $|z|^{1/3}$; this is consistent with the mild steepening of bands with depth seen in figure \ref{fig:dipole_overview}. Finally, consistent with the shallowing of bands with increasing stratification in figure \ref{fig:band_strat}, the slope in \eqref{dzdx_pred} is proportional to $N^{-2/3}$.

\subsection{Validation from analytical and numerical solutions}

Do these heuristic predictions survive closer scrutiny? They do:  appendix I summarizes a leading-order analytical wave solution, \eqref{la_sol}, obtained from the linearized YBJ equation. This analytical solution generalizes the heuristic predictions to encompass arbitrary, but slowly-varying vorticity gradient $\nabla \zeta$. Near the jet center, where $\nabla \zeta \approx -\gamma\hat{\mathbf{x}}$, the heuristic predictions for the horizontal and vertical wavenumbers, \eqref{kt} and \eqref{mt}, are precisely recovered. As a bonus, the analytical derivation predicts not only the wavevector, but also the wave amplitude.

\begin{figure*}[h!]
\centering
\includegraphics[width=.32\textwidth]{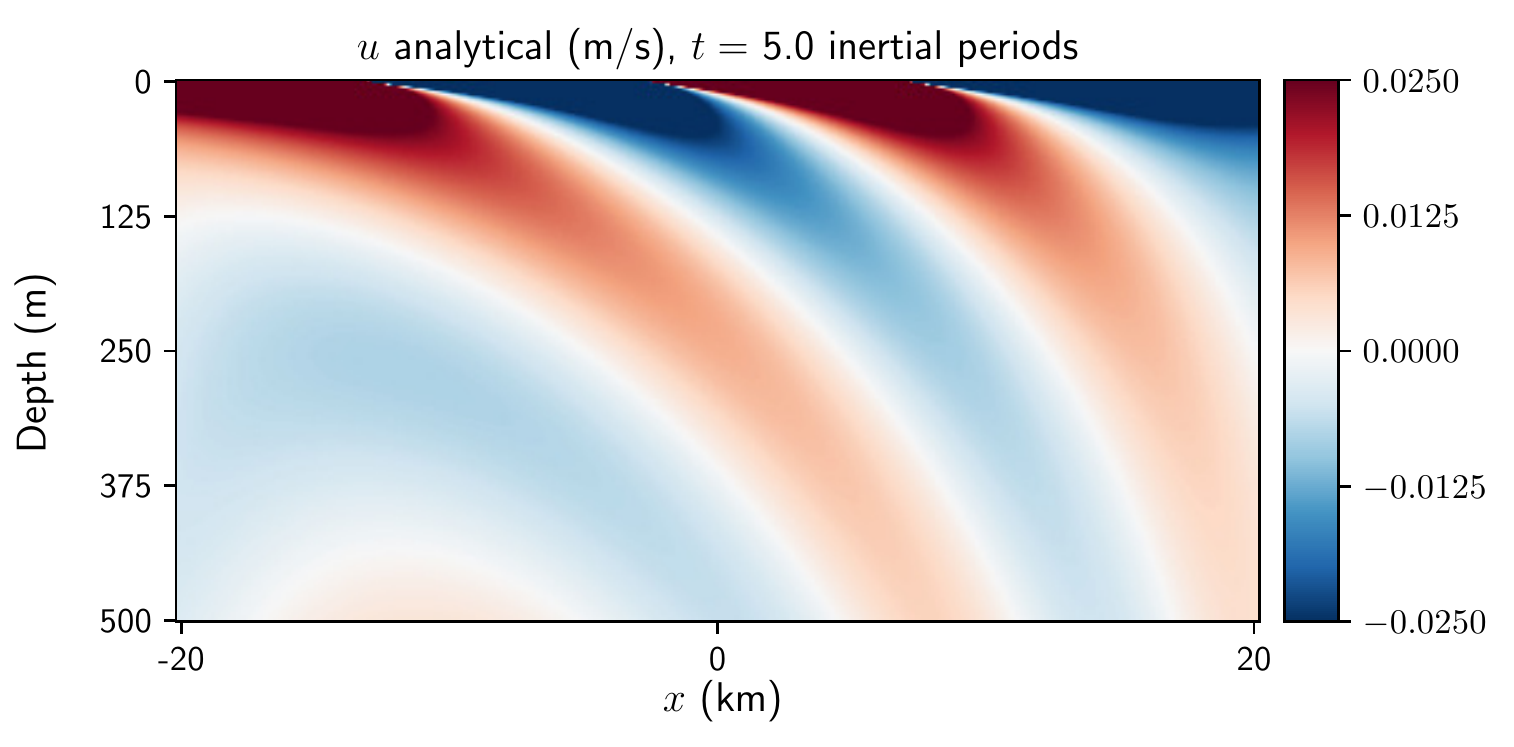}
\includegraphics[width=.32\textwidth]{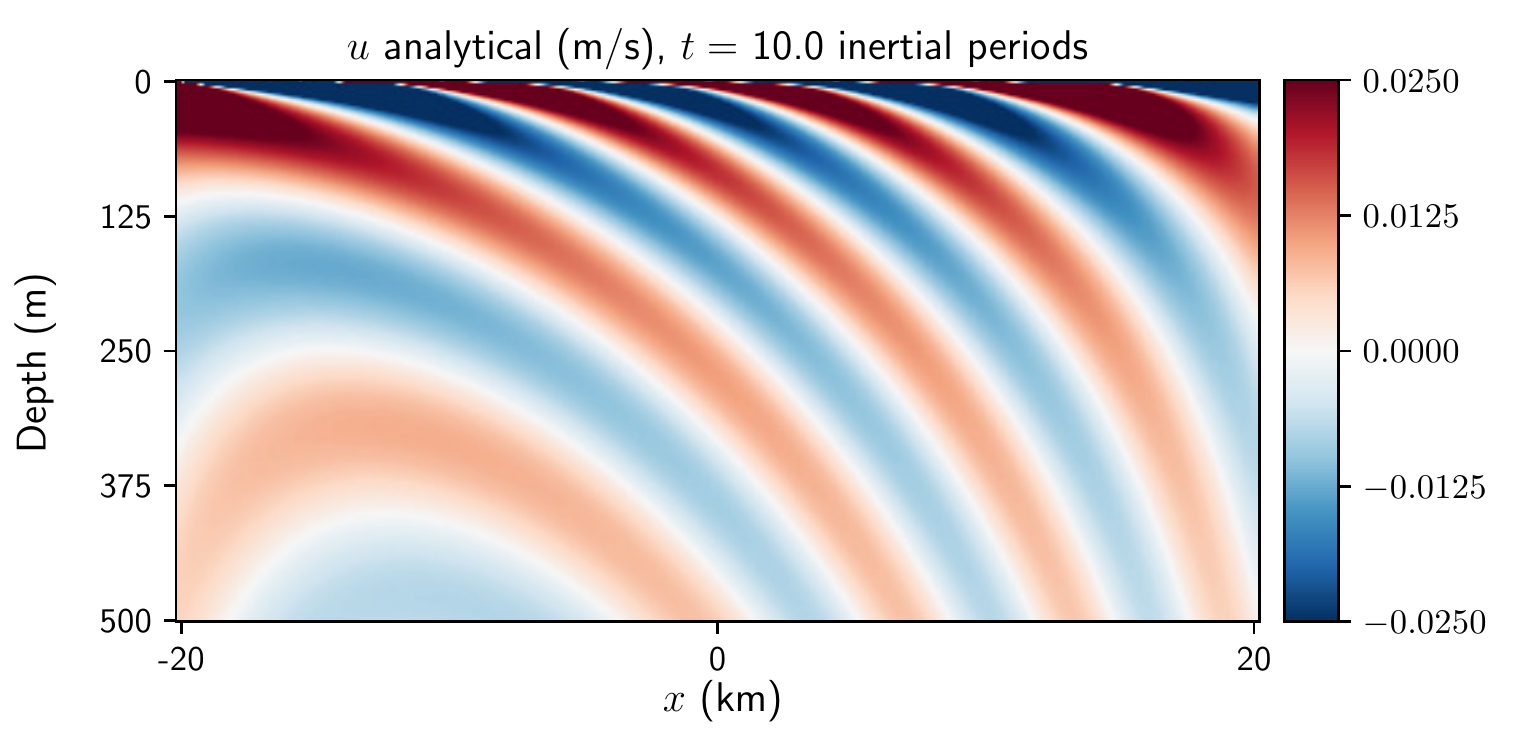}
\includegraphics[width=.32\textwidth]{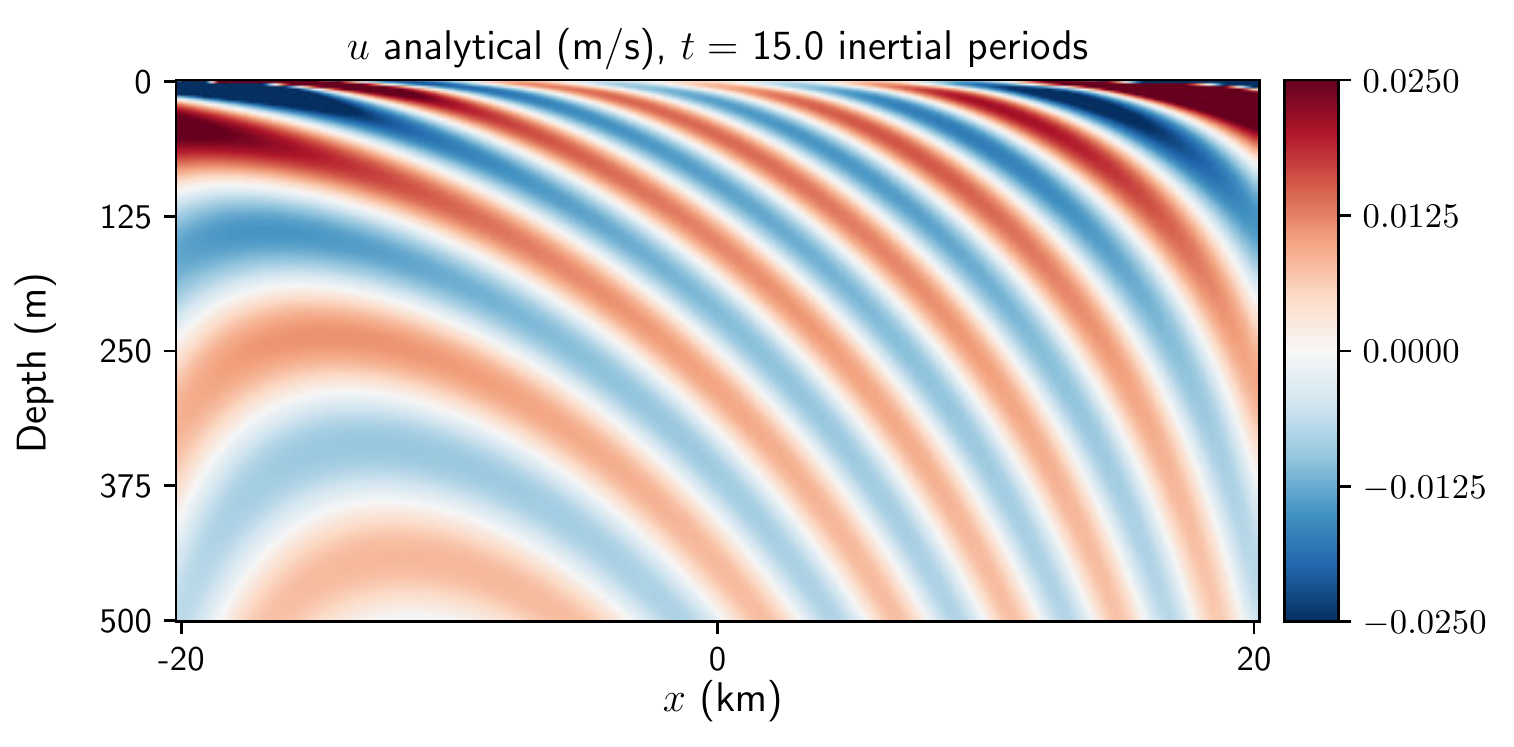}
\includegraphics[width=.32\textwidth]{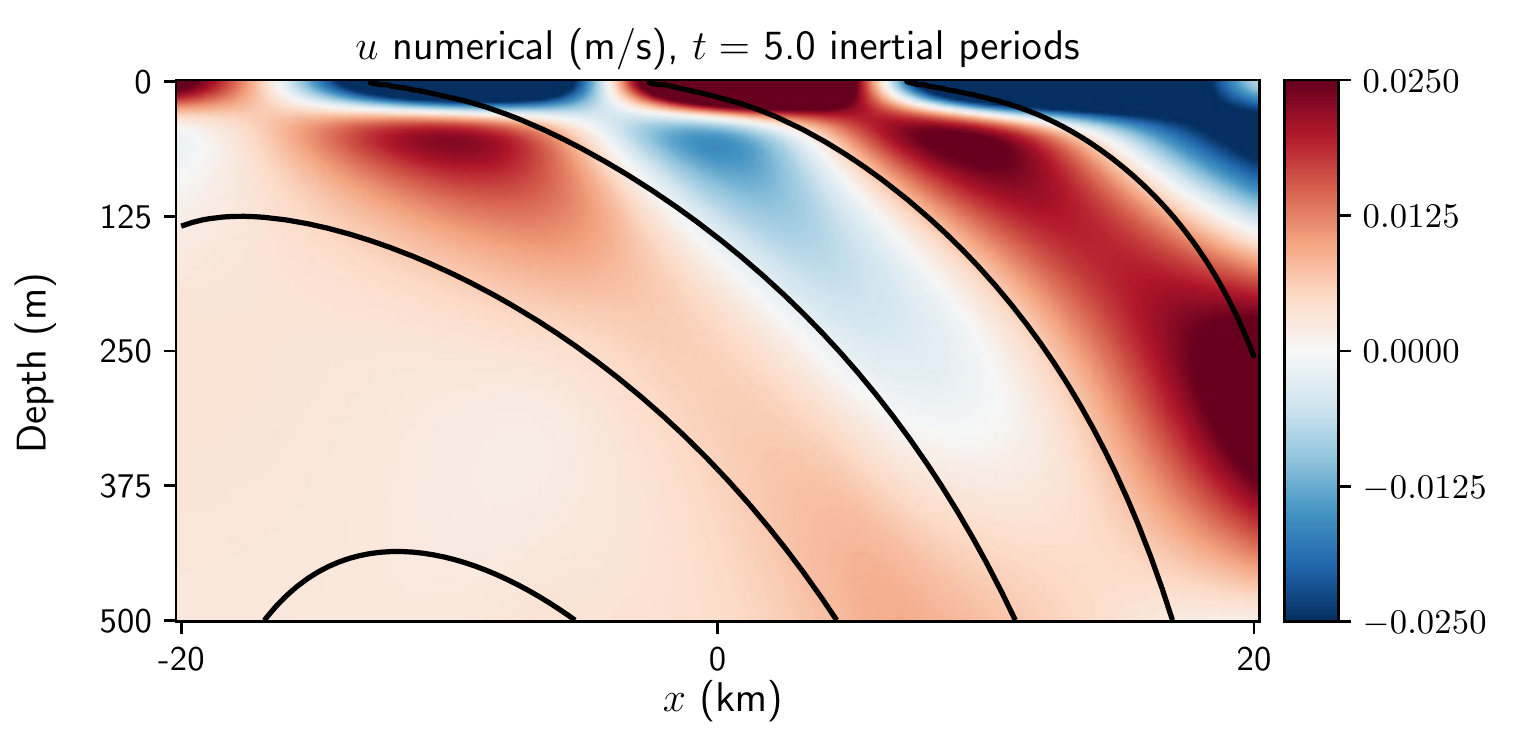}
\includegraphics[width=.32\textwidth]{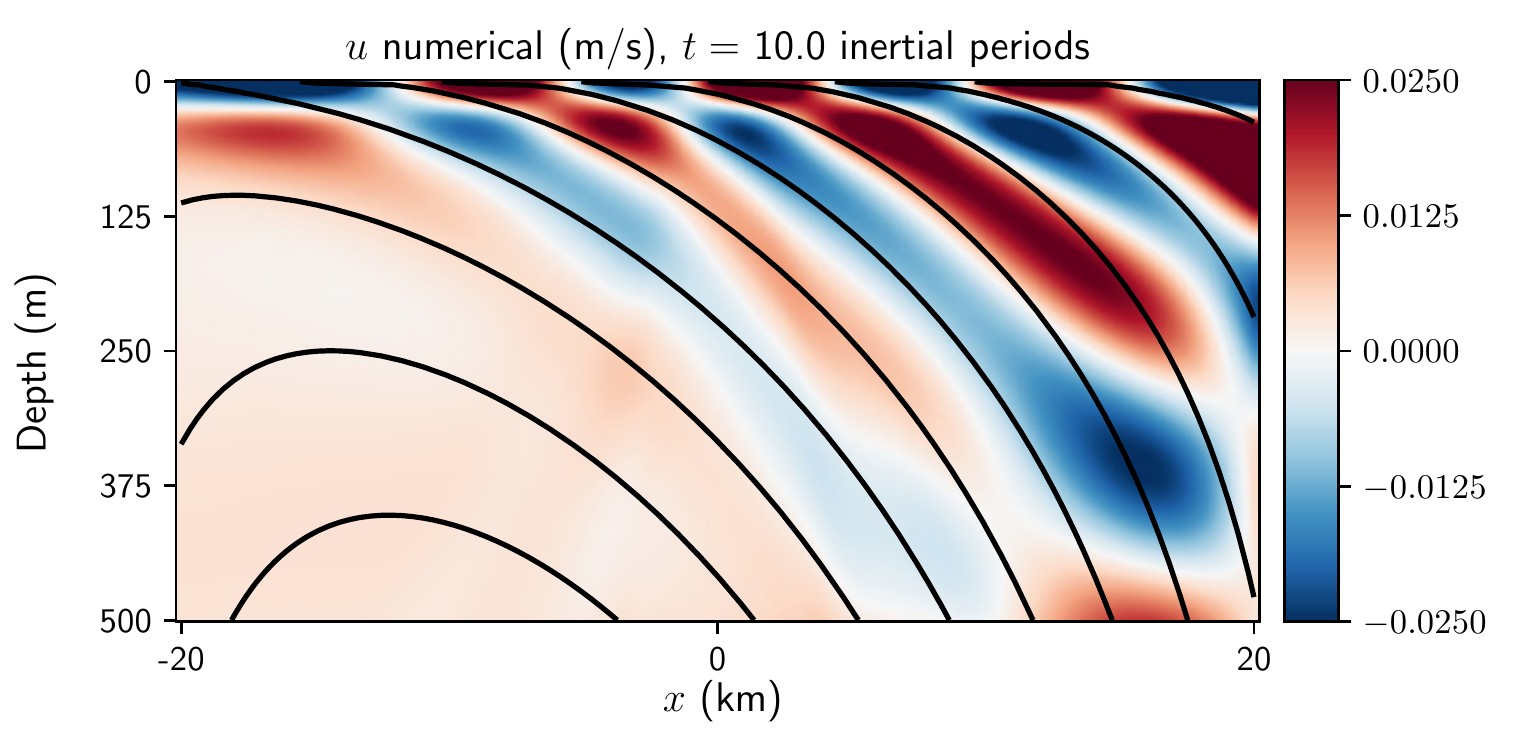}
\includegraphics[width=.32\textwidth]{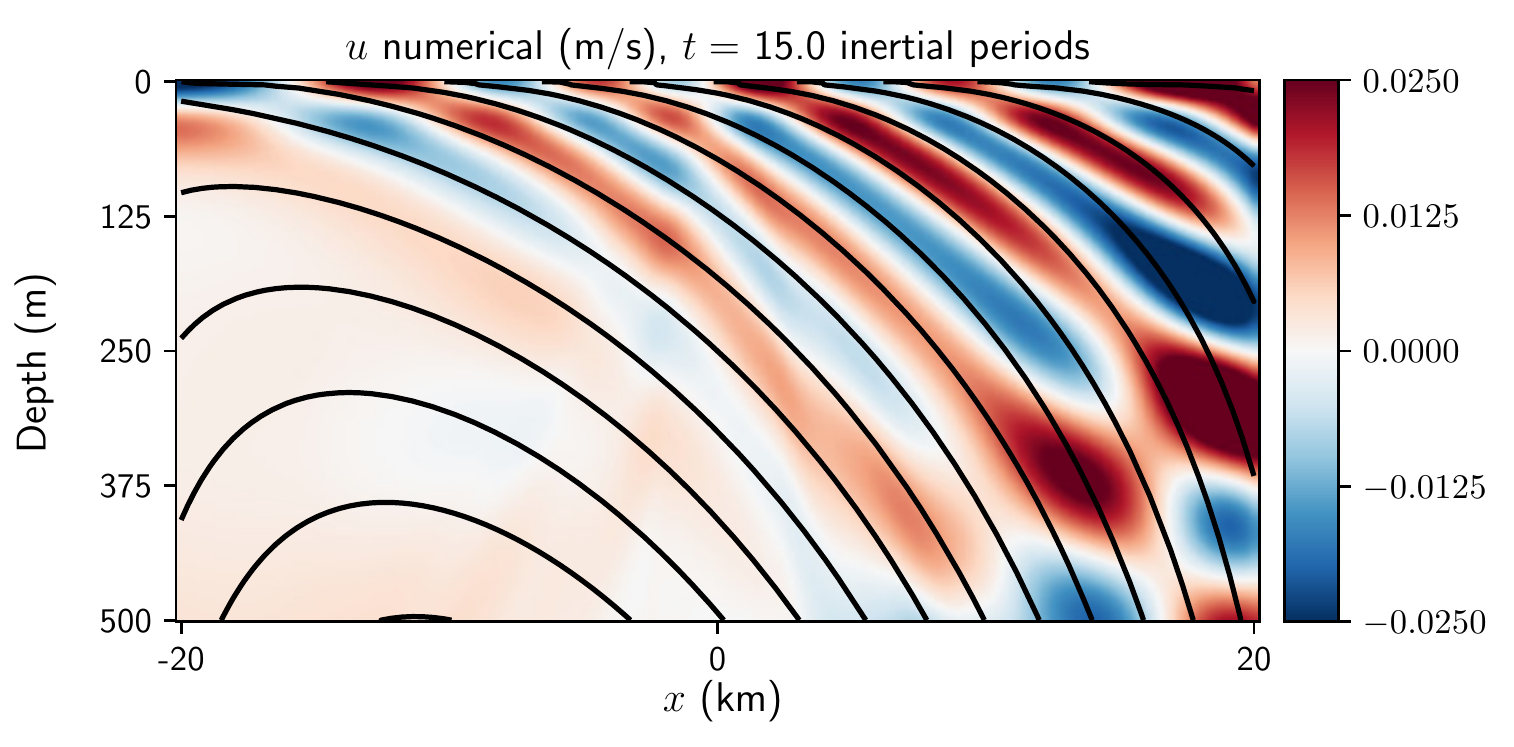}
\caption{Wave velocity after 5, 10 and 15 inertial periods along $y=0$. Top: leading-order analytical solution, \eqref{la_sol}, using the $x$-dependent $\nabla \zeta$. Bottom: solution of the full numerical model with $u=0$ contours from the analytical solution.}
\label{fig:la_comp}
\end{figure*}

The analytical solution is compared to the numerical solution of the full YBJ equation in figure \ref{fig:la_comp}. Contours of the analytical solution are added to the numerical solution to facilitate comparison. The analytical solution provides a useful cartoon of the salient wavy patterns near the surface and in the jet region. In particular, it captures the multiplication and shrinking of bands, their steepening with depth and approximately time-invariant shape, which were predicted via the heuristic approach. 

The analytical solution, however, fails to capture the cyclone-anticyclone energy gradient. This is due to the assumption of slowly-varying vorticity gradient, \textit{i.e.}, the wave amplitude feels a uniform vorticity gradient such that energy concentration remains horizontally uniform. The analytical solution  also fails as one approaches near the anticyclonic vortex core because rays emitted from the other side of the anticyclone interfere with the locally-emitted rays (see  figure \ref{fig:band_strat}).

\begin{figure*}[h!]
\centering 
\includegraphics[trim = 0 20 0 0, clip, width=.345\textwidth]{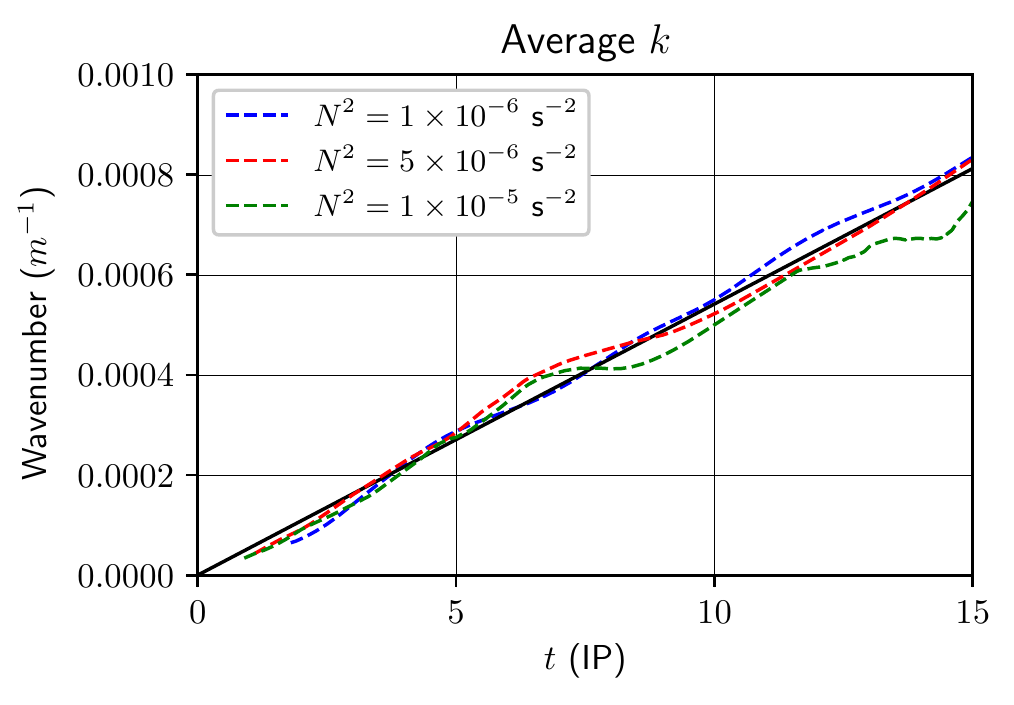}
\includegraphics[trim = 20 20 0 0, clip, width=.31\textwidth]{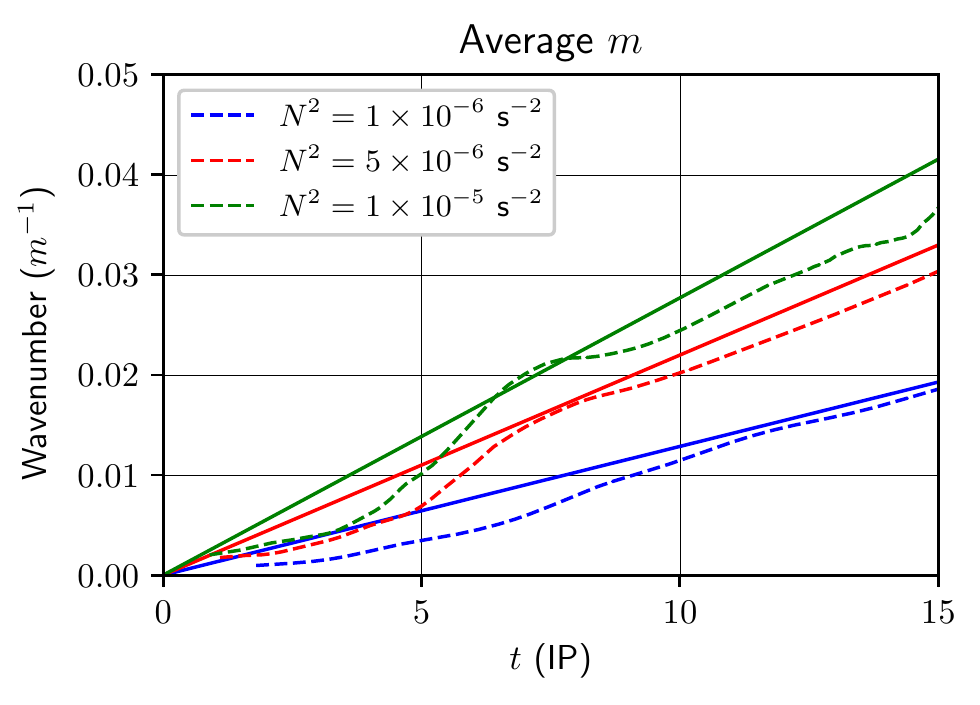}
\includegraphics[trim = 10 20 0 0, clip, width=.31\textwidth]{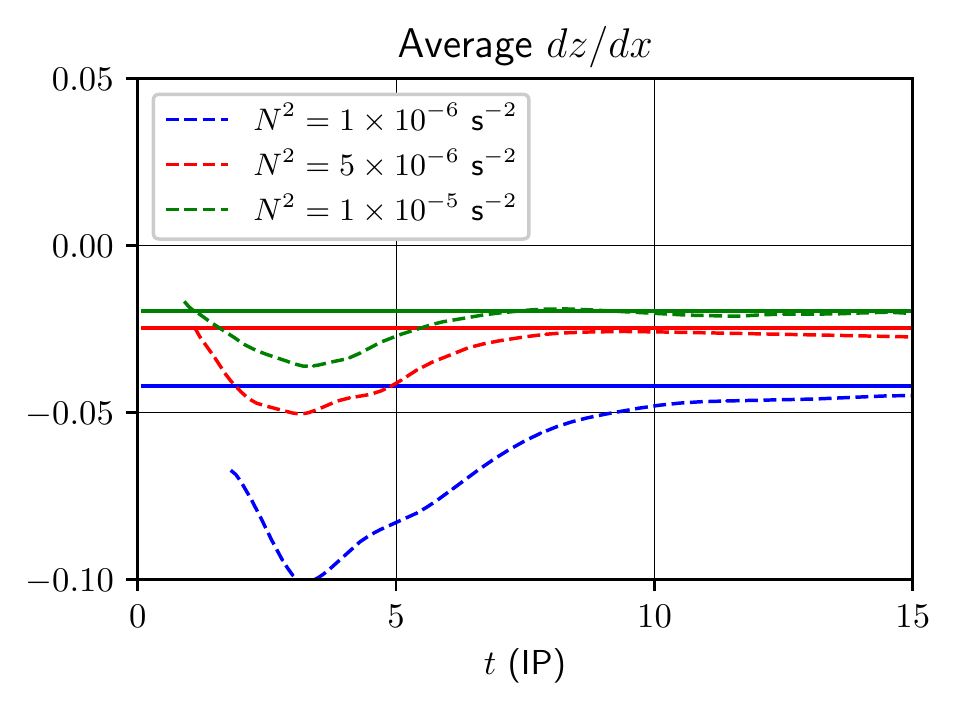}
\caption{Spatially-averaged (dotted) and predicted (solid) horizontal \eqref{k_pred} and vertical \eqref{m_pred} wavenumbers (left and middle) and wave band slope \eqref{dzdx_pred} (right) over the jet center region, $x \in [5,15]$ km, $y=0$, and $z \in -[100, 300]$ m. Different colors indicate different stratifications. To reduce noise only regions with wave energy above $10^{-5}$ (m/s)$^2$ are included in the average.}
\label{fig:km1}
\end{figure*}

In figure \ref{fig:km1}, we compare the horizontal and vertical wavenumbers estimated from the model output (dotted; see Appendix II) with the predicted values (solid). Wavenumbers are averaged over the region $x \in [5,15]$ km and $z \in -[100, 300]$ m, where shear bands appear most clearly. Following the analytical prediction of Appendix I, $\gamma$ is replaced with the locally-averaged vorticity gradient in the heuristic formulae, \eqref{kt} and \eqref{mt}. These formulae correctly predict the linear growth and dependence on stratification. That is, stratification leaves $k$ essentially unchanged, but higher stratification leads to a more rapid growth of $m$. In turn, shear bands are steeper in regions of weak stratification as in figure \ref{fig:band_strat}. The right panel shows that the wave band slope asymptotes to its predicted constant value, \eqref{dzdx_pred}. 

\subsection{Back-tracking wave bands} \label{sec:back}

The conservation of the wave frequency allows one to back-track the surface origin of wave bands observed at depth. The full dispersion relation of the $f$-plane YBJ equation \eqref{ybj} is:
\beq
\omega =  \bU \cdot \bk + \frac{f}{2} \left( Ro + Bu\right) \label{omega-f},
\eeq
where $\omega$ is the wave frequency in excess of $f$, and we defined the vorticity-based Rossby number and the wave Burger number:
\beq
Ro \defn \frac{\zeta}{f}, \qquad Bu \defn \left( \frac{N|\bk|}{fm}  \right)^2.
\eeq
In a steady flow such as the dipole, the Eulerian frequency $\omega$ in \eqref{omega-f} is conserved following a ray. Furthermore, we have seen that Doppler shift is negligible near the jet center such that the intrinsic frequency,
\beq
\omega_i =  \frac{f}{2} \left( Ro + Bu\right) \label{omegai},
\eeq
is also conserved in its vicinity.

Are the heuristic predictions consistent with the conservation of $\omega_i$? Using \eqref{dzdx_pred} one obtains an expression for the wave Burger number,
\beq
Bu = \left( \frac{Nk}{fm}\right)^2=  \left( \frac{3 N \gamma |z| }{2f^2} \right) ^{2/3} \label{bu_pred}.
\eeq
Near the jet region, where $\kappa x \ll 1$ and $\kappa y \ll 1$, 
\beq
Ro \approx - \frac{\gamma x}{f}.
\eeq
Assume that a wave is launched from the surface, $z=0$, and not too far from the jet center, $x=x_0$ with $\kappa x_0 \ll 1$. At inception, the wave experiences $Ro= -\gamma x_0$ and $Bu=0$, assuming that wind generation implies $\bk=0$. The beam trajectory is obtained by integrating \eqref{dzdx_pred}:
\begin{align}
x(z) - x_0 = \left(\frac{3N|z|}{2} \right)^{2/3} \left( f \gamma \right)^{-1/3}  \label{traj}
\end{align}
Upon multiplying the above with $-\gamma$, one obtains the change in Rossby number,
\beq
\Delta Ro =  - \left( \frac{3 N \gamma |z| }{2f^2} \right) ^{2/3}, 
\eeq
which, from \eqref{bu_pred}, is precisely $-Bu=-\Delta Bu$. Therefore, the sum of $Bu$ and $Ro$ remains constant as the wave propagates, as required by the conservation of the intrinsic frequency, \eqref{omegai}.

This leads to a result of practical utility. If wave bands are observed in a region with  $Ro^*=\zeta^*/f$ with scales corresponding to $Bu^*$, then the wave must have originated from a region where $Ro=Ro^* + Bu^*$. Thus one can back-track wind-generated wave bands observed at depth.

\subsection{Wave energy penetration}

The heuristic solution provides an approximate time scale for wave energy radiation below the mixed layer. The time taken for a disturbance with a vertical wavenumber $m$ to propagate to a depth $|z|$ is obtained by re-arrangement of \eqref{mt}:
\beq
t = \left( \frac{12 f |z|}{N^2 \gamma^2}\right)^{1/3}m. \label{tt}
\eeq
In a barotropic flow, the vertical wavenumber spectrum of wave energy is time-independent, and thus can be obtained from  a vertical Fourier transform of the wave initial condition, \eqref{wave_IC}, yielding
\begin{align}
\mathcal{E}(m) &= \half |\widehat{\LL A}_0(m)|^2 \propto \exp \left( - m^2 \sigma^2/2\right). 
\label{FTIC7}
\end{align}
After a given time $t$, disturbances with wavenumbers from $m=0$ up to a cut-off wavenumber $m_c(t)$ have reached the mixed layer base or below. We estimate $m_c(t)$ by replacing $|z|$ with the mixed layer depth $\sigma$ in \eqref{mt}:
\beq
m_c(t) = \left( \frac{N^2 \gamma^2}{12 f \sigma}\right)^{1/3} t\per \label{mc}
\eeq
 One  thus estimates the fraction of wave energy radiated out of the mixed layer by integrating $\mathcal{E}(m)$ from $m=0$ to $m_c(t)$: 
\beq
\frac{\WE_{rad}(t)}{\WE_{tot}} \approx  \frac{\int_0^{m_c(t)} \mathcal{E}(m)dm}{\int_0^{\infty} \mathcal{E}(m) dm } =  \text{erf}\left( \frac{m_c(t) \sigma}{\sqrt{2}}\right), \label{rad}
\eeq
where erf stands for the error function.

\begin{figure}
\centering
\includegraphics[trim = 0 0 0 0, clip, width=.4\textwidth]{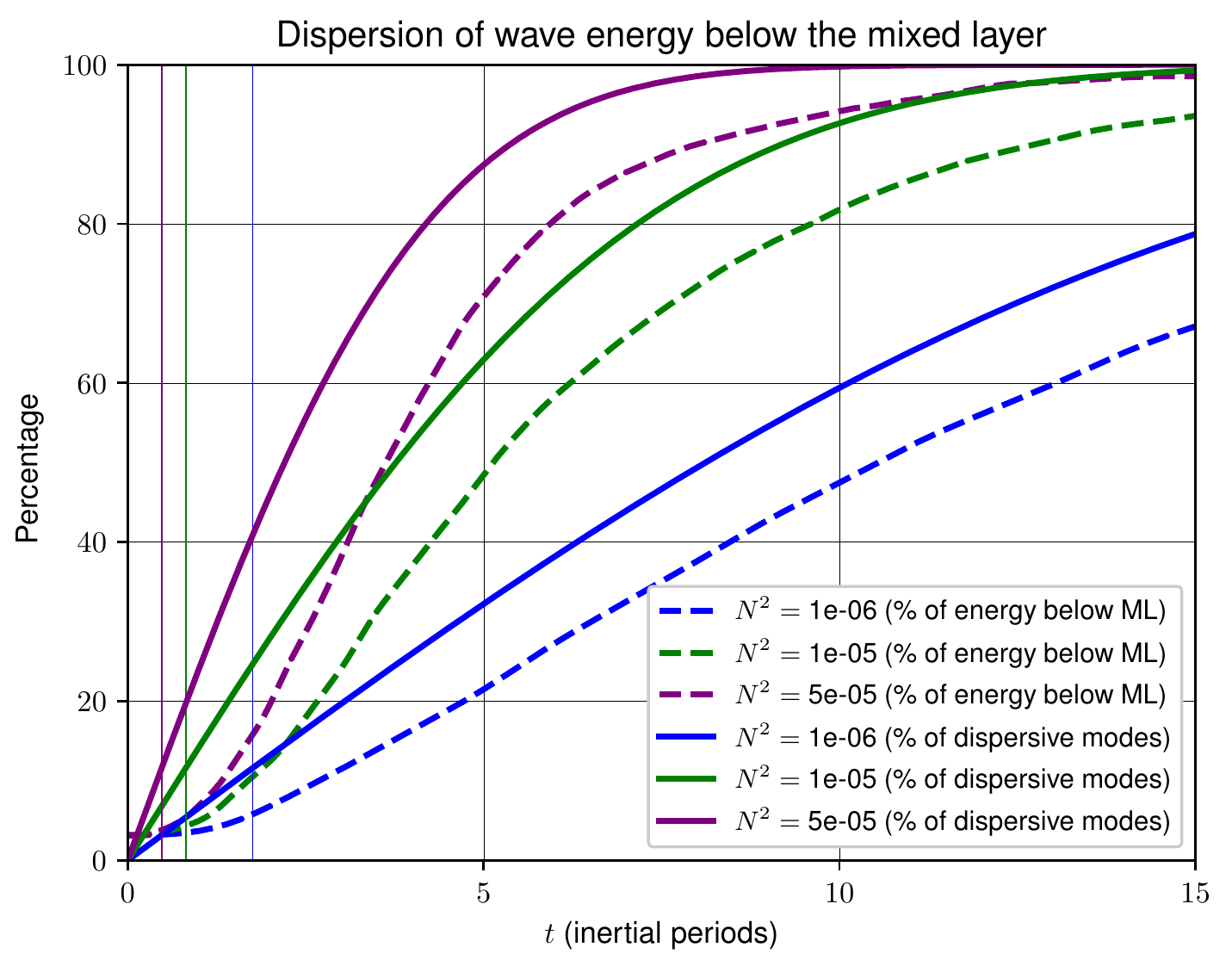}
\caption{Fraction of energy in wavenumbers $m< m_c(t)$ estimated from the heuristic solution (solid curves) versus the fraction of wave energy below the mixed-layer from the numerical  solution (dashed curves). Vertical lines corresponds to Gill's estimate of the decay  time scale of mixed-layer near-inertial oscillations \eqref{tg}. Different colors show different buoyancy frequencies. }
\label{fig:time}
\end{figure}

In figure \ref{fig:time}, we compare this estimate of wave energy radiation (dashed lines) with the fraction of energy below the mixed layer in the full numerical solution (solid lines). The heuristic prediction overestimates the rapidity of downward radiation. A rationalization of  this discrepancy is that the heuristic calculation uses the maximum vorticity gradient $\gamma$ in \eqref{mc} and thus overestimates  the cut-off wavenumber $m_c(t)$.
The heuristic estimate also assumes that once a disturbance reaches the mixed layer base, its energy is fully located below the mixed-layer. 


How does this calculation compare with previous estimates of the mixed-layer decay time scale? \cite{gill1984} proposed that the decay scale can be estimated as the time needed for the gravest vertical mode to undergo a phase change of 90 degrees. \cite{dasaro1989} adapted this idea to estimate the decay scale due to the $\beta$-effect. For a constant-$N$ flow, Gill's estimate of the  decay time scale is
\beq
t_G = \left( \frac{3 \pi^3 f}{N^2 H^2 \beta^2}\right)^{1/3}, \label{tg}
\eeq
which is equivalent to \eqref{tt} after substituting $\gamma$ with $2\beta$ and setting $z=H=\pi/m$. The colored vertical lines in figure \ref{fig:time} show the values of $t_G$ when the relevant vorticity gradient is considered, \textit{i.e.}, $\beta \to \half \gamma$ in \eqref{tg}. In our setup, where stratification is constant and vertical wavenumbers are distributed as in \eqref{FTIC7} with  $\sigma^{-1} \approx 0.03 $ m$^{-1}$, $t_G$  underestimates the time needed for a significant loss of the mixed layer energy. 

\section{Strain} \label{sec:strain}

\begin{figure*}
\centering
\includegraphics[trim = 60 0 50 0, clip, width=.4\textwidth]{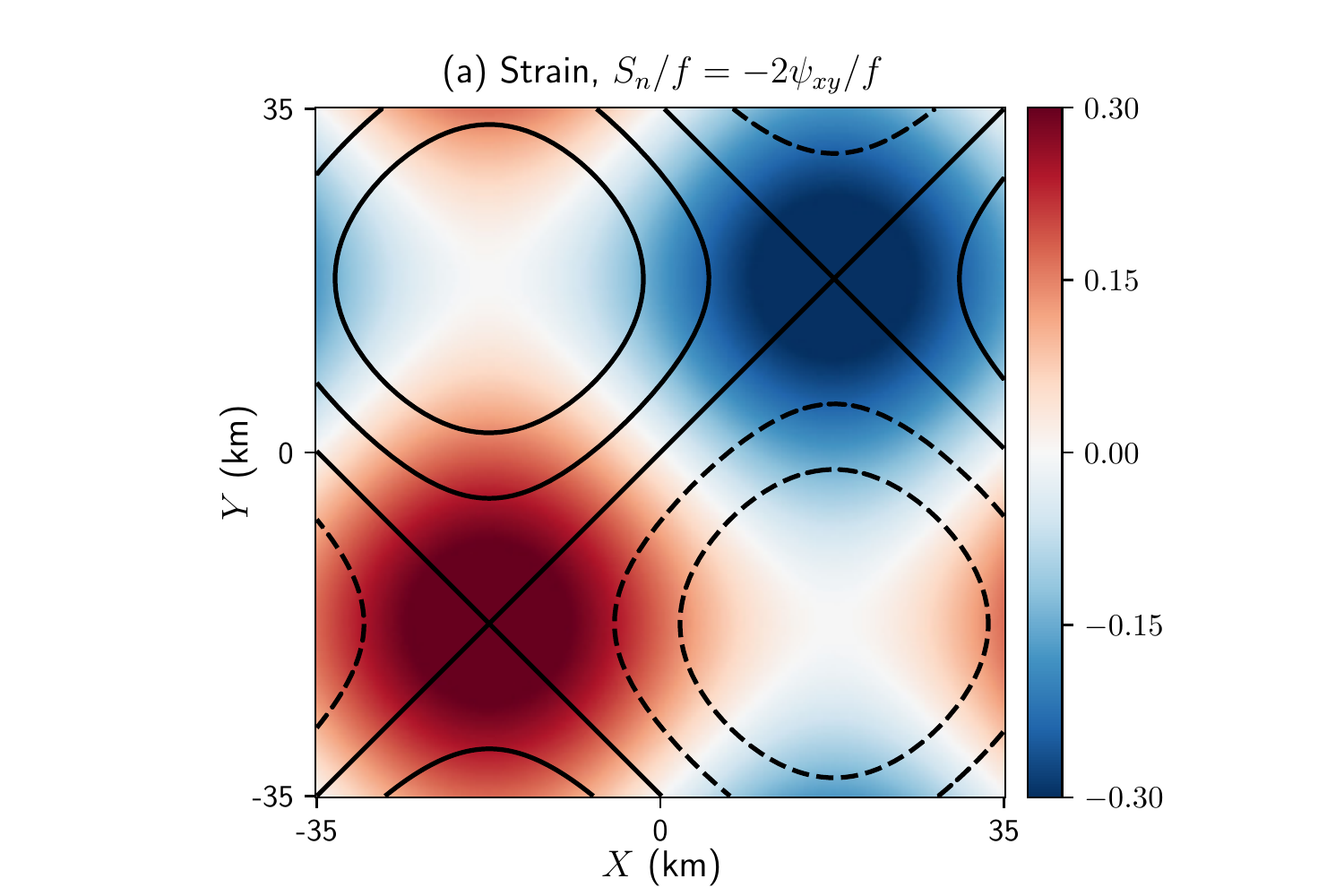} \qquad
\includegraphics[trim = 60 0 50 0, clip, width=.4\textwidth]{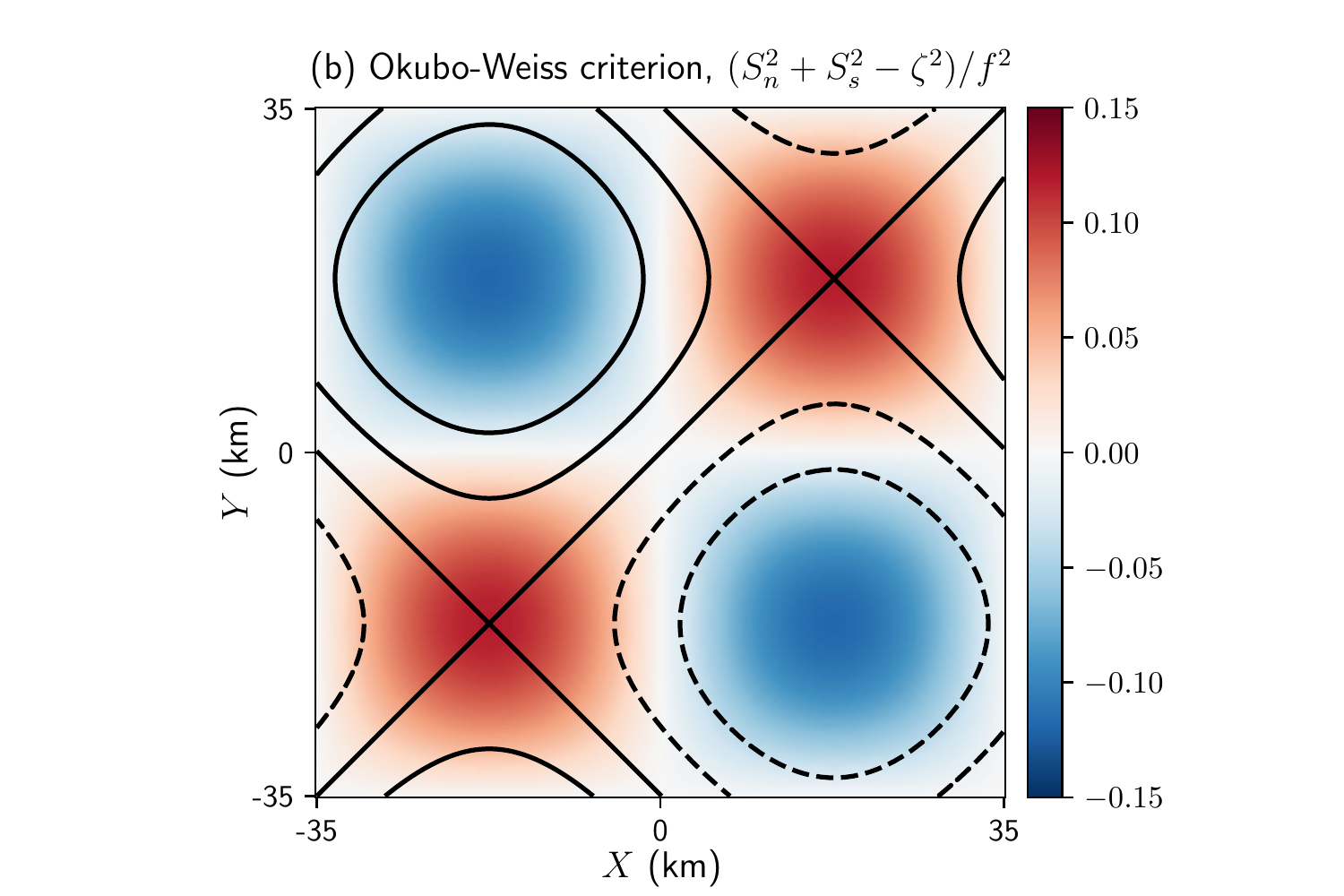}
\caption{(a) Normalized strain (note that the other component, $S_s$, is zero for the dipole) and Okubo-Weiss criterion (b), which indicates strain- ($>0$) versus vorticity-dominated ($<0$) regions of the domain. Local values of the Okubo-Weiss criterion are indicated in the confluent (purple), jet (golden) and diffluent (green) regions. Vorticity contours of 0, $\pm 0.1 f$ and $\pm 0.2f$ are overlaid.}
        \label{fig:strain}
\end{figure*}

Thus far we have considered wave evolution close to the jet center. In this special location, the strain vanishes, \textit{i.e.} $\psi_{xy}=\psi_{xx} - \psi_{yy}=0$ (figure \ref{fig:strain}a). The Doppler shift, $\bU \bcdot \bk$,  is also zero because $\bk$ is anti-parrallel to $\grad \zeta$, and thus perpendicular to $\bU$. What about strain-dominated regions, such as the dipole's confluent and diffluent regions? How does strain affect the wavevector $\bk$?

\cite{bm2005} examined straining by a steady flow with constant spatial gradients of velocity (thus forbiding $\zeta$-refraction, which relies on second-order spatial derivatives of velocity). In this case, analytical solutions can be found for the evolution of the packet-following three-dimensional wavevector \citep{jones1969}. In a barotropic flow, the strain tensor  in \eqref{eq1} can be separated into two distinct contributions:
\beq
\begin{pmatrix}  U_x & V_x \\ U_y & V_y\end{pmatrix}
= \frac{1}{2}\begin{pmatrix}
S_n & S_s \\ 
S_s & -S_n \\ 
\end{pmatrix}   
+ \frac{1}{2}\begin{pmatrix}
0 & \zeta \\ 
-\zeta & 0 \\ 
\end{pmatrix} \com   
 \label{strain_mat}
\eeq
where $S_n \defn U_x - V_y=-2\psi_{xy}$ and $S_s \defn U_y + V_x=\psi_{xx}-\psi_{yy}$ are the normal and shear components of strain. The first term of \eqref{strain_mat} tends to increase $|\bk|$ exponentially in time; the second term, on its own,  rotates the $\bk$ with a frequency $\zeta/2$. In regions of positive Okubo-Weiss criterion (shown for the dipole in \ref{fig:strain}b),
\beq
OW = S_n^2 + S^2_s - \zeta^2,
\eeq
the wavevector $\bk$ undergoes exponential growth with an e-folding time scale $OW^{-1/2}$ \citep{bm2005}. Is this the case in the dipole solution?  


\begin{figure*}
\centering 
\includegraphics[trim = 0 20 0 0, clip, width=.345\textwidth]{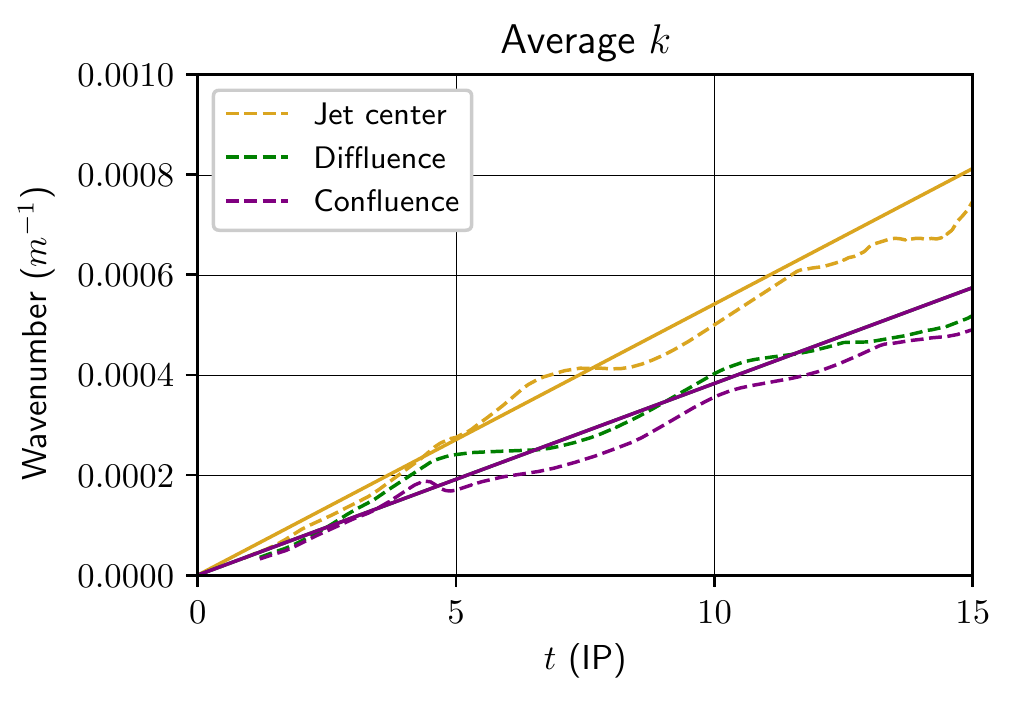}
\includegraphics[trim = 20 20 0 0, clip, width=.31\textwidth]{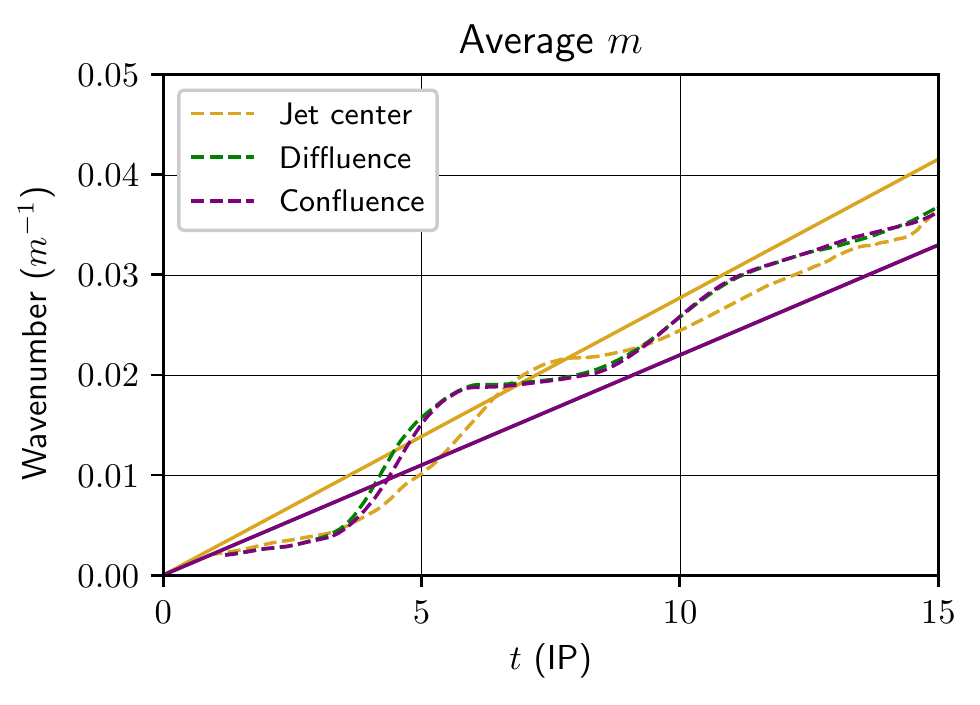}
\includegraphics[trim = 10 20 0 0, clip, width=.31\textwidth]{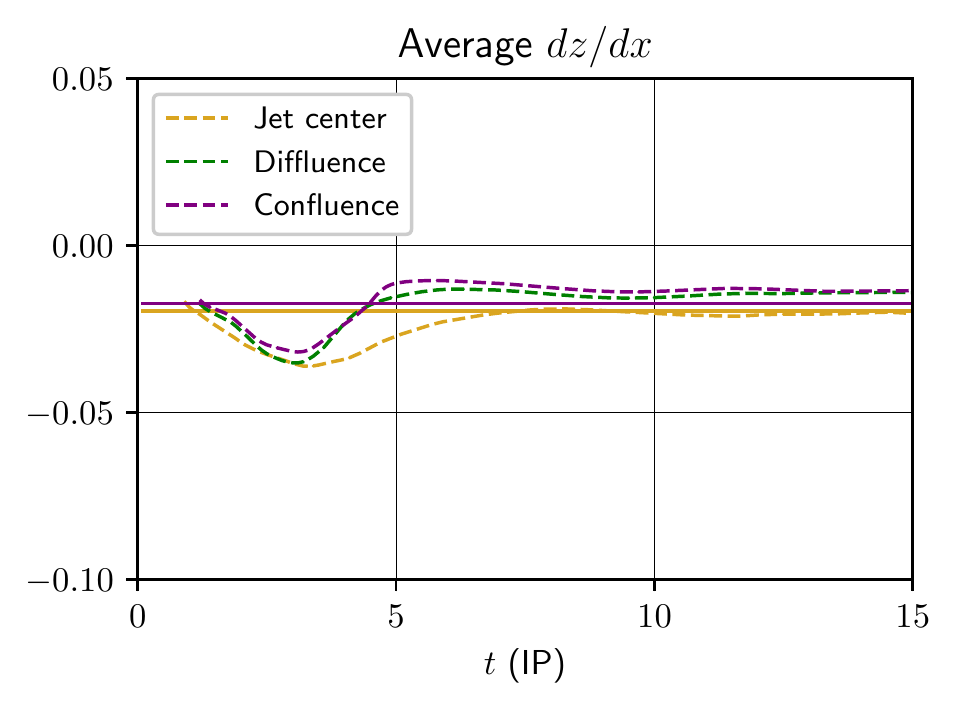}
\caption{Spatially-averaged (dotted) horizontal and vertical (left and middle) and wave band slope (right) over the region $x \in [5,15]$ km and $z \in -[100, 300]$ m of the dipole. Colors specify the location on the $y$ axis, \textit{i.e.}, the confluent ($\kappa y = -\pi/4$, purple), jet center ($y=0$, golden) and diffluent ($\kappa y = +\pi/4$, green) regions shown as dots in figure \ref{fig:strain}b.  Superimposed in solid lines are the $\zeta$-refraction predictions, \eqref{kt}-\eqref{mt}, but based on the local vorticity gradient instead of $\gamma$.  Note that the solid purple line depicts the identical predictions for the confluence and diffluence regions. To reduce noise only regions with wave energy above $10^{-5}$ (m/s)$^2$ are included in the average.}
\label{fig:km2}
\end{figure*}


The lower panel of figure \ref{fig:km2} shows time series of the wavevector in the confluent ($\kappa y= - \pi/4$, purple), diffluent ($\kappa y=\pi/4$, green) and jet center ($y=0$, golden) regions with $N^2=10^{-5}$ s$^{-2}$. Colored dots in figure \ref{fig:strain}b indicate the location of these regions, as well as their local $OW$ value. As in figure \ref{fig:km1}, wavenumbers are averaged over the region $x \in [5,15]$ km and $z \in -[100, 300]$ m, where shear bands appear most clearly.

Although the confluent and diffluent regions are characterized by $OW/f^2 = \pm 0.06$, corresponding to an e-folding time scale of $OW^{-1/2} \approx 0.7$ inertial period, there is no sign of a wavevector exponential growth, even on a range of 15 inertial periods. In fact, the refraction-only prediction (solid lines), \eqref{kt} and \eqref{mt} with $\gamma$ replaced by the local $\nabla \zeta$, is equally good at predicting $k$ and $m$ in the confluent and diffluent regions as it is in the jet center (figure \ref{fig:km1}). In other words, $\zeta$-refraction alone accounts for the wavevector evolution, and strain is ineffective. Why is that?

A first clue comes from the work of \cite{cesar}. In a barotropic flow, vertical modes are uncoupled. For any given vertical mode, the horizontal group velocity,
\beq
\mathbf{c}^h_g = \hbar \bk,
\eeq
where $\hbar \defn N^2/fm^2$ is the wave dispersivity, grows proportionally with $\bk$. As straining exponentially compresses wave crests, horizontal group velocity also increases exponentially. Thus, if $\hbar$ is large enough, the wave rapidly escapes the straining region and its growth is no longer exponential. This is consistent with low-$m$ disturbances leaving the jet region to accumulate in the anticyclone. 

But what about weakly-dispersive waves, for which $\hbar$ is small and escape is slow? Why don't we observe exponential growth for the weakly-dispersive modes left in the jet region? To answer these questions, we now consider the combined effects of refraction and strain in the limit of weak dispersion.

\section{Weak Dispersion} \label{sec:all}

The two previous sections considered refraction and strain separately, focusing on the case of the dipole flow. Along the jet's confluent, diffluent and center regions, $\bk \approx -t\nabla \zeta/2$. This is consistent with $\zeta$-refraction, but inconsistent with straining. In this section, we consider the evolution of $\bk$ under the combined effects of refraction and strain, but also $\bk$-advection, which has so far received limited attention. The limit of weak dispersion proves enlightening not only to explain why strain is impotent in the dipole, but also by yielding a remarkably simple analytical wave solution in arbitrary barotropic flows.  


\subsection{General wavevector evolution}

Let's first consider general near-inertial wave dynamics for an arbitrary barotropic flow. We begin by decomposing the wave field into vertical normal modes:
\begin{align}
\LL A (x,y,z,t) & = - \sum_{n=1}^{\infty} r_n^{-2}  A_n(x,y,t) g_n(z) \label{vmode},
\end{align}
where $g_n$ is the vertical eigenmode and $r_n$ the Rossby radius of the $n$'th vertical mode. 
Subtituting \eqref{vmode} into the YBJ equation \eqref{ybj} yields decoupled equations for each vertical mode:
\beq
\partial_t A_n + J(\psi,A_n) + i \left( \beta y + \frac{\zeta}{2} \right) A_n - \frac{\ii}{2} \hbar_n \lap A_n = 0\com  \label{ybj_vmode}
\eeq
where $\hbar_n \defn r_n^2/f$ is the dispersivity of mode $n$.
Following \cite{klein2004} we  write the wave envelope as $A_n=R_n\text{e}^{i\theta_n}$, where $R_n$ and $\theta_n$ are the real-valued amplitude and phase. Substituting  into $\eqref{ybj_vmode}$ and separating  the real and imaginary parts:
\begin{align}
\partial_t R + J(\psi,R) =& -\frac{\hbar}{2} \left( 2\nabla R\cdot \nabla \theta + \lap \theta R \right) \label{r_eq}, \\
\partial_t \theta + J(\psi,\theta) =& -\left( \beta y + \frac{\zeta}{2}\right)  + \frac{\hbar}{2}\left( \frac{\lap R}{R} - |\bk|^2  \right)\per  \label{theta_eq} 
\end{align}
Above we have lightened notation by dropping the  mode index $n$. The  above equations are an exact reformulation of the YBJ equation for an arbitrary barotropic flow and arbitrary stratification. Note that  \eqref{theta_eq} is a generalized version of the dispersion relation \eqref{omega-f}, where $\theta_t \equiv -\omega$ and $J(\psi,\theta)$ is the Doppler shift.

The evolution of the wavevector is obtained by taking the horizontal gradient of \eqref{theta_eq}:
\begin{align}
\partial_t \bk + \underbrace{J(\psi,\bk)}_{\text{$\bk$-advection}} &+ \,\, \underbrace{J(\nabla \psi, \theta)}_{\text{strain}} = - \!\!\! \!\!\!\!\!\!\underbrace{\beta\hat{y}}_{\beta\text{-refraction}} - \!\!\! \!\!\! \underbrace{\tfrac{1}{2} \nabla \zeta}_{\zeta\text{-refraction}}  \nonumber \\
& +\underbrace{ \frac{\hbar}{2}\, \nabla \left( \frac{\lap R}{R} \right) - \left( \mathbf{c}^h_g \cdot \nabla \right) \bk}_{\text{dispersive effects}}. \label{k_full}
\end{align}
Equation \eqref{k_full}  incorporates all processes discussed so far --- $\beta$-refraction, $\zeta$-refraction and strain --- and  also group velocity propagation  and $\bk$-advection, which appear  in the  ray-derivative on  the left-hand side of  \eqref{eq1}:
\begin{align}
\frac{\ddg \bk}{\dd t} \ & \defn \partial_t \bk  +  J(\psi,\bk) + \left( \mathbf{c}^h_g \cdot \nabla \right) \bk. \label{ddg}
\end{align}
The WKB approximation of \eqref{k_full}   in \eqref{eq1} neglects only  the dispersive  term  $\hbar\grad (\lap R/2R)$. 


In section \ref{sec:refraction} we limited attention to the jet center, where both strain and the Doppler shift (and thus $\bk$-advection) are zero. The weakly-dispersive modes remain at the jet center and $\partial_t \bk = -\half \nabla \zeta$ (figure \ref{fig:km1}). That is, $\ddg /\dd t$ is well approximated by the Eulerian time derivative $\partial_t$ in \eqref{ddg}. But this is not the case in the confluence and diffluence regions, where neither strain nor $\bk$-advection is zero. Yet, $\partial_t \bk = -\half \nabla \zeta$ in these regions too (figure \ref{fig:km2}). The key to this apparent paradox is that strain and $\bk$-advection largely cancel each other in steady barotropic flows. To see this, we consider the limit of weak dispersion, $\hbar \to 0$.

\subsection{Weakly-dispersive limit: an analytical solution}

In the weakly-dispersive wave limit, $\hbar \rightarrow 0$, only the local, instantaneous vorticity gradient affects the wavevector of an initially-uniform inertial wave. Crucially, this holds true not only for the dipole case, but for \textit{any} barotropic quasi-geostropic flow, steady or unsteady. 

To see this, we note that in the weakly-dispersive limit, the phase-amplitude formulation of the $f$-plane YBJ in  \eqref{r_eq} and \eqref{theta_eq}, simplifies to:
\begin{align}
\partial_t R + J(\psi,R) & = 0\com  \label{amp_eq} \\
\partial_t \theta + J(\psi,\theta) & = -\tfrac{1}{2}\zeta \per \label{t_eq}
\end{align} 
In the numerical solution the equations above are solved alongside the barotropic quasi-geostrophic equation:
\beq
\partial_t \zeta + J(\psi,\zeta) =0. \label{zeta_eq}
\eeq
The exact solution for \textit{arbitrary} barotropic flow, verified by substitution, is:
\begin{align}
R &= R_0, \\ 
\theta & =\theta_0 -\frac{t}{2} \zeta(x,y,t) \com    \label{ez_sol_theta} \\ 
\bk & = -\frac{t}{2} \nabla \zeta(x,y,t),   \label{ez_sol}
\end{align}
where $R_0$ and $\theta_0$ are the uniform amplitude and phase of the initial condition. It is remarkable that the evolution of vorticity fully dictates the wave phase at all times. During the initial stages of evolution, $\zeta$-refraction imprints the vorticity onto the phase, and the phase and vorticity are subsequently advected by the same streamfunction. As a result, the spatial structure of the  phase $\theta$ is slaved to the vorticity field.

We emphasize that this solution retains all processes in \eqref{k_full} except dispersion (and the $\beta$-effect, which is  weak). In particular, \eqref{k_full} includes $\zeta$-refraction, strain and $\bk$-advection and holds for an arbitrary barotropic flow, \textit{i.e.}, the flow is neither assumed steady nor spatially uniform. Moreover, the solution gives the evolution of $\bk$ at a fixed point in space, \textit{not} following the wave packet. At all points in space and time, the  wavevector $\bk$ is determined by the local instantaneous vorticity gradient.
 In a steady flow such as the dipole, $\bk$ grows linearly in time, consistent with the behavior in the jet center (figure \ref{fig:km1}) and in the confluent and diffluent regions (figure \ref{fig:km2}).

\subsection{Validation}

\subsubsection{Dipole flow}

\begin{figure*}[h!]
\centering
\includegraphics[trim = 0 0 0 0, clip, width=.32\textwidth]{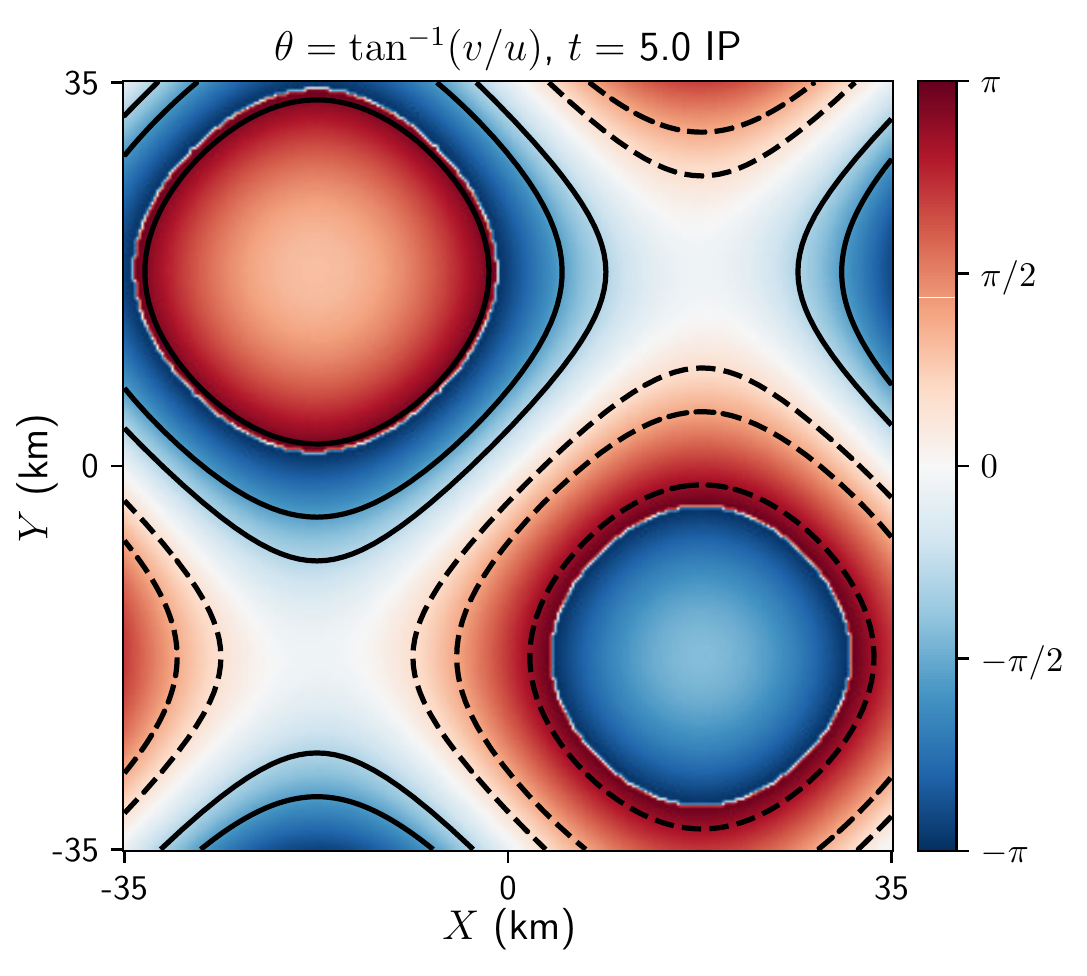}
\includegraphics[trim = 0 0 0 0, clip, width=.32\textwidth]{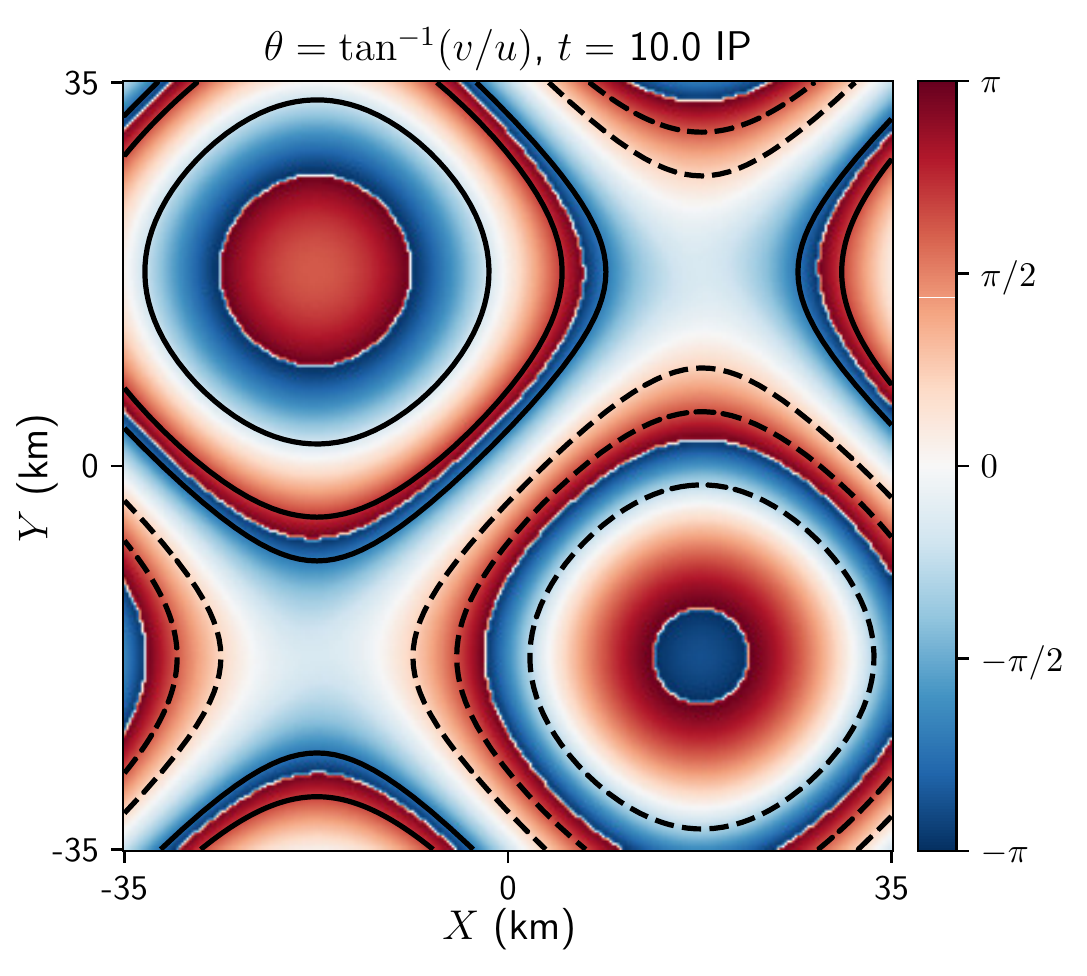}
\includegraphics[trim = 0 0 0 0, clip, width=.32\textwidth]{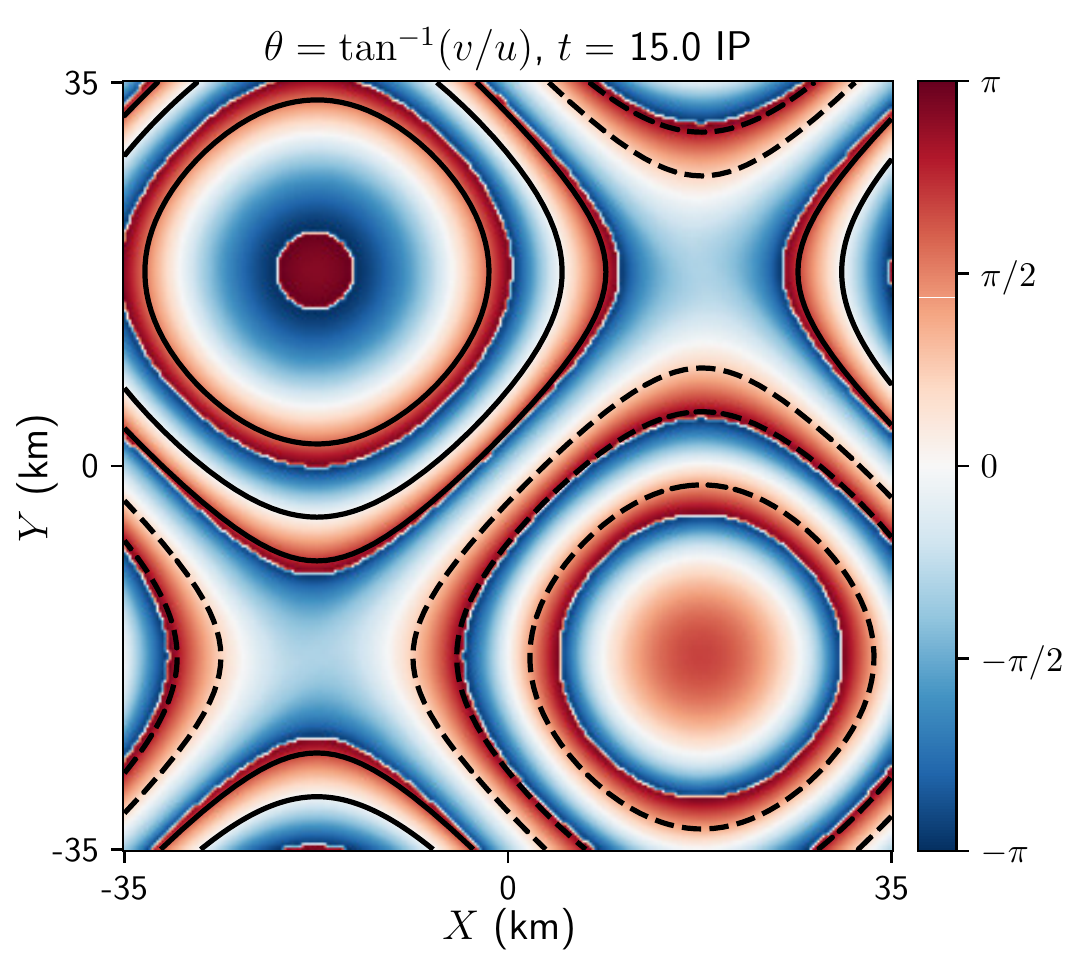}
\includegraphics[trim = 0 0 0 0, clip, width=.32\textwidth]{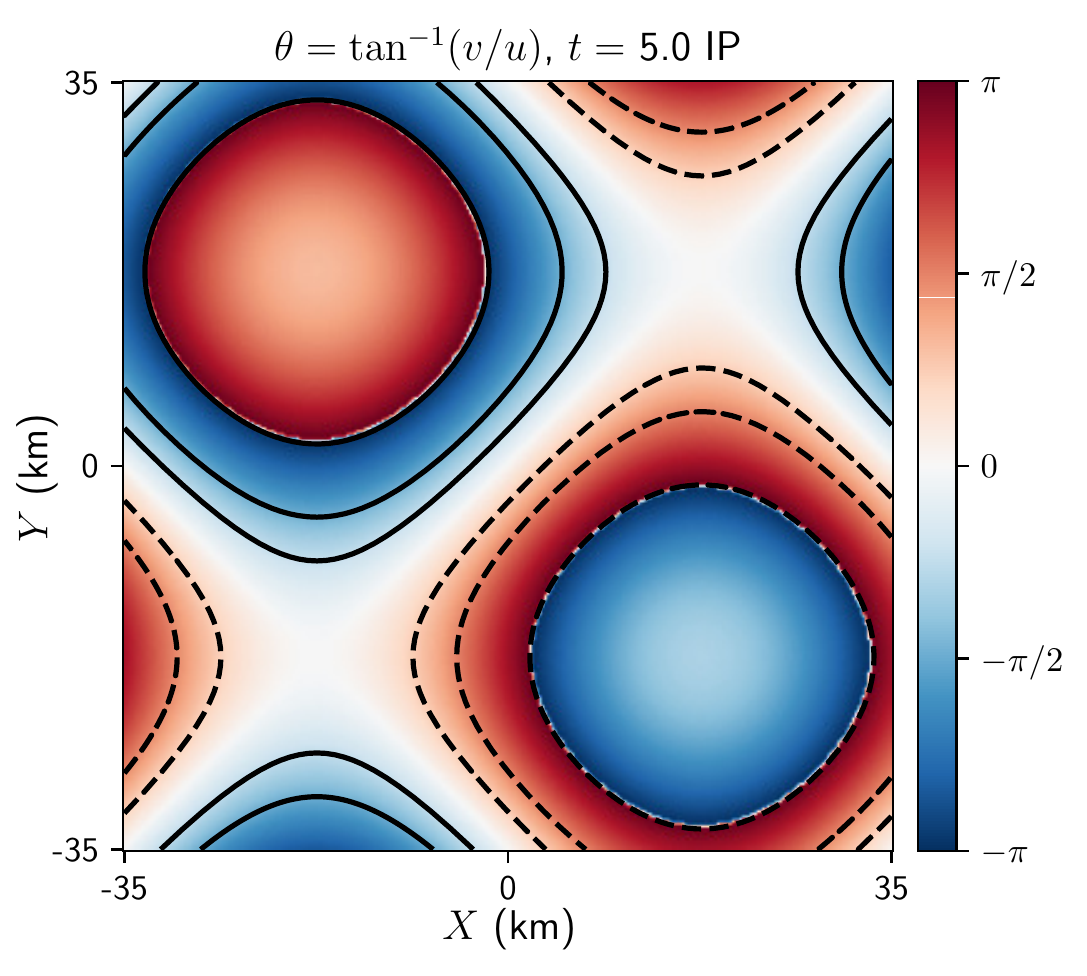}
\includegraphics[trim = 0 0 0 0, clip, width=.32\textwidth]{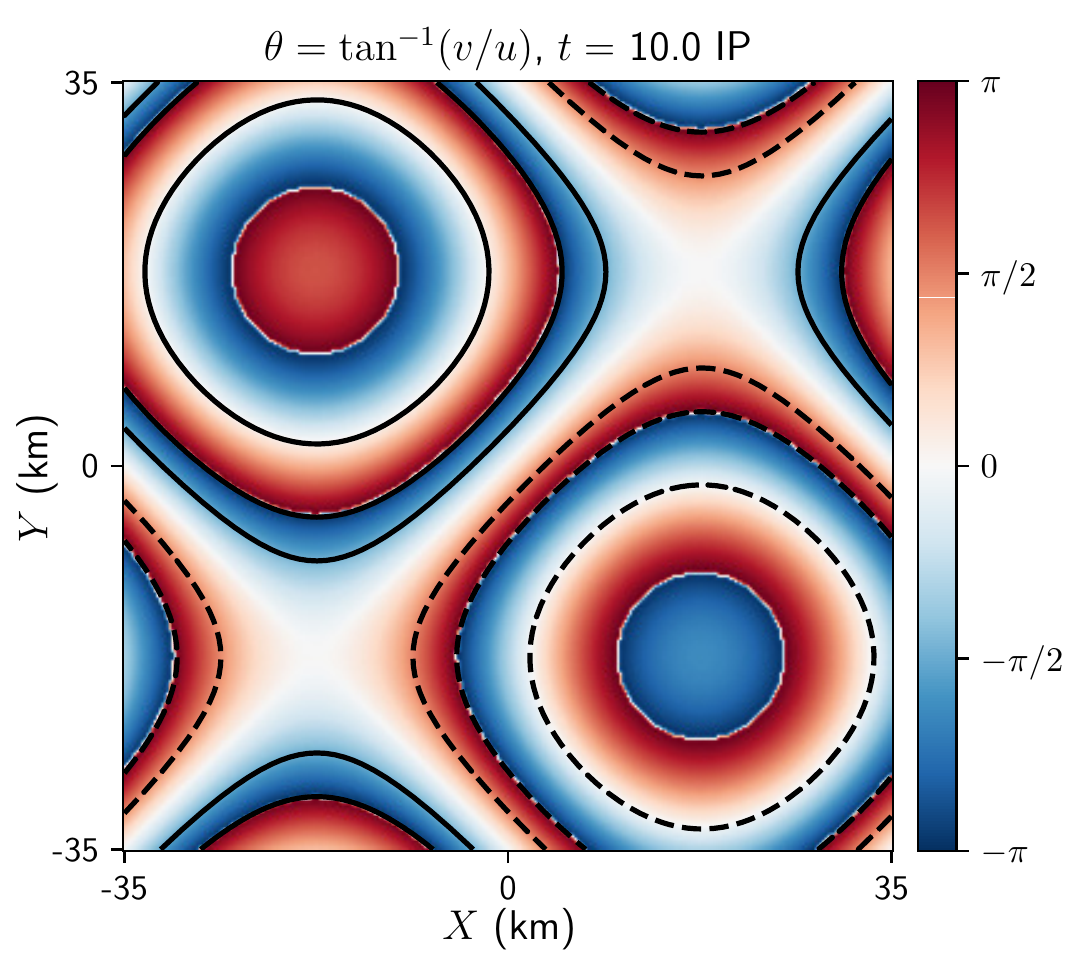}
\includegraphics[trim = 0 0 0 0, clip, width=.32\textwidth]{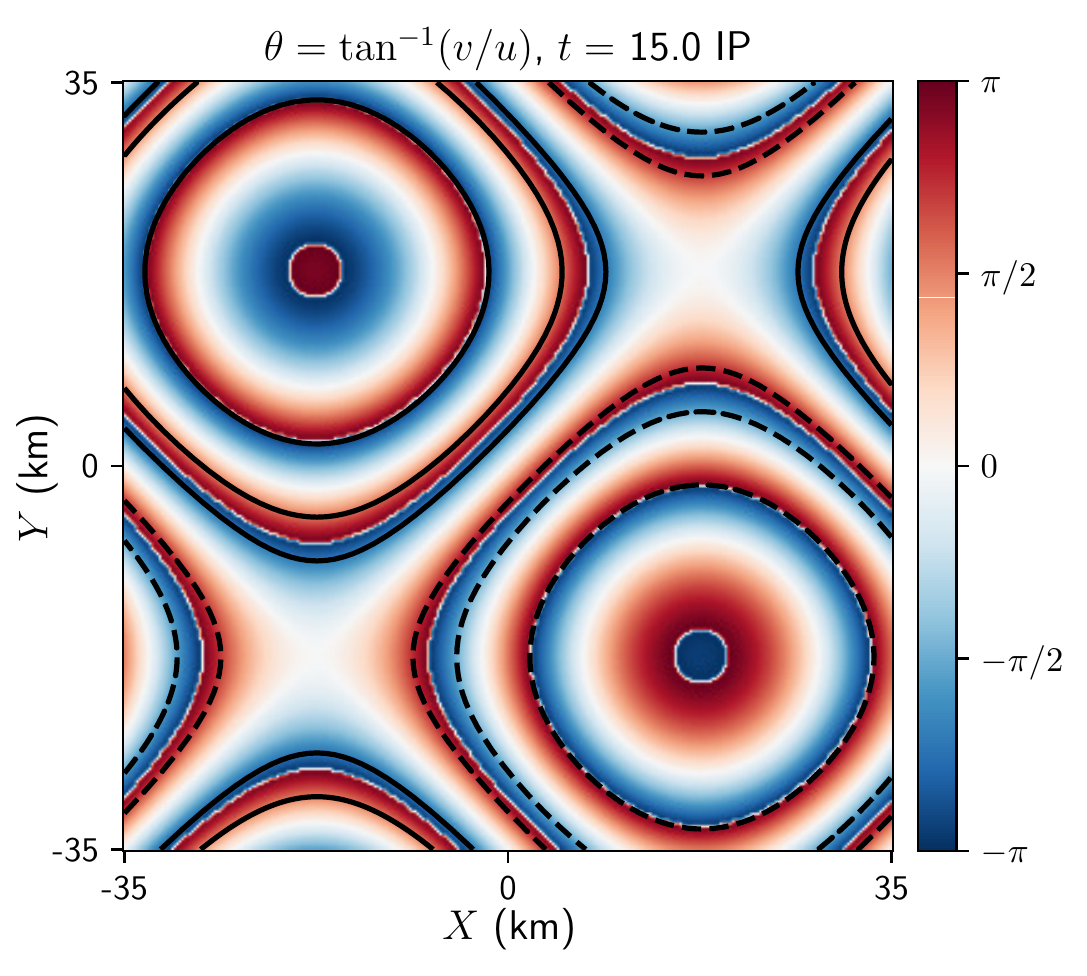}
\includegraphics[trim = 0 0 0 0, clip, width=.32\textwidth]{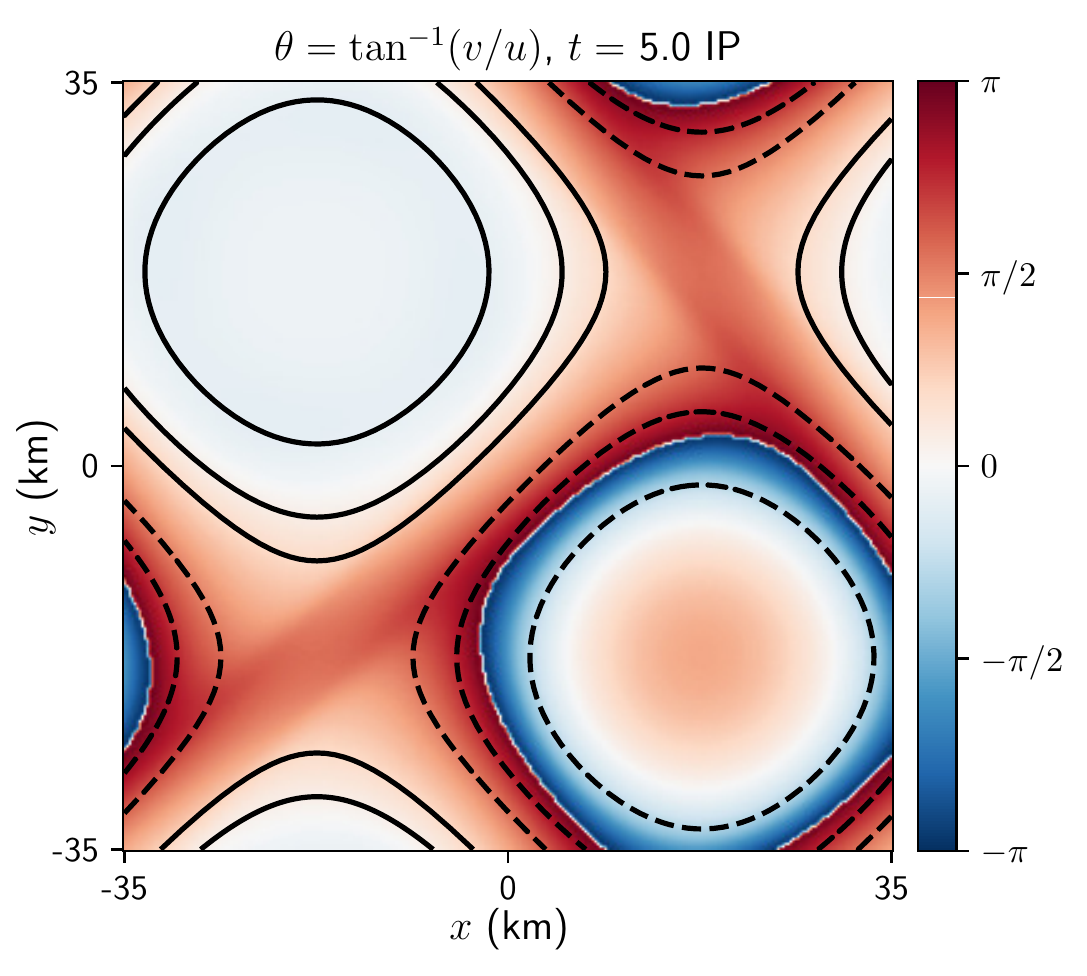}
\includegraphics[trim = 0 0 0 0, clip, width=.32\textwidth]{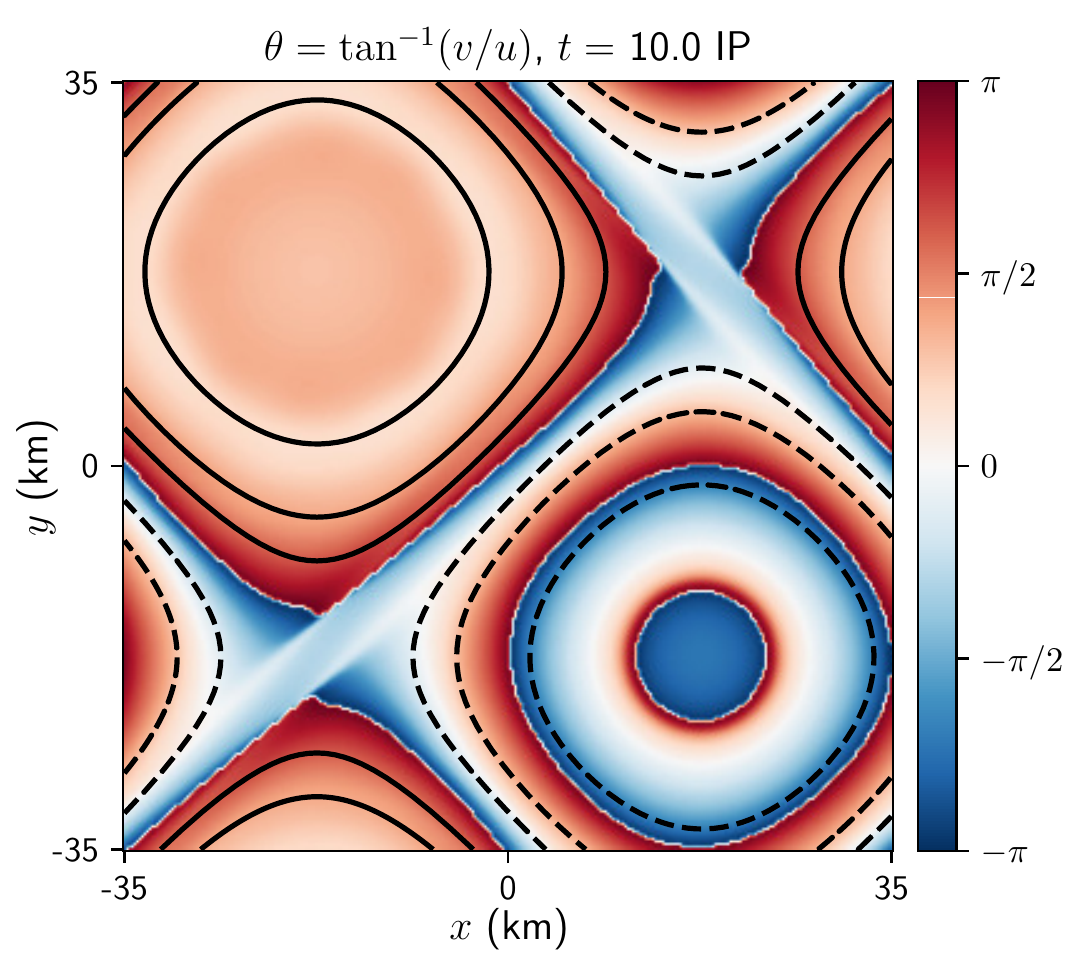}
\includegraphics[trim = 0 0 0 0, clip, width=.32\textwidth]{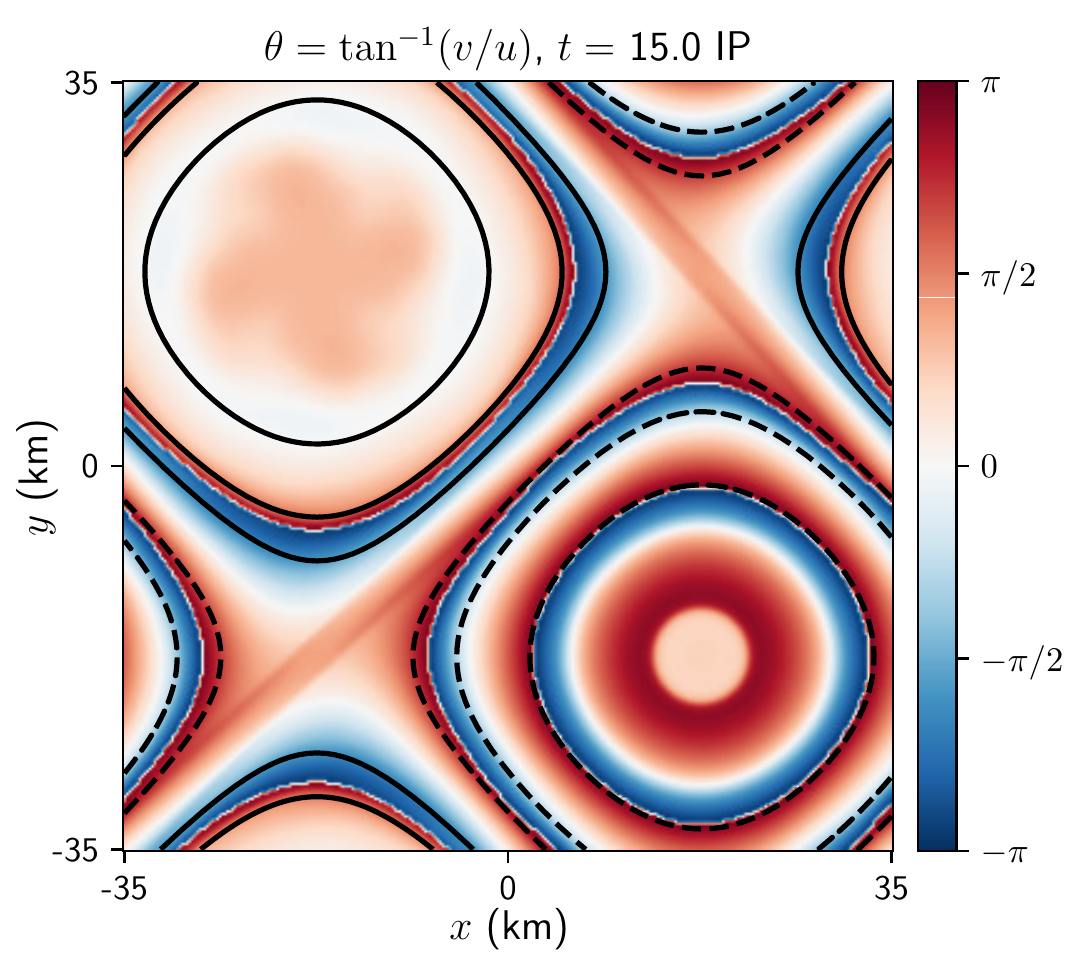}
\caption{Evolution of the wave phase ($\theta$ in \eqref{decomp}; here shown in colors) in the dipole flow. Top: sea-surface $\theta$ from the full, dispersive YBJ system, \eqref{ybj}, with $N^2=10^{-5}$ \pss. Middle: sea-surface $\theta$ non-dispersive YBJ solution, \eqref{ybj} with the $\lap A$ term set to zero. Bottom: $\theta$ at a depth of 200 meters in the full, dispersive YBJ system, \eqref{ybj}, with $N^2=10^{-5}$ \pss. Contours are for vorticity values of 0, $\pm 0.05$, $\pm 0.1 f$ and $\pm 0.2f$.}
        \label{fig:phase_dipole}
\end{figure*}

In figure \ref{fig:phase_dipole}, we compare the sea-surface wave phase ($\theta$, colors) for the dipole solution with (top) and without (middle) dispersion. The dispersive and non-dispersive solutions look qualitatively similar. The wave phase grows proportionally to the local $\zeta(x,y)$, such that vortex-shape annuli form, and with time shrink and multiply. In accordance to \eqref{ez_sol_theta}, the phase of the non-dispersive solution (bottom) is perfectly antisymmetric between the cyclonic and anticyclonic cores. Over time, this antisymmetry is broken in the dispersive solution (top). In particular, the term $-\hbar|\bk|^2$ in \eqref{theta_eq} is negative everywhere, causing a quicker decrease of the dispersive phase in the cyclone and a slower increase of the phase in the anticyclone. 

No wave energy propagation occurs without dispersion: the weakly-dispersive solution is agnostic about what happends below the mixed-layer, where wave energy remains confined. That said, the wave phase solution at a depth of 200 meters (bottom panels) does exhibit some similarities with the non-dispersive solution (middle panels) in the anticyclone. In particular, phase contours largely align with streamlines, indicating weak Doppler shift. Outside of the anticyclone, there is almost no wave energy and the wave phase is not a meaningful quantity to compare.

\begin{figure}[h]
\centering
\includegraphics[width=.45\textwidth]{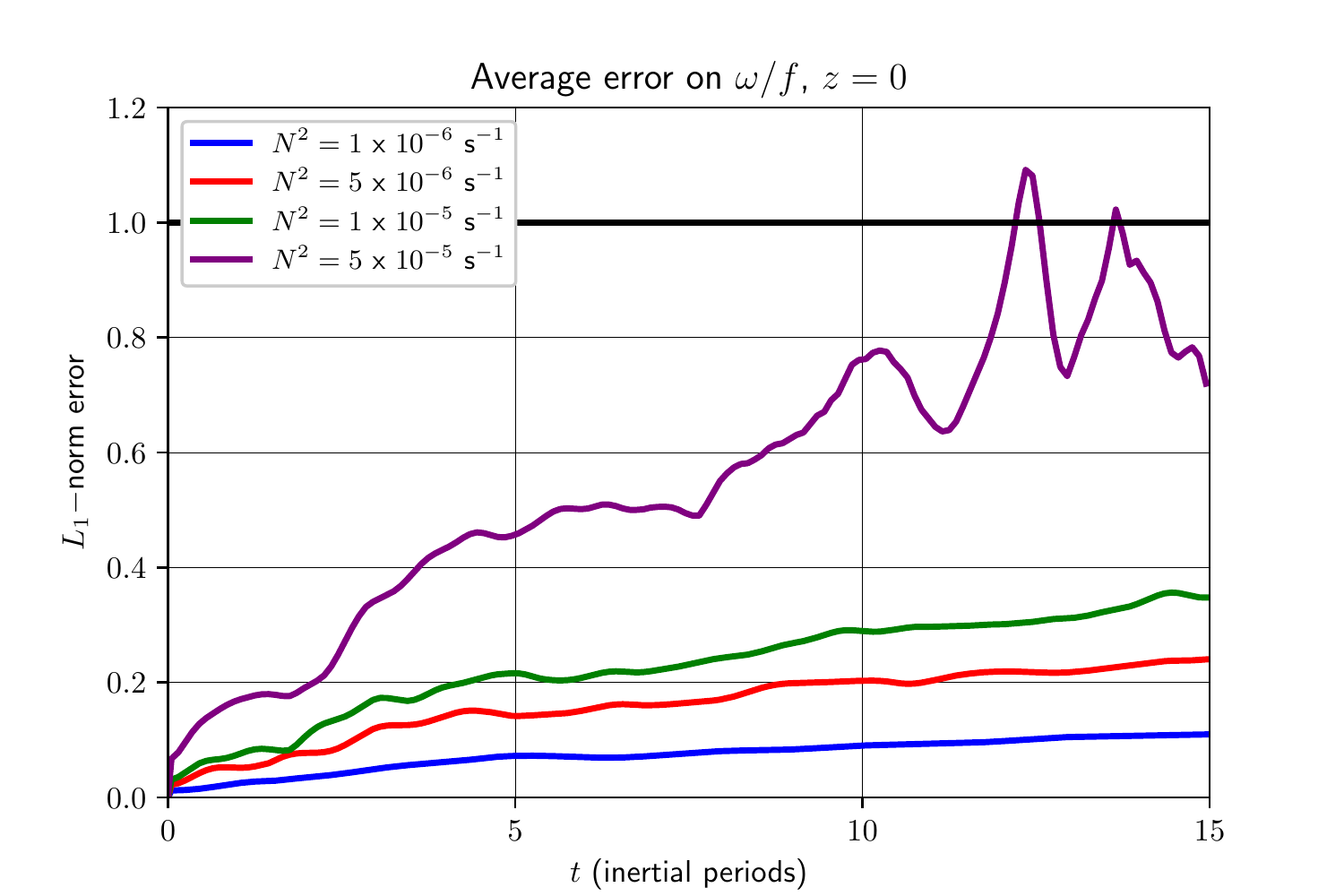}
\caption{Surface-averaged error on frequency for the dipole flow.}
\label{fig:dipole_error}
\end{figure}

Figure \ref{fig:dipole_error} shows the time series of the normalized sea-surface error on wave frequency,
\beq
e_\omega(t) \defn \frac{\laa| \omega(x,y,0,t)-\half \zeta(x,y) |\raa}{\laa \half \zeta(x,y) \raa},
\eeq
where $\omega \defn -\theta_t$ is the wave frequency computed from dispersive YBJ numerical solutions and brackets denote spatial averaging. As expected, the error grows monotonically with stratification, or dispersivity $\hbar$. The error nevertheless remains small for longer than expected from figure \ref{fig:time}, which gave estimates of the fraction of wave energy in dispersive modes after a given time. For instance, 50\% of the energy is in dispersive modes after only 4 inertial periods when $N^2=10^{5}$ \pss{} (green). The error in figure \ref{fig:dipole_error} stays small for longer, consistent with downward propagation of the low dispersive modes that do not obey the weakly-dispersive solution and the retention of the high modes which are better captured by it.





\subsubsection{Realistic flow}

\begin{figure*}
\centering
\includegraphics[trim = 0 0 0 0, clip, width=.32\textwidth]{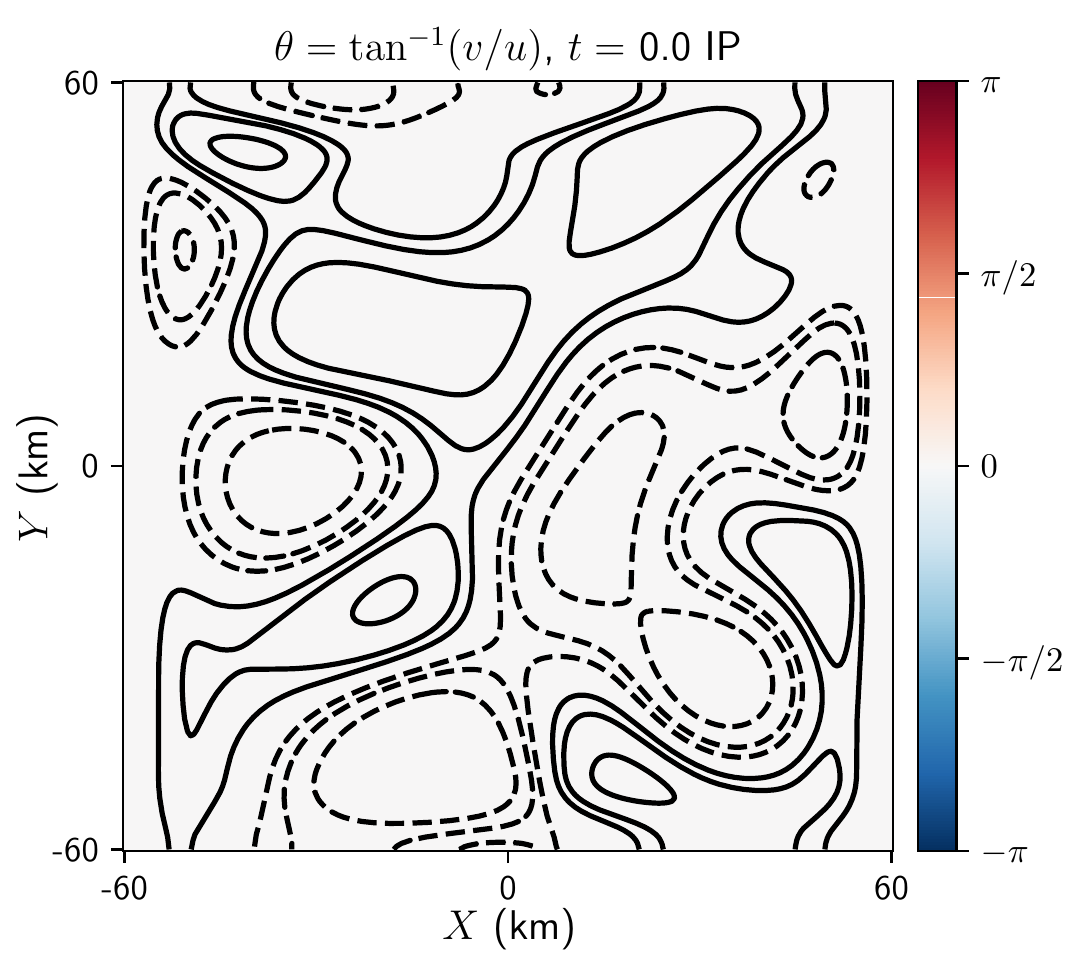}
\includegraphics[trim = 0 0 0 0, clip, width=.32\textwidth]{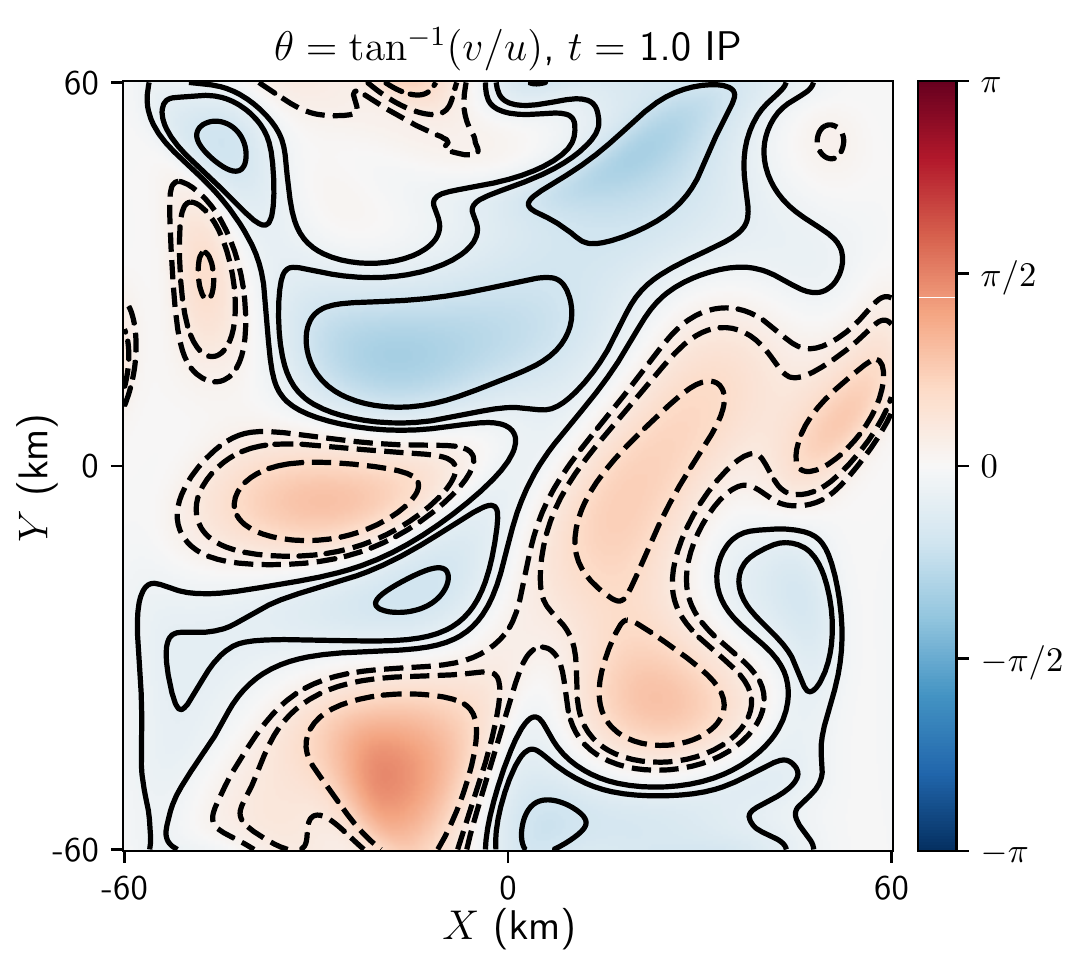}
\includegraphics[trim = 0 0 0 0, clip, width=.32\textwidth]{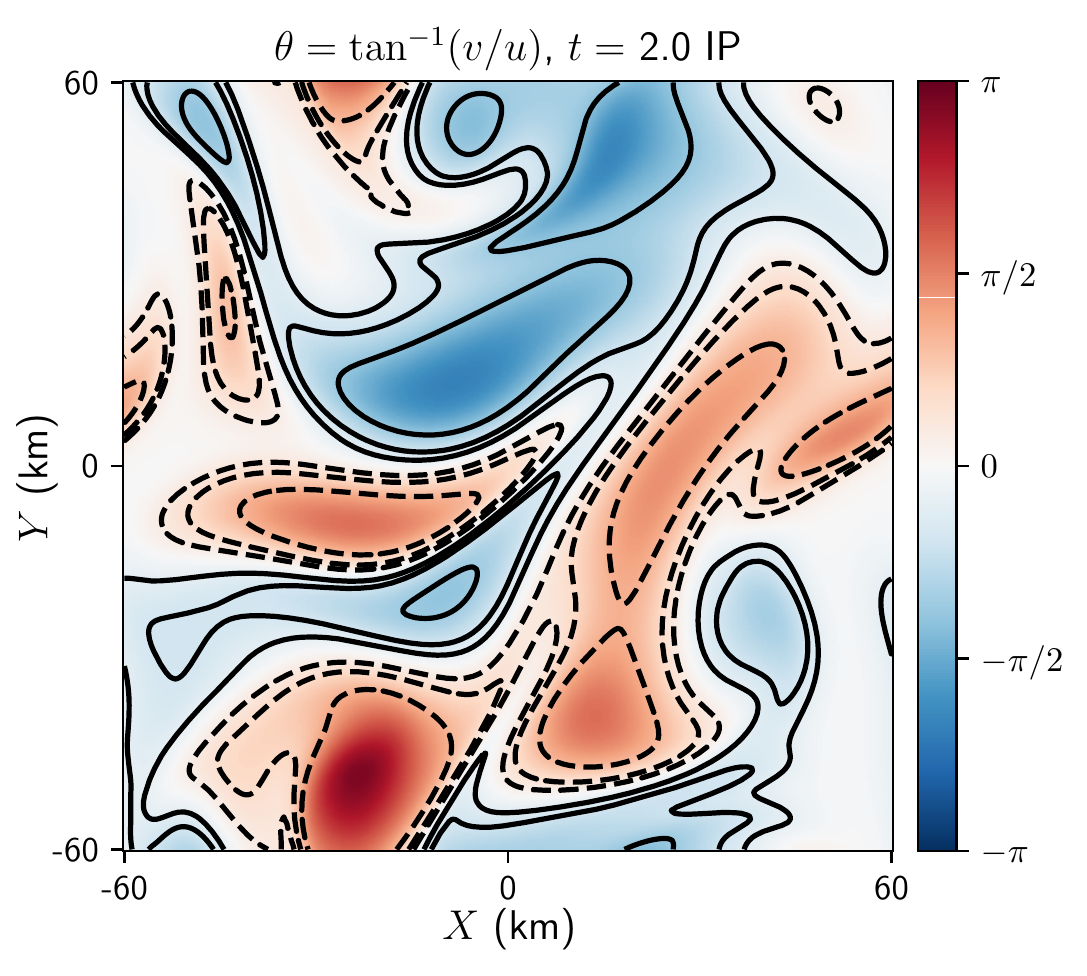}
\includegraphics[trim = 0 0 0 0, clip, width=.32\textwidth]{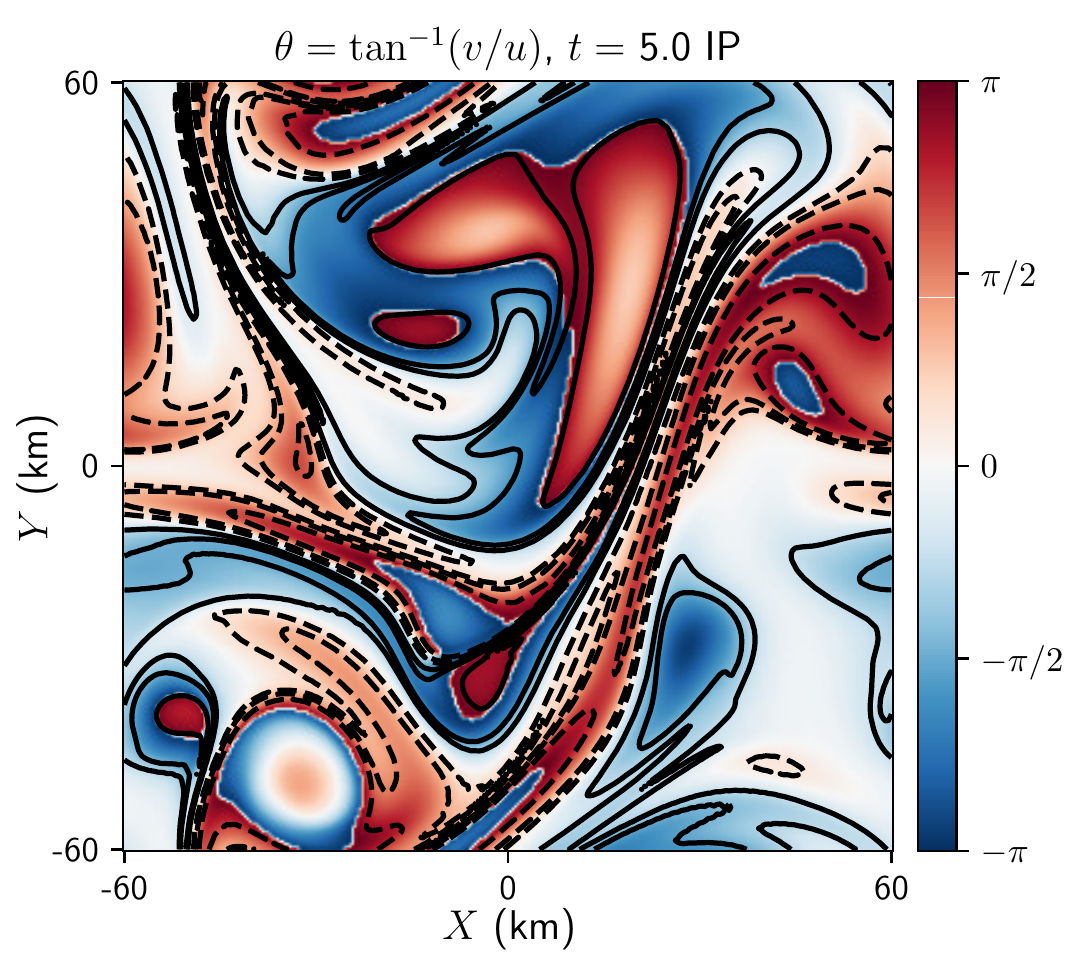}
\includegraphics[trim = 0 0 0 0, clip, width=.32\textwidth]{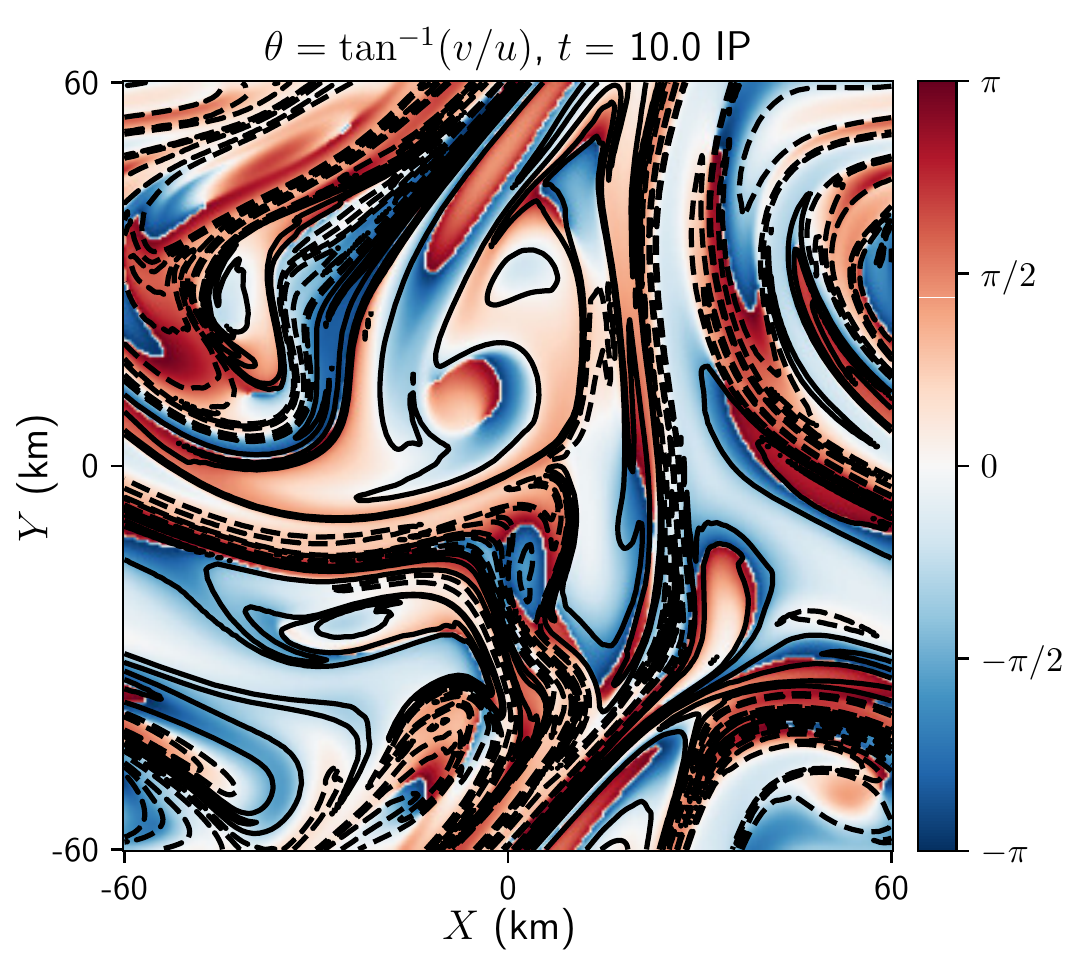}
\includegraphics[trim = 0 0 0 0, clip, width=.32\textwidth]{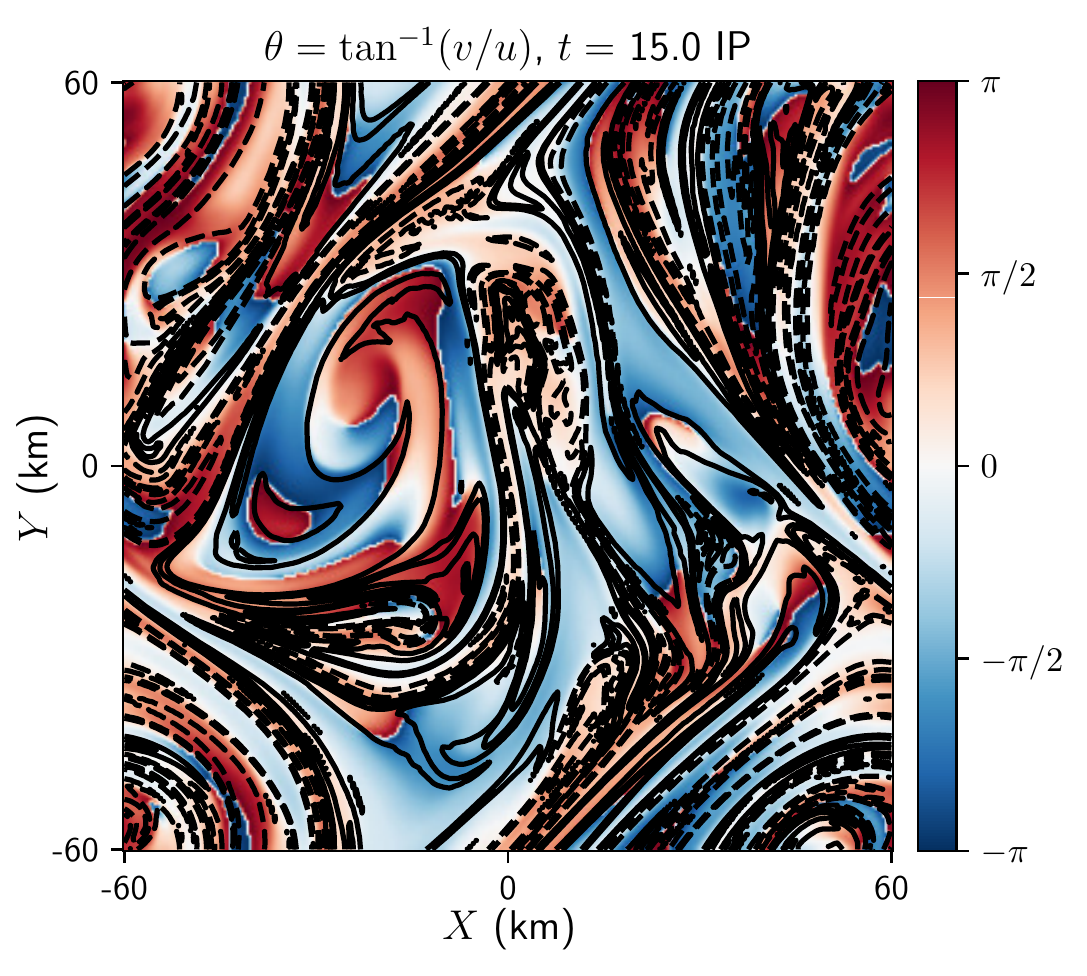}
\includegraphics[trim = 0 0 0 0, clip, width=.32\textwidth]{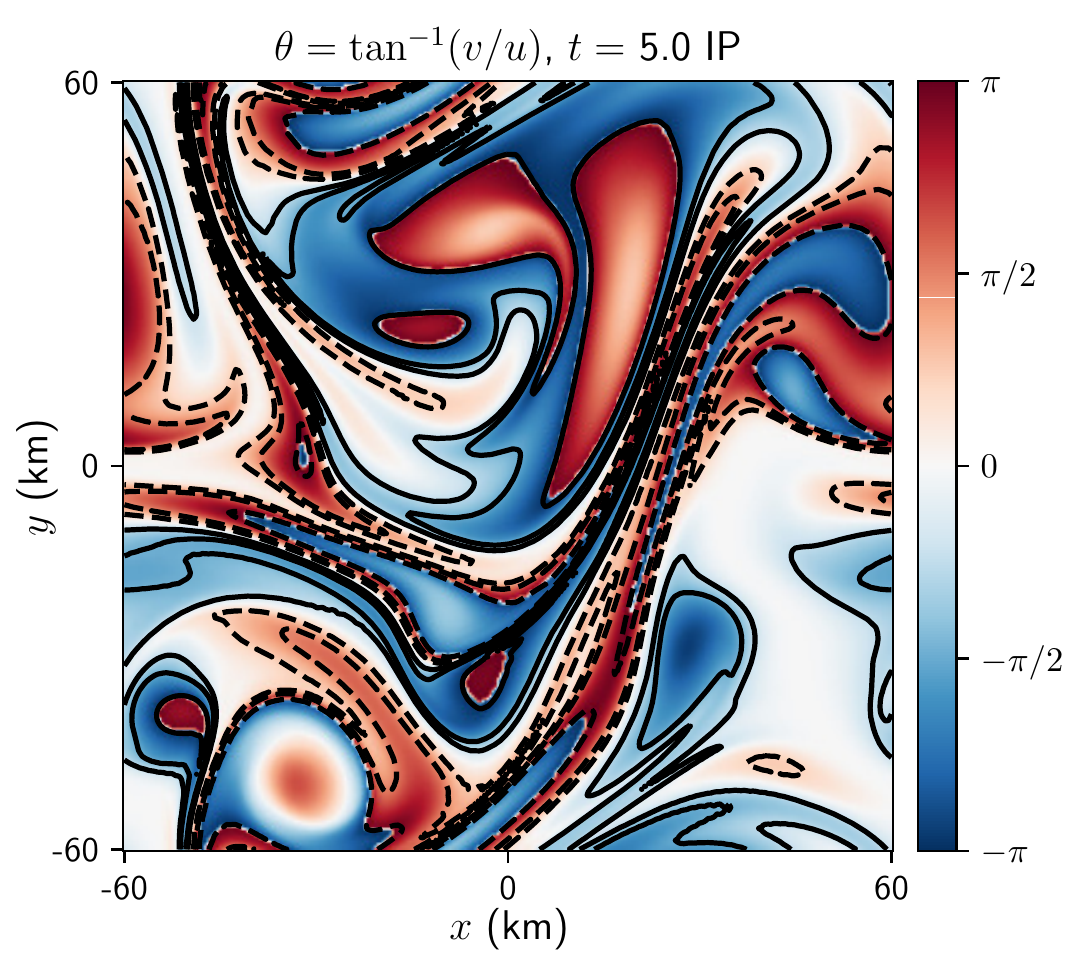}
\includegraphics[trim = 0 0 0 0, clip, width=.32\textwidth]{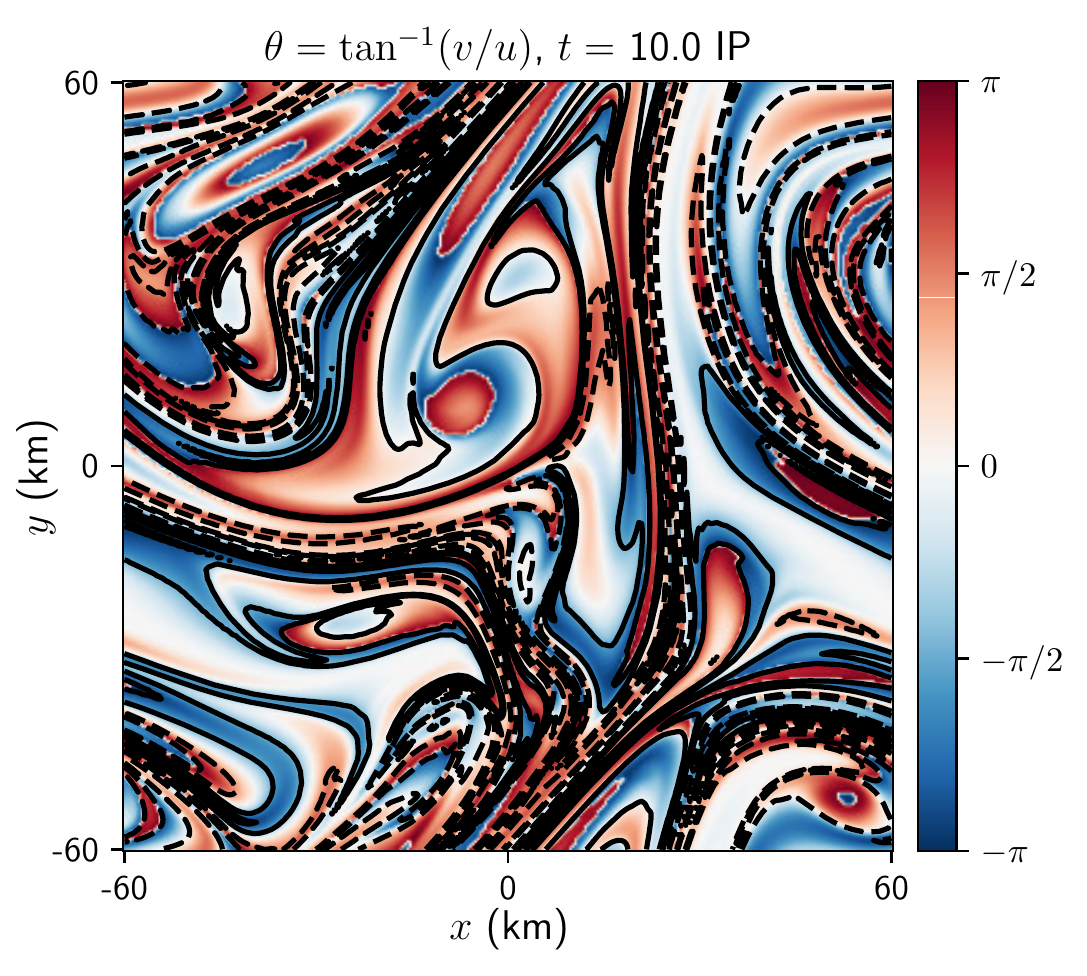}
\includegraphics[trim = 0 0 0 0, clip, width=.32\textwidth]{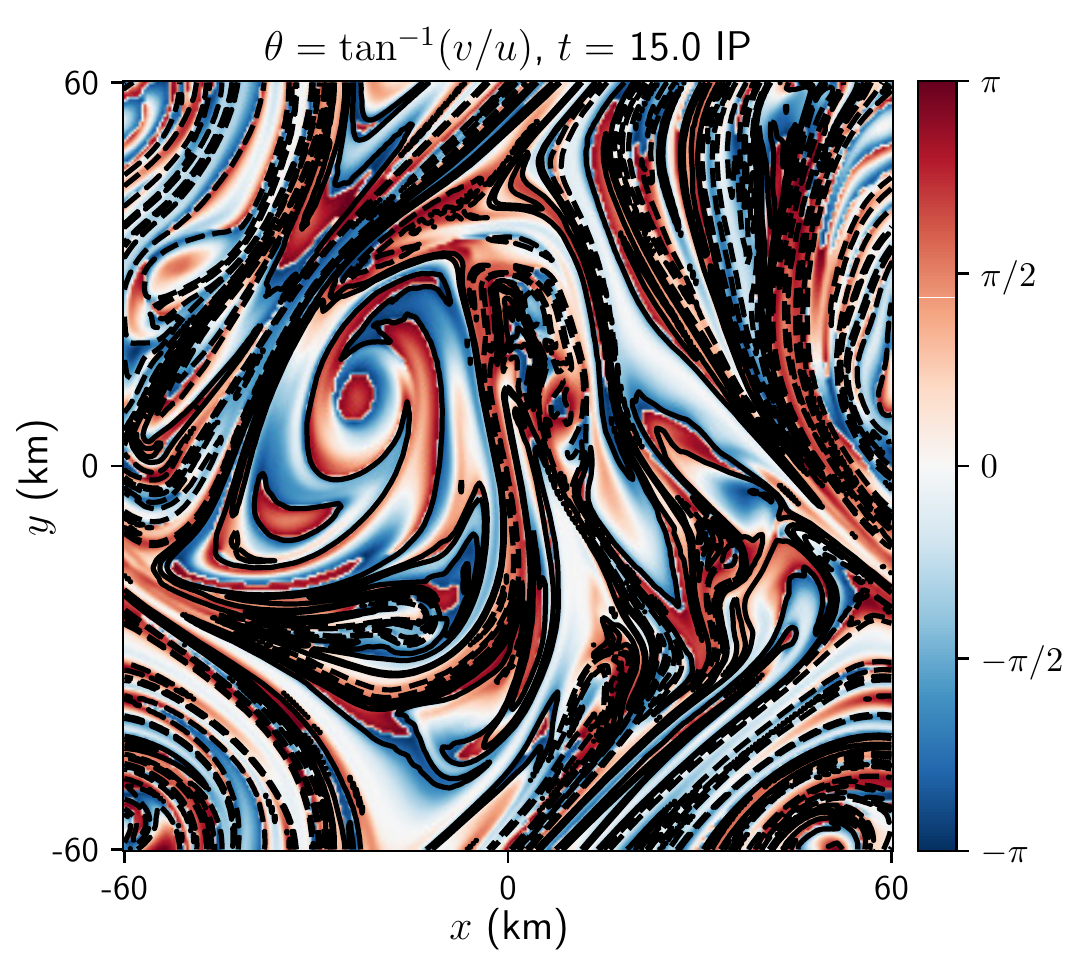}
\caption{Evolution of the sea-surface wave phase ($\theta$ in \eqref{decomp}; here shown in colors). The two upper panels show the solution of the full, dispersive YBJ system, \eqref{ybj}, with $N^2=10^{-5}$ \pss{} coupled with a flow evolving according to the barotropic QG equation, \eqref{zeta_eq}. The NISKINe observed flow (figure \ref{fig:vort}a) is used as an initial condition. Contours are for vorticity values of 0, $\pm 0.05$, $\pm 0.1 f$ and $\pm 0.2f$. For comparison, the bottom panels show the solution of the non-dispersive YBJ system for the same times as the middle panels: 5, 10 and 15 inertial periods.}
        \label{fig:phase}
\end{figure*}

As a more challenging test of the weakly-dispersive solution, the  YBJ equation is integrated with the usual wave initial condition, but coupled with the realistic NISKINe flow field (figure \ref{fig:vort}a). Contrary to the dipole flow, which is a steady solution of \eqref{zeta_eq}, the NISKINe flow is not steady. This is clear from the rapid evolution of the vorticity contours in figure \ref{fig:phase}. Within a few inertial periods, isolines of vorticity are squeezed into filaments as enstrophy cascades forward and the flow becomes rapidly unrecognizable. 

Overlaid on top of these vorticity contours are color maps of the sea-surface wave phase, $\theta$. Vorticity contours engulf regions of relatively uniform wave phase, consistent with phase being slaved to vorticity. Strikingly, the two top rows of figure \ref{fig:phase} show the solution of full \textit{dispersive} YBJ. The lower row of  figure \ref{fig:phase} shows the non-dispersive YBJ solution. Again, the dispersive and non-dispersive solutions look qualitatively similar, although they begin to diverge significantly after 10 and 15 inertial periods.



\begin{figure}
\centering
\includegraphics[width=.45\textwidth]{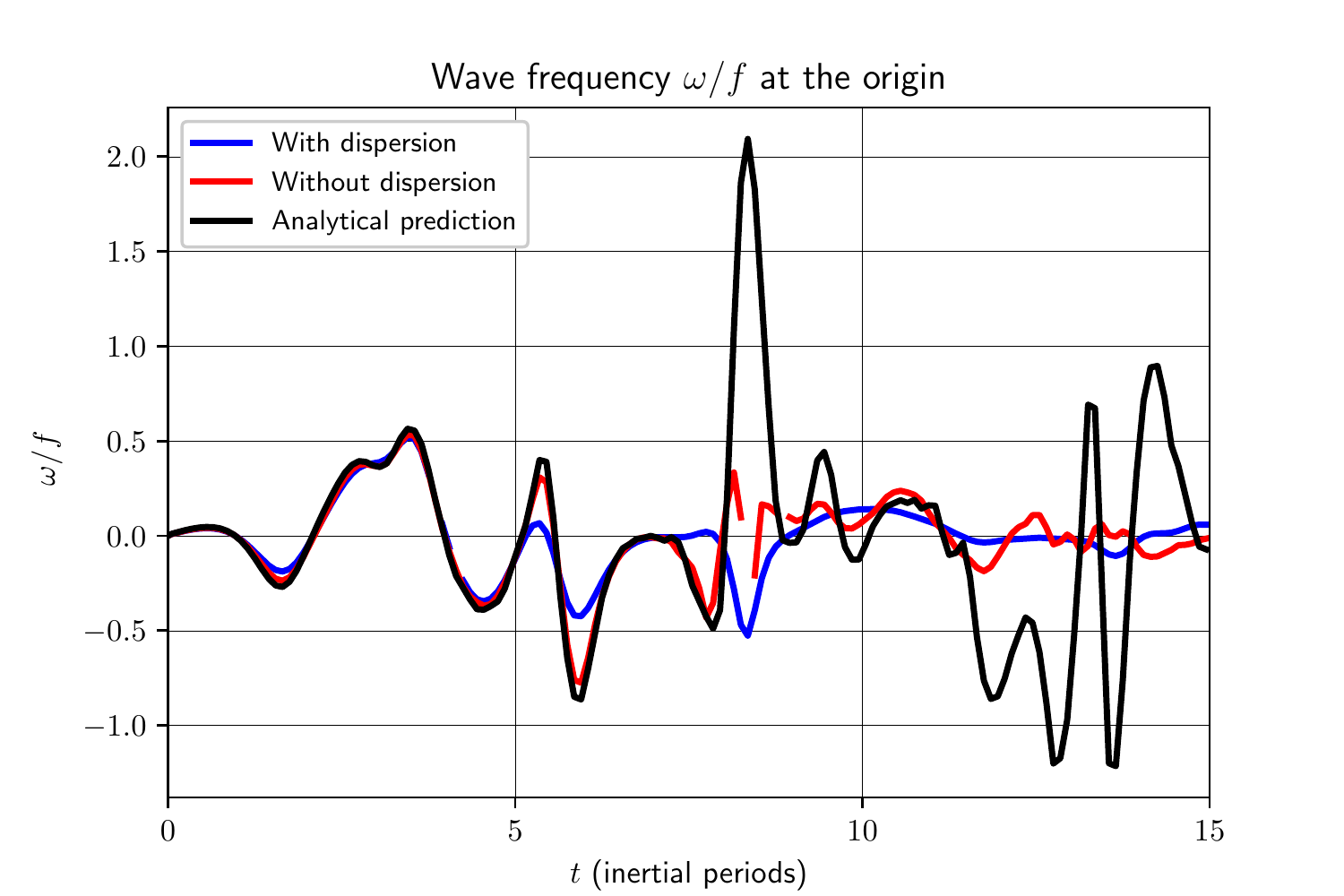}
\caption{Case: unsteady NISKINe flow with $N^2=10^{-5}$ s$^{-2}$. Wave frequency at the domain origin in the full numerical solution (blue/red: with/without dispersion) compared with the weakly-dispersive analytical solution (black).}
\label{fig:sigma}
\end{figure}

In figure \ref{fig:sigma}, we compare time series of the normalized wave frequency predicted from the weakly-dispersive analytical solution (black) with numerical solutions of the dispersive YBJ equation (blue; $N^2=10^{-5}$ \pss) and of a non-dispersive control run (red; YBJ with the $\lap A$ term set to zero). Even when dispersion is included, the numerical solution (blue) remains close to the analytical prediction for about 5 inertial periods. The non-dispersive numerical solution (red) is almost perfectly described by the analytical solution, \eqref{ez_sol}, except when sharp filaments pass by resulting in  significant dissipation \textit{e.g.}, at around 8 inertial periods.

\subsection{Interpretation}

\subsubsection{Steady Flow}

We finally return to one of the central question of this paper: why is strain ineffective in the dipole flow? The key is that both strain and $\bk$-advection originate  from the gradient of the Doppler shift (or phase advection):
\beq
\nabla \!\!\!\! \underbrace{J(\psi,\theta)}_{\text{Doppler shift}} = \underbrace{J(\psi,\bk)}_{\text{$\bk$-advection}} + \underbrace{J(\nabla \psi, \theta)}_{\text{Strain}}. \label{grad_theta_adv}
\eeq
In a barotropic steady flow such as the dipole, contours of vorticity align with streamlines: $J(\psi,\zeta)=0$ in \eqref{zeta_eq} if $\partial_t \zeta=0$. In the weakly-dispersive limit, wave phase is slaved to vorticity \eqref{ez_sol_theta} such that  wave phase contours also align with streamlines (figure \ref{fig:phase_dipole}). Thus, the Doppler shift remains zero at all times:
\beq
J(\psi,\theta) = -\half t J(\psi,\zeta) = 0. \label{theta_adv}
\eeq
Consequently, $\bk$-advection must cancel strain everywhere and at all times to satisfy \eqref{grad_theta_adv}-\eqref{theta_adv}:
\beq
\bk\text{-advection} = - \text{strain}. 
\eeq
This is true provided that (a) the barotropic flow is steady, and (b) the initial phase of the wave is uniform. Condition (b) is the usual assumption that the near-inertial wave is quickly generated by atmospheric forcing with large horizontal scale.

We can finally answer the question posed above: strain and $\bk$-advection are both ineffective in the dipole because they cancel each other. The jet region of the dipole is dominated by weakly-dispersive modes which have not escaped yet to the anticyclonic core. In accordance to \eqref{k_full}, the Eulerian wavevector increases linearly with the local vorticity gradient, 
\beq
\partial_t \bk = -\half \nabla \zeta(x,y), \label{steady_k}
\eeq
consistent with figures \ref{fig:km1} and \ref{fig:km2}. Only refraction modifies the wavevector.

Finally, we note that since the Doppler shift vanishes everywhere for steady flows in the weakly-dispersive limit \eqref{grad_theta_adv}, the \textit{intrinsic} wave frequency \eqref{omegai} is conserved along ray trajectories. Following section \ref{sec:refraction}\ref{sec:back}, wave bands can be back-tracked not only near the jet center, but anywhere.

\subsubsection{Unsteady Flow}

In unsteady flows, the wavevector is also only determined by the local instantaneous vorticity gradient, \eqref{ez_sol}. This seems to suggest that only refraction is acting. But this is not the case: strain also shapes the wavevector. Unlike the steady case, Doppler shift is \textit{not} zero in an unsteady flow:
\beq
J(\psi,\theta) = -\frac{t}{2} J(\psi,\zeta) = \frac{t}{2} \partial_t \zeta.  
\eeq
The gradient of $J(\psi,\theta)$ still results in  both $\bk$-advection and strain via \eqref{grad_theta_adv}, but these processes no longer perfectly cancel:
\beq
\bk\text{-advection} \neq - \text{strain}. 
\eeq
 In unsteady flows, the rate of change of the wavevector is the sum of the local instantaneous vorticity gradient (pure refraction) plus a term due to the time-dependence of vorticity, which encapsulates the non-canceling effects of strain and $\bk$-advection:
\beq
\bk_t = -\underbrace{\frac{1}{2} \nabla \zeta(x,y,t)}_{\text{refraction}} - \underbrace{\frac{t}{2} \partial_t  \nabla \zeta(x,y,t)}_{\text{strain}}. \label{unsteady_k}
\eeq
If the flow is steady, only refraction is effective and one recovers \eqref{steady_k}.



\section{Energetics}

\begin{figure*}
\centering
\includegraphics[trim = 0 0 0 0, clip, width=.4\textwidth]{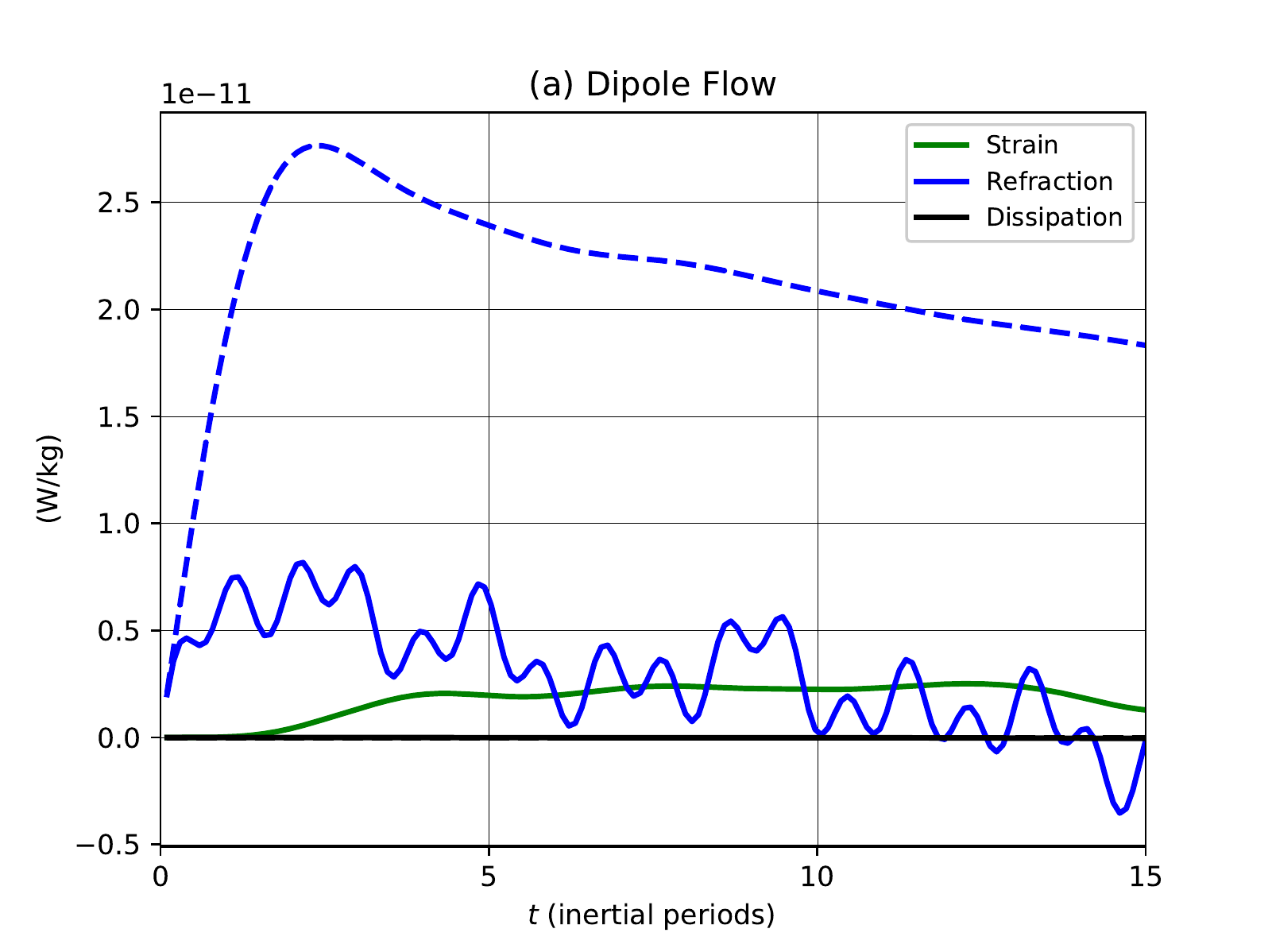} \quad
\includegraphics[trim = 0 0 0 0, clip, width=.4\textwidth]{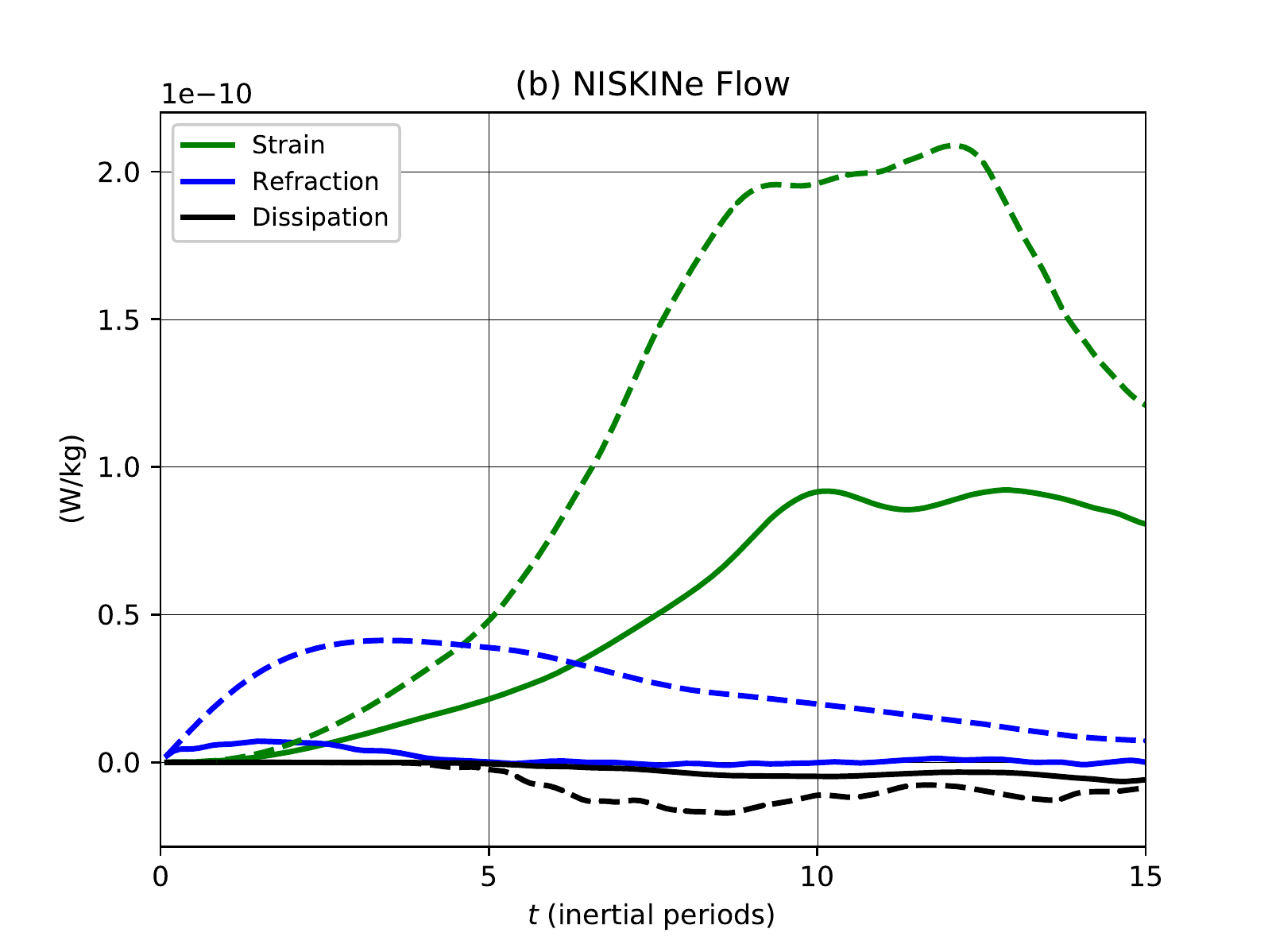}
\caption{Wave potential energy production via refraction \eqref{Gamma_r} and strain \eqref{Gamma_s}, and loss via dissipation, for the dipole (a) and NISKINe (b) flows, with (solid) and without (dashed) dispersion.}
\label{fig:wpe}
\end{figure*}


To quantify the roles of refraction and strain in eddy-wave energy transfers, we introduce the volume-averaged refraction- ($\Gamma_r$) and strain-induced ($\Gamma_s$) production of wave potential energy \citep{cesar},
\begin{align}
\Gamma_r & \defn- \laa \frac{1}{2f} \nabla \zeta \cdot \mathbf{F} \raa \label{Gamma_r}, \\ 
\Gamma_s & \defn  - \frac{1}{4} \frac{f^2}{N^2} \laa
 \begin{pmatrix} A^*_{xz} & A^*_{yz}
		\end{pmatrix} \begin{pmatrix}
S_n & S_s \\ 
S_s & -S_n \\ 
\end{pmatrix}   \, \begin{pmatrix} A_{xz} \\ A_{yz}
		\end{pmatrix}
\raa, \label{Gamma_s}
\end{align}
where $\mathbf{F}$ is the wave energy flux. Refraction produces wave potential energy ($\Gamma_r > 0$) when the wave energy flux goes against vorticity gradients, \textit{i.e.}, as wave energy propagates towards more anticyclonic regions. Straining produces wave potential energy ($\Gamma_s > 0$) when wave gradients are enhanced via geostrophic straining.

Figure \ref{fig:wpe} (a) shows wave potential energy production for the dipole flow, with (solid lines) and without (dashed lines) dispersion. As predicted, the non-dispersive dipole has exactly zero contribution from strain (green dashed line is zero). With dispersion, refraction (solid blue) dominates the early-time creation of gradients as wave energy accumulates in the anticyclonic region. Straining (solid green) does kick in after a few inertial periods, but its magnitude remains relatively weak, compared with expectations from passive-scalar picture of straining. This is consistent with figure \ref{fig:phase_dipole}, in which wave phase lines remain nearly parallel to streamlines, such that $J(\psi,\theta)\approx 0$ and strain is largely cancelled by $\bk$-advection. 

Strain is more potent in the NISKINe flow (b) than in the dipole (a). This is because letting the flow evolve allows the generation of stronger flow gradients via the forward enstrophy cascade. Note that straining is not, like for the dipole, eliminated by removing dispersion (green dashed). Quite the contrary: strain is much stronger in the non-dispersive than in the dispersive case for the NISKINe flow. In such case, waves cannot disperse and wave escape is impossible. Dissipation is correspondingly larger as waves are strained into oblivion.

\section{Discussion and conclusions} \label{sec:disc}

In this paper we examined the evolution of wind-generated near-inertial waves in steady and unsteady barotropic flows, with an eye on how refraction and strain shape the wavevector. We now assemble the main findings of the previous sections and discuss their implications and limitations.



\subsection{Wave bands along the dipole jet}

Nearly monochromatic wave bands appear along the dipole jet as wave energy propagates downwards and towards the anticyclonic drainpipe (figure \ref{fig:dipole_overview}). With the knowledge of stratification and vorticity gradient alone, one can predict the bands' horizontal \eqref{kt} and vertical \eqref{mt} wavenumbers, their slope \eqref{dzdx_pred}, Burger number \eqref{bu_pred}, and the trajectory of the beams. Although these results followed from analysis at the jet center, their validity extends along the axis of the jet. Conservation of the intrinsic frequency \eqref{omega-f} allows one to back-track the surface origin of wave bands observed at depth in the jet vicinity. Finally, if the distribution of vertical wave modes is known, the radiation of wave energy out of the mixed layer can be estimated (figure \ref{fig:time}).


\subsection{Strain is ineffective in steady barotropic flows}

The weakly-dispersive limit helps explain why wind-generated inertial waves do not experience a strain-induced exponential growth of $|\bk|$ in barotropic steady flows. If the primordial wave has a horizontal scale larger than that of eddies, $\zeta$-refraction is the only process acting initially. Vorticity imprints its scales onto wave phase, and both fields are subsequently advected by the same streamfunction. In barotropic steady flows, vorticity advection, and thus phase advection is zero everywhere \eqref{theta_adv}. This means that strain and $\bk$-advection, which both originate from the gradient of phase advection \eqref{grad_theta_adv}, must cancel each other.  As a result, the $\bk=-\grad \zeta(x,y)t/2$ (figures \ref{fig:km1} and \ref{fig:km2}). Note that this is true only of the weak-dispersion limit. When dispersion is included, strain operates, albeit more weakly than anticipated from the passive-scalar analogy (figure \ref{fig:wpe}).


\subsection{Strain is effective in unsteady barotropic flows}

In the limit of weak dispersion, the wavevector grows proportionally to the local instantaneous vorticity gradient \eqref{ez_sol}. This result  does \textit{not} imply that the wavevector is only modified by refraction in unsteady flows. Unsteady vorticity gradients are associated with straining \eqref{unsteady_k}. In the unsteady flow considered, strain is actually more effective than refraction in producing wave gradients (\textit{c.f.} figure \ref{fig:wpe}). Only for steady flows such as the dipole is strain cancelled by $\bk$-advection. 

\subsection{Forward cascade of wave phase}

Unsteady quasi-geostrophic flows, by analogy with two-dimensional flows, promote forward cascades of potential vorticity variance \citep{kraichnan1967,charney1971}. For barotropic flows, this implies a relentless enhancement of vorticity gradients via squeezing and stretching of filaments until statistical stationarity is attained (figure \ref{fig:phase}). In the weakly-dispersive limit, the wave phase is slaved to vorticity \eqref{ez_sol_theta}. One therefore also expects a forward cascade of wave phase variance --- in fact even faster than that of vorticity variance because of the additional $t^2$ factor in \eqref{ez_sol_theta}. In other words, the wave phase rapidly becomes  decoherent as a consequence of quasi-geostrophic turbulence (\textit{c.f.} figure \ref{fig:phase}).

As wave phase gradients ($\nabla \theta = \bk$) are enhanced by this forward cascade, $Bu$ increases and strengthens dispersive effects. \cite{ybjp} show that dispersion eventually halts the forward cascade. This is consistent with the right panel of figure \ref{fig:wpe}, which shows shear production by strain (green) and loss via dissipation (black) in the unsteady NISKINe flow. The dispersive solution (solid) suffers less straining and dissipation than the non-dispersive solution (dashed) --- wave escape upsets wave capture \citep{cesar}.

\subsection{Predictability of the wave phase in a barotropic flow}

The weakly-dispersive solution \eqref{ez_sol_theta} promises strong predictive powers over the sea-surface wave phase for both steady and unsteady barotropic flows. According to this solution, one can deduce the wave phase anywhere and anytime from the local instantaneous flow vorticity and time elapsed since the wave inception. The horizontal wavevector or wave frequency can similarly be predicted given the local instantaneous vorticity gradient or tendency.

It is remarkable that wave dynamics do not explicitly depend on the \textit{history} of the evolution of the barotropic flow --- whatever happened between the wave inception and the time of measurement --- but only on its \textit{instantaneous} state. The wave phase also depends only on \textit{spatially-local} measurements of vorticity. The spatiotemporal locality of the weakly dispersive solution makes it powerful for interpreting spatially- and temporally-sparse observational data.

With great predictive powers come great limitations. The weakly-dispersive solution crucially depends on three restrictive assumptions: \textit{(i)}  the flow must be barotropic, otherwise the wave phase and vorticity will diverge from one another; \textit{(ii)}  the primordial horizontal scale of the wave must be much larger than eddies, otherwise the initial refractive imprinting will be imperfect; \textit{(iii)}  the waves must be weakly dispersive. This last assumption limits the validity of the weakly-dispersive solution to early times following the wave inception, \textit{i.e.}, before a significant fraction of the wave energy is radiated away by the dispersive modes (figure \ref{fig:time}). That said, figures \ref{fig:phase_dipole}-\ref{fig:sigma} suggest that the prediction's validity extends far beyond what is expected. Finally, the weakly-dispersive solution is restricted to shallow depths. No wave energy propagation is permitted in the non-dispersive solution, and thus no prediction is given below depths where energy is initially located.

The weakly-dispersive solution is useful despite its limitations. \cite{leifobs} use the weakly-dispersive solution \eqref{ez_sol} to explain the wavevector evolution observed in the unsteady NISKINe flow. In two distinct regions of the flow, timeseries of the wavenumber closely follow the observed local instantaneous vorticity gradient for several inertial periods. This is a strong test of the weakly-dispersive solution, where $\grad \zeta$ varies both in space and time.







\acknowledgments  O.A. thanks Eric Kunze for insightful discussions. This work was supported by the National Science Foundation Award  OCE-1657041 and by the Office of Naval Research award N00014-18-1-2803. We are grateful for computer resources provided by  the Extreme Science and Engineering Discovery Environment (XSEDE), which is supported by National Science Foundation grant number ACI-1548562.

\section*{Appendix I: Jet region solution}

We seek a solution of the linear YBJ equation in the vicinity of the jet region, $\kappa x\ll 1$, and along the line joining the vortex cores ($y=0$). Assuming constant $f$ and $N$, YBJ \eqref{ybj} melts down to:
\beq
A_{zzt} + \frac{i}{2} \zeta A_{zz} + i\frac{N^2}{2f}  A_{xx} = 0 \label{ybj_simple}.
\eeq
Following \cite{moehlis2001}, we seek for solutions of the form $A(x,z,t)=B(x,z,t)\exp[-i\zeta t/2]$. Substituting this ansatz into \eqref{ybj_simple} yields
\beq
B_{zzt}   + \frac{N^2 \zeta_{x}}{2f} t B_{x}  = i \frac{N^2 \zeta_{x}^2 }{8f} t^2 B - \frac{N^2\zeta_{xx}}{4f} t B  - i \frac{N^2}{2f} B_{xx} . \label{beq_full}
\eeq
It is insightful to non-dimensionalize \eqref{beq_full} with
\begin{gather}
\! \! \! X= x \kappa, \quad  Z = \frac{z}{\sigma}, \quad T = \frac{t}{\tau}, \quad
 \tau = \left( \frac{f}{N^2\gamma^2 \sigma^2} \right)^{1/3}\! \! \! \! \! \! \! \! \!.
\end{gather}
Then, \eqref{beq_full} becomes:
\begin{align}
B_{ZZT} - & \tfrac{1}{2} \eta T \cos X B_{X} =  \nonumber\\
& \tfrac{i}{8} T^2  \cos^2 X B -  \tfrac{1}{4} \eta  T \sin X  B - \tfrac{i}{2} \chi B_{XX}  \label{beq_full_nd}.
\end{align}
Two dimensionless numbers emerge; using the typical dipole values we obtain:
\beq
\eta = \frac{\kappa}{\gamma \tau} \approx 0.2, \qquad \chi = \left( \frac{N^2 \kappa^2 \sigma^2}{f^2} \right) (\tau f) \approx 0.03. 
\eeq
where $\tau \approx 2.5$ inertial periods. As such, the leading-order solution is dominated by:
\beq
B_{zzt} \approx i \frac{N^2 \zeta_{x}^2 }{8f} t^2 B  \label{beq_1st_order}.
\eeq
To solve the above equation, we first take a Fourier transform in the vertical,
\beq
\hat{B}(m) = \int_{-\infty}^{\infty} \!\! B(z) \ee^{-  \ii m z}\,  \dd z,
\eeq
where we assumed an infinite domain, \textit{i.e.}, the solution is largely concentrated near the surface and the bottom boundary can neglected. Note that we took an even extension of $B$ into positive $z$. The resulting ordinary differential equation has the solution: 
\beq
\hat{B}=\hat{B}_0 \exp \left(- \ii \frac{ N^2  \zeta_{x}^2 t^3}{24 f m^2} \right)\per  \label{Bsol}
\eeq
To recover the solution in terms of $A$, we have to perform the inverse Fourier transform of \eqref{Bsol}: 
\beq
B(z,t) = \frac{1}{2\pi}\int_{-\infty}^\infty \!\! \hat{B}_0(m) \ee^{\ii \Phi} \, \dd m, \label{Bint}
\eeq
where we introduced the phase
\beq
\Phi \defn - \frac{N^2 \zeta_{x}^2 t^3}{24 f m^2} + mz.
\eeq
We can find an approximate solution to this integral using the method of stationary phase (\textit{e.g.}, \cite{whitham2011} and references therein). The idea is that the largest contributions to the integrand happen when the phase is stationary, \textit{i.e.},  when $\Phi_m=0$. This is the case for:
\beq
m^* =  - \left( \frac{N^2 \zeta_{x}^2 }{12 f z} \right)^{\frac{1}{3}} t \label{mstar}.
\eeq 
Next, we expand the phase around $m^*$:
\begin{gather}
\Phi(m) \approx \Phi(m^*) + (m-m^*) \underbrace{\Phi_m(m^*)}_{=0} \nonumber  \\ + \half (m-m^*)^2 \Phi_{mm}(m^*) + \ldots
\end{gather}
Coming back to \eqref{Bint}, we pull all $m$-independent terms out of the integral:
\begin{align}
B(z,t)  \approx & \, \frac{\hat{B}_0(m^*)  \exp  i \Phi(m^*)}{2\pi} \, \times \nonumber \\ 
& \int_{-\infty}^\infty  \exp \left[ \halfi (m-m^*)^2 \Phi_{mm}(m^*) \right] \dd m. \label{Bint}
\end{align}
Noting that $\Phi_{mm}=-N^2 \zeta_{x}^2 t^3/4fm^4$ is always negative, we can change the variable $m$ to $n= (m-m^*) \sqrt{|\Phi_{mm}(m^*)|/2}$:
\beq
B(z,t) \approx  \frac{\hat{B}_0(m^*)  \exp  i \Phi(m^*)}{\pi \sqrt{2|\Phi_{mm}(m^*)|}}  \int_{-\infty}^\infty  \exp \left(-i n^2 \right) \dd n. \label{Bint2}
\eeq
These are Fresnel integrals and can be evaluated exactly:
\begin{gather}
\int_{-\infty}^\infty  \exp \left(-i n^2 \right) dn = \sqrt{\pi} \text{e}^{ -i\pi/4}. 
\end{gather}
Consequently,
\beq
B(z,t) \approx    \frac{  \hat{B}_0(m^*)  \exp  i (\Phi(m^*)-\pi/4)}{\sqrt{2 \pi |\Phi_{mm}(m^*)|}} . \label{Bint3}
\eeq 
To obtain an expression for $\hat{B}_0(m^*)$ we Fourier-transform the wave initial condition \eqref{wave_IC}:
\begin{align}
\widehat{\LL A}_0 & = u_0 \int_{-\infty}^\infty \exp{\left( -(z/\sigma)^2 - imz\right)} \dd z  \nonumber
\\ & =  u_0 \sigma \sqrt{\pi} \exp \left( -m^2 \sigma^2/4\right) 
\end{align}
Then, we can evaluate $\hat{B}^0(m^*)$, which is  $-(N/fm)^{2} \widehat{\LL A}_0 $ evaluated at $m^*$. We plug back all expressions into the solution \eqref{Bint3} and bring back the $x$-dependent part of the solution:
\beq
A(x,z,t) \approx  \frac{\sqrt{2} u_0 N \sigma}{\zeta_{x} (ft)^{\tfrac{3}{2}} } \exp \left[ i (\Phi^* - \tfrac{\pi}{4} - \tfrac{\zeta }{2}t) - \tfrac{m^{*2} \sigma^2}{4}\right], \label{full_sol}
\eeq
where $\Phi^* \defn \Phi(m^*)     = \tfrac{3}{2} m^* z$.
Finally, we can apply the $\LL$ operator on \eqref{full_sol} to get the solution in terms of the back-rotated wave velocity, 
\beq
\LL A= F(x,z,t)  \exp \left[ i (\Phi^* - \tfrac{\pi}{4} - \tfrac{\zeta }{2}t) - \tfrac{m^{*2} \sigma^2}{4}\right], \label{la_sol} 
\eeq
with
\begin{align}
F(x,z,t) &= - \frac{u_0 \sigma}{\sqrt{6z m^{*3}}} \left[ \Gamma_{zz} + \left( \Gamma_z \right)^2 \right]  \label{la_amp}
\end{align}
Above, $\Gamma$ is the bracketed expression in \eqref{la_sol}, whose derivatives are
\begin{align}
\Gamma_z &= i m^*  +  \frac{1}{6} \frac{\sigma^2 m^{*2}}{z} , \\
\Gamma_{zz} &= - i \frac{m^*}{3z} - \frac{5}{18} \left( \frac{\sigma m^*}{z} \right)^2.
\end{align}


One can predict the wavenumbers characterizing the monochromatic leading-order solution by differentiating the phase of \eqref{la_sol}:
\begin{align}
k  &\defn \frac{\partial}{\partial x} \left(\Phi^* - \tfrac{\zeta}{2} t \right) = - \tfrac{1}{2} \zeta_{x} t + \frac{\zeta_{xx}}{\zeta_{x}} m^* z  \approx  - \tfrac{1}{2} \zeta_{x} t\com   
\label{k_pred}   \\
m & \defn \frac{\partial}{\partial z} \left(\Phi^* - \tfrac{\zeta}{2} t \right)  = m^* = \left( \frac{N^2 \zeta_{x}^2 }{12 f |z|} \right)^{\frac{1}{3}} t\per
 \label{m_pred}
\end{align}
At the jet center, $\zeta_{xx}=0$ and we recover the heuristic predictions, \eqref{kt} and \eqref{mt}, except that here $\zeta_{x}$ may be an arbitrary, albeit slowly-varying function of $x$. 

\section*{Appendix II: Estimating wavenumbers}

\noindent We express the back-rotated velocity as 
\beq
\LL A = R \text{e}^{i\theta}, \label{decomp}
\eeq 
where $R$ and $\theta$ are both real. For any space or time variable $\alpha$,
\beq
\frac{\LL A_\alpha}{\LL A} = \frac{R_\alpha}{R} + i \theta_\alpha \quad \Longrightarrow \quad \theta_\alpha =  \Im \left( \frac{\LL A_\alpha}{\LL A} \right). \label{feff}
\eeq
This rule permits the calculation of local Eulerian frequency and wavenumbers at every point in space and time, which in the WKB framework are defined as:
\beq
\omega \defn - \theta_t, \qquad \bk \defn \theta_\mathbf{x}.
\eeq

\bibliographystyle{ametsoc2014}
\bibliography{bibliography}



\end{document}